\documentclass[]{aa}
\usepackage{graphicx}
\usepackage{multirow}
\usepackage{color}
\usepackage{booktabs}      
\usepackage[varg]{txfonts}

\newcommand{\kms}{km\,s$^{-1}$}
\newcommand{\rstar}{R$_{\star}$}

\begin{document}
\title{An observational study of dust nucleation in Mira ($o$\,Ceti):}
\subtitle{II. Titanium oxides are negligible for nucleation at high temperatures}

\author{T. Kami\'nski\inst{\ref{inst1},\ref{inst2}}, H.S.P. M\"uller\inst{\ref{inst3}}, M.R. Schmidt\inst{\ref{inst4}}, I. Cherchneff\inst{\ref{inst5}}, K.T. Wong\inst{\ref{inst6}}, \\
         S. Br{\"u}nken\inst{\ref{inst3}}, K.M. Menten\inst{\ref{inst6}}, J.M. Winters\inst{\ref{inst7}}, C.A. Gottlieb\inst{\ref{inst2}}, N.A. Patel\inst{\ref{inst2}}
        }
\institute{\centering 
            ESO, Alonso de Cordova 3107, Vitacura, Casilla 19001, Santiago, Chile, \email{tomasz.kaminski@cfa.harvard.edu}\label{inst1},
       \and Harvard-Smithsonian Center for Astrophysics, 60 Garden Street, Cambridge, MA, USA \label{inst2}
       \and I. Physikalisches Institut, Universit\"at zu K\"oln, Z\"ulpicher Strasse 77, 50937 K\"oln, Germany \label{inst3} 
       \and Nicolaus Copernicus Astronomical Center, Polish Academy of Sciences, Rabia{\'n}ska 8, 87-100 Toru\'n \label{inst4}
       \and Departement Physik, Universit\"at Basel, Klingelbergstrasse 82, 4056, Basel, Switzerland\label{inst5}
       \and Max-Planck-Institut f\"ur Radioastronomie, Auf dem H\"ugel 69, 53121 Bonn, Germany\label{inst6} 
       \and IRAM, 300 rue de la Piscine, Domaine Universitaire de Grenoble, 38406, St. Martin d'H{\'e}res, France \label{inst7}
       }


\date{Received; accepted}
\abstract 
{The formation of silicate dust in oxygen-rich envelopes of evolved stars is thought to be initiated by formation of seed particles that can withstand the high temperatures close to the stellar photosphere and act as condensation cores farther away from the star. Among the candidate species considered as first condensates are TiO and TiO$_2$.}
{We aim to identify and characterize the circumstellar gas-phase chemistry of titanium that leads to the formation of solid titanium compounds in the envelope of $o$\,Ceti, the prototypical Mira, and seek an observational verification of whether titanium oxides play a major role in the onset of dust formation in M-type asymptotic-giant-branch (AGB) stars.}
{We present high angular-resolution (145\,mas) ALMA observations at submillimeter (submm) wavelengths supplemented by APEX and {\it Herschel} spectra of the rotational features of TiO and TiO$_2$. In addition, circumstellar features of TiO and \ion{Ti}{I} are identified in optical spectra which cover multiple pulsation cycles of $o$\,Ceti.} 
{The submm ALMA data reveal TiO and TiO$_2$ bearing gas within the extended atmosphere of Mira. While TiO is traceable up to a radius (FWHM/2) of 4.0 stellar radii (\rstar), TiO$_2$ extends as far as 5.5\,\rstar\ and unlike TiO appears to be anisotropically distributed. Optical spectra display variable emission of \ion{Ti}{I} and TiO from inner parts of the extended atmosphere (<3\,\rstar).}
{Chemical models which include shocks are in general agreement with the observations of gas-phase titanium-bearing molecules.  It is unlikely that substantial amounts of titanium is locked up in solids because the abundance of the gaseous titanium species is very high. In particular, formation of hot titanium-rich condensates is very improbable because we find no traces of their hot precursor species in the gas phase. It therefore appears unlikely that the formation of dust in Mira, and possibly other M-type AGB stars, is initiated by titanium oxides.}

        
\titlerunning{Ti-bearing gas and dust in Mira}
\authorrunning{Kami\'nski et al.}
\maketitle
\section{Introduction}\label{sec-intro}
Asymptotic giant branch (AGB) stars lose a significant part of their mass through stellar winds that are rich in dust \citep[e.g.][]{gehrz,agbdust}. The dust enters the interstellar medium where it is processed and later becomes an important ingredient of the material from which new stars and planets form. In a galaxy like ours, AGB stars contribute the most mass to the overall dust production, mostly in the form of carbonaceous and silicate dust \citep[e.g.][]{ferrarotti}. The silicate-dust production is initiated in oxygen-rich (O-rich) stars by formation of condensates of small metal oxides. Based on infrared interferometric observations, the transition from molecular to solid form must occur at distances of $\lesssim$2 stellar radii \citep[e.g.][]{danchi,norris,karovicova,zhao}. Because temperatures in this region are high (roughly 1000--2500\,K), these first condensates must be formed from very refractory species. In most theoretical studies attempting to explain the nucleation process it has been assumed that circumstellar molecules and dust form under chemical- and thermal-equilibrium (TE) conditions \citep{GS86,GS98,tielens91,jeong}. These models indicate that the dust production is likely taking place through \emph{heterogeneous nucleation}. In this scenario, the first hot condensates are produced from inorganic material other than silicates. At lower temperatures, roughly <600\,K, where most of the mass of silicate dust is build up, they act as condensation cores ({\it seeds}) that make the grain growth more efficient \citep[e.g.][]{hofner}. Likely candidates for the first condensates are titanium oxides (mainly TiO$_2$) and alumina dust (Al$_2$O$_3$) \citep[e.g.][]{GS98,jeong}. However, because atmospheres of AGB stars are affected by shock waves triggered by pulsations, it is questionable whether the equilibrium conditions can prevail in the dust-formation regions \citep{willacy,cherchneff2006}. In consequence, it has been unclear if the heterogeneous nucleation indeed takes place around pulsating AGB stars. Since recently, however, the role of titanium- and aluminum- oxides as seeds can be tested observationally. The gas-phase precursors of these solids can now be observed directly by optical, infrared, and millimeter/submillimeter (mm/submm) spectroscopy. In \citet{kami_mira1} (hereafter Paper\,I), we performed a first observational verification of whether aluminum-bearing  species that are necessary precursors of alumina seeds are present in O-rich AGB stars. Although a full quantitative analysis was not possible, all observations of the prototypical M-type AGB star, $o$\,Ceti, are consistent with aluminum depletion within 2--4 stellar radii. However, not all Al is consumed from the gas phase in the dust-formation region and the Al depletion is an anisotropic process. In this paper, we focus on the role of titanium oxides in the formation of dust at high temperatures.

The important role of titanium oxides, particularly of TiO$_2$, for inorganic-dust formation has been chiefly promoted on theoretical grounds \citep[e.g.][]{GS98,jeong,jeong1999}. There is also observational evidence that Ti-species are involved in nucleation. Studies of presolar grains show that titanium oxide grains, although scarce, form in AGB winds \citep[e.g.][]{Nittler2008}. It has been also long known that titanium is considerably depleted in the interstellar medium presumably because it is effectively locked into dust grains \citep{stokes,churchwell}. Titanium oxides are also considered to play an important role in the condensation of solids in the atmospheres of planets and brown dwarfs, although in different temperature and pressure regimes than prevailing around evolved stars \citep[e.g.][]{BS99,lee}.

On the other hand, the role of titanium oxides in the seed formation process has often been questioned, mainly because titanium is not an abundant metal. It is 35 times less abundant than aluminum whose production of refractory alumina dust is an alternative scenario to Ti seeds. Very efficient nucleation of TiO$_2$ would be necessary for titanium oxides to be responsible for the observed dust-production rates in O-rich evolved stars \citep{jeong1999}. Moreover, solid titanium compounds have not been conclusively identified in stellar spectra although optical constants have been measured for a number of them, e.g. TiO$_2$, CaTiO$_3$, and MgTiO$_3$ \citep{posch,tamanai}. Studies of presolar grains do not identify titanium compounds as forming the grain cores that would be expected in a heterogeneous condensation sequence \citep{stroud,Nittler2008}. Moreover, \citet{kami_tio} showed that Ti condensation is inefficient in some circumstellar environments. They identified emission features of \ion{Ti}{i}, TiO, and TiO$_2$ in the spectra of the red supergiant VY\,CMa and found very high abundances of the gas phase species at large distances from the star. These results are in stark contrast with the nucleation scenario in which titanium oxides play a major role in the dust production close to the star \citep[see also][]{beck}. It was realized, however, that the environment of VY\,CMa may be too violent to be considered representative for the majority of Galactic dust-producers, such as AGB stars \citep{kami_tio}.    

It was recently proposed that the onset of dust formation may be controlled solely by SiO \citep{gailSiO,gail2016}. Revised laboratory data extrapolated to conditions characteristic of the upper atmosphere indicate that SiO seeds can grow under equilibrium conditions at higher temperatures (but still $\lesssim$800\,K) than it was assumed when the heterogeneous nucleation scenario was first introduced \citep[][and references therein]{gailSiO}. Models based on chemical kinetics show, however, that the formation of dust is induced by non-equilibrium processes pertaining to the shocked gas and that dimerization of SiO under non-equilibrium conditions is very slow \citep{cherchneff2006,cherchneff12,gobrecht,bromley2016}. Observational test of the heterogeneous nucleation is necessary.

In a parallel approach to Paper\,I \citep[see also][]{kami_tio}, we attempt here to characterize all accessible gas-phase titanium-bearing species in the envelope of $o$\,Ceti, to test whether they are efficiently depleted into solids and whether they are the likely candidates of the first condensates in the winds of M-type stars. An introduction to the nucleation process, a description of the structure of a circumstellar envelope of an AGB star, and a characterization of our primary target, Mira, are given in Paper\,I and are not repeated here. In Sect.\,\ref{sec-species}, we first review all known Ti tracers and identify species of our main interest. The observational material is briefly presented in Sect.\,\ref{sec-obs} and includes millimeter/submillimeter-wave (submm/mm) data (Sect.\,\ref{sec-obs-submm}) and multi-epoch optical spectra (Sect.\,\ref{sec-opt}). To provide a context to these observations, in Sect.\,\ref{sec-env} we briefly analyze the state and structure of Mira and its extended atmosphere at the time of the observations. Then, in Sect.\,\ref{sec-mm}, the Ti-bearing gas is characterized based on submm/mm features of TiO and TiO$_2$. In the following Sect.\,\ref{opt}, we identify and analyze optical signatures of Ti-bearing gas, including those of \ion{Ti}{II}, \ion{Ti}{I}, TiO, and TiO$_2$. The relevance of the results on the chemistry and dust-nucleation in M-type AGB stars is discussed in Sect.\,\ref{discussion}.

\subsection{Ti-bearing species in circumstellar envelopes}\label{sec-species}
The first solid particles based on titanium oxides are thought to be TiO$_2$ (titania), and can be followed by formation of higher oxides, including Ti$_2$O$_3$, Ti$_3$O$_5$, and Ti$_4$O$_7$. Other titanium compounds, e.g. CaTiO$_3$ (perovskite) and MgTiO$_3$, are also potentially produced by evolved stars \citep{sharp,GS98,plane2013}. However, the  formation of solids from  TiO$_2$ clusters is considered the simplest plausible scenario for the grain growth, in particular because TiO$_2$ is thought to be present in the gas phase. TiO$_2$ has several natural solid forms, e.g. crystalline anatase, brookite and rutile. Those solids are formed directly out of the gaseous TiO. The solid form of TiO also exists and can be produced in a reaction of H$_2$ with gaseous TiO$_2$. The TiO$_2$ gas, in turn, forms directly from TiO in a reaction with OH \citep{plane2013,gobrecht} and with water \citep{GS98}, while TiO is formed from atomic titanium, \ion{Ti}{I}. Atomic titanium can exist in neutral and ionic forms in the atmospheres and envelopes of cool evolved stars. In chemical equilibrium, the ionization fraction is mainly dependent on the photospheric temperature \citep[e.g.][]{tsuji} but in real pulsating stars this fraction can be strongly influenced by the shock strength and the level of penetration by the shock-induced ultraviolet radiation. The equilibrium chemistry of the Ti-bearing species is discussed in great detail in \citet{GS98}.

In order to trace the different chemical forms of titanium in the gas phase, we aim here at identifying and characterizing features of atomic titanium and the two simplest oxides, TiO and TiO$_2$. The electronic bands of TiO have long been known in the photospheres of late-type O-rich stars but only a few astronomical objects manifested the presence of circumstellar TiO-bearing gas in those bands \citep[][and references therein]{kami_tio}. In Sect.\,\ref{opt}, we attempt to identify all forms of circumstellar Ti-bearing species in the optical spectra of Mira. Millimeter/submillimeter observations of circumstellar TiO and TiO$_2$, through their pure rotational lines, are also very rare and have become possible only recently \citep{kami_tio}. We successfully identify the rotational spectra of both species in Mira in Sect.\,\ref{sec-mm}. In addition, TiO has its fundamental vibration bands near 10\,$\mu$m which so far have been identified in one S-type AGB star, NP\,Aur \citep{smolders}. Owing to their extreme rarity, the mid-infrared bands do not seem to be a good tracer of TiO in stars. Additionally, in observations that do not spatially resolve the extended atmosphere of Mira, the TiO feature would be completely buried under the much stronger silicate feature. We therefore do not attempt to trace TiO in these infrared bands.      

In addition to the oxides and atomic forms, among other possible carriers of Ti in envelopes of O-rich stars are the hydride, TiH, and simple compounds of other abundant metals, e.g. TiN, TiS, TiF, and TiF$_2$ \citep[cf.][]{tsuji}. In M-type AGB stars and in TE, their abundances are expected to be orders of magnitude less than TiO, and observations of their spectral features were considered to be very unlikely. In particular, TiH is not expected to be present under TE conditions in the low-pressure environments of AGB stars but non-equilibrium chemistry might produce traceable quantities. In Paper\,I, we reported a tentative discovery of electronic absorption bands of AlH, a molecule also not expected in TE, implying that TiH is more likely. Also, TiH is a non-negligible carrier of Ti in dwarf stars of spectral types M and later, especially those with low metallicity \citep{TiH}. Electronic bands of TiS were observed in a near-infrared spectrum of one S-type AGB star \citep{TiSopt}. Although they are not expected to be important for dust formation, we included these species for completeness.

Titanium has five stable isotopes, $^{46,47,48,49,50}$Ti, whose solar and terrestrial relative abundance ratios are 11.2:10.1:100.0:7.3:7.0, respectively \citep{lodders,composition_elements_2009}. Most spectroscopic studies to date have focused on the dominant isotope, $^{48}$TiO (hereafter simply TiO). Deriving observational constraints on the abundance of the rare isotopologues based on electronic bands is very challenging \citep{Lambert1977} and the only attempt to derive them in Mira resulted in upper limits that are generally consistent with the terrestrial/solar values \citep{Wyckoff1972}. A tentative observation of an enhanced abundance of $^{50}$TiO was reported for Mira in \citet{Lambert1977} but has never been confirmed. Such an enhancement is virtually possible in AGB stars through the slow-neutron capture process ($s$-process) and the third dredge-up \citep[e.g.][]{sprocessTi}. The $s$-process is expected to enhance the amounts of the rare isotopes relative to $^{48}$Ti resulting in: a very modest overabundance of $^{46}$Ti; considerable overabundance of $^{47}$Ti and $^{49}$Ti; and largest increase of $^{50}$Ti \citep{clayton2003,Howard1972,sprocessTi}. Considering that even in unprocessed matter the rare isotopes add up to $\sim$24\% of the total titanium content, they are important tracers of Ti-bearing species and therefore are included in our study as well. In Sect.\,\ref{sec-lab}, we describe the spectroscopic information required to identify these species at the submm wavelengths. 

\section{Observations}\label{sec-obs}

\subsection{Submillimeter/millimeter and FIR data}\label{sec-obs-submm}
Most of the mm--far-infrared observations used in our analysis have been presented in detail in Paper\,I. An extensive search for Ti-bearing species was performed on all available data from Atacama Pathfinder Telescope (APEX), Atacama Large submillimeter/Millimeter Array (ALMA), and Herschel Space Observatory ({\it Herschel}). Below we briefly describe only the most crucial data-sets.

\begin{table*}
\caption{ALMA observations used in this work.}\label{tab-ALMAdata}
\centering\small
\begin{tabular}{cc cc cc}
\hline
Date            & PI       & Frequency ranges                    & Resolution & Angular    & Reference \\
obs.            &          & (GHz)                      & (\kms) & resolution &   \\
\hline
24 Feb 2014     & Ramstedt & 330.2--334.0, 342.3--346.1 &  0.4   & 1\farcs6$\times$0\farcs5 & \citet{ramstedt} \\ 
03 May 2014     & Ramstedt & 330.2--334.0, 342.3--346.1 &  0.4   & 0\farcs5$\times$0\farcs4 & \citet{ramstedt} \\
12--15 Jun 2014 & Planesas & 330.4--330.7, 345.6--345.9 &  0.1   & 0\farcs3$\times$0\farcs3 & \citet{planesas} \\
                &          & 331.1--332.8, 343.7--345.5 & 13.6   & & \\           
21 Jul 2015     &Kami\'nski& 342.2--345.9, 354.2--357.9 &  0.9   & 0\farcs16$\times$0\farcs13 & Paper\,I\\[5pt]
16 Jun 2014 & Planesas & 679.11--682.77 &  6.9   & 0\farcs23$\times$0\farcs16 & \citet{planesas} \\
  
\hline
\end{tabular}
\end{table*}

\begin{itemize}
\item Observations carried out with APEX in 2013--2015 utilized the FLASH$^+$ \citep{flash} and SHeFi-1 \citep{shfi} receivers to cover multiple spectral ranges between 222 and 500\,GHz (see Table\,A.1 in Paper\,I). Several spectral setups were arranged to target lines of TiO and resulted in serendipitous discovery of TiO$_2$ features. The observations were obtained with telescope beams of 13\arcsec--28\arcsec\ FWHM.
\item One APEX spectral setup, centered on 222.7\,GHz and first observed in December 2013, was re-observed on 1, 5, and 7 August 2016, i.e. after the publication of Paper\,I. The combined data of 2013 and 2016 resulted in a nominal rms noise level of 1.1\,mK ($T_A^*$) per 3.8\,MHz. The total observation time on source was 7.5\,h. Those observations were designed to detect two $J$=7--6 lines of TiO.    
\item All archival spectra obtained with the HIFI instrument \citep{hifi} on {\it Herschel} were examined, including those of the HIFISTARS key programme (PI: V. Bujarrabal) and Performance-Verification data. The spectra cover narrow frequency ranges between about 556.6\,GHz and 1902.4\,GHz, however spectra at the longest wavelengths were obtained with a very low sensitivity. The telescope beam is in the range 20\arcsec--38\arcsec\ FWHM for the part of data used here.
%
\item Mira was observed by ALMA in Band\,7 on four different observing runs that are summarized in Table\,\ref{tab-ALMAdata}. The ALMA data obtained on 21 July 2015 are our primary source of information. The complex gains of those interferometric observations were (self-)calibrated on the strong continuum sources of the Mira A/B binary system. This procedure resulted in a nominal sensitivity of about 1.7\,mJy/beam per 0.97\,\kms, making them the most sensitive submm observations of Mira to date.   
\item We found TiO and TiO$_2$ features in a Mira spectrum obtained with ALMA in Band\,9. Those observations are described in detail in \citet{planesas} (see also Table\,\ref{tab-ALMAdata}).
%
\end{itemize} 

In Paper\,I, we used ALMA Band\,6 data obtained within the Long Baseline Science Verification Campaign \citep{alma}. Their angular resolution of $\sim$30\,mas is the highest among all mm/submm data available for this source. The data do not cover any TiO or TiO$_2$ features at a sufficiently high signal-to-noise ratio (S/N) to be analyzed here. Our best angular resolution of $\sim$145\,mas was achieved in the Band\,7 data from 2015 in which the stellar photosphere is not resolved. Many features of interest, however, are observed at a very high S/N allowing size measurements even for regions smaller than the beam size.

Interferometric data were processed and imaged in CASA 4.6 \citep{casa} in the same way as described in Paper\,I. In particular, all spectral data were processed in CLEAN after continuum subtraction in the $u\varv$ domain (cf. Appendix\,B of Paper\,I). 

\subsection{Optical spectroscopy}\label{sec-opt}
We use the optical spectra presented in Paper\,I. These are thirty high-resolution ($R$>5000) observations extracted from archives and covering the time span from December 1965 to August 2012.  They come from a dozen of different observatories and instruments. We supplement these Paper\,I data with spectra obtained at the Mercator telescope and kindly provided to us by H. van Winckel. All spectra were acquired with the high-resolution mode (HRF) of the HERMES instrument \citep{hermes} resulting in the coverage 377--900\,nm at $R\!\sim$\,85\,000. Twelve epochs were covered between July 2009 and August 2014, but most of the data come from 2009. In Table\,\ref{tab-mercator}, we list the observation dates and the corresponding visual phases and visual magnitudes calculated in the same way as in Paper\,I. The exposure times were in the range 6--900\,s which, combined with different variability phases, resulted in a very broad range of signal-to-noise ratios. All the HERMES spectra were reduced with the default pipeline. 

Most of the collected spectra are not calibrated in flux. The spectra spatially correspond to a region of the size of the astronomical seeing on the given night but unfortunately the information about seeing is not available for most of the data. Even with those shortcomings, all the optical spectra collected by us for Mira constitute the richest and longest-spanning spectroscopic record of variability in a Mira star to date.

\section{The state and structure of Mira's atmosphere}  
\label{sec-env}

The main source of information on titanium oxides here is the dataset obtained with ALMA in 2015 at visual phase 0.21, i.e. close to the highest bolometric brightness, which occurs at $\varphi \approx 0.17$. At approximately this phase, the pulsation shock arises from below the photosphere and the star appears hottest. Characterized below different parts of Mira's envelope at this epoch. As in Paper\,I, we define R$_{\star}$=14.4\,mas (2.4$\cdot$10$^{13}$\,cm=331\,R$_{\sun}$) as an idealized typical optical/infrared stellar radius which we use here as a characteristic scale that is independent of the pulsation phase. Also following definitions introduced in Paper\,I (Sect.\,1.2), we refer here to the {\it extended atmosphere} as the region with a radius of 3--4\,\rstar\ where the pulsation shock strongly influences the kinematics of circumstellar gas; above this region is the {\it dust-formation zone} where the bulk of dust mass is built up from silicates and where the wind acceleration takes place; at about 10\,\rstar, the material is accelerated to the terminal velocity and we treat this region as the inner boundary of the stellar {\it wind}.     

\subsection{Continuum}
The continuum emission of Mira\,A at submm wavelengths is dominated by the radio photosphere \citep{RM97,RM07}. We measured the submm-continuum source by fitting models to calibrated visibilities. A Gaussian fit resulted in a FWHM of 33.3$\times$32.8 ($\pm$0.3) mas and a corresponding photospheric temperature of 4805\,K. However, a uniform disk is thought to better represent the radio photosphere \citep{Vle,lynn,wong}. Our data can be reproduced by an elliptical disk with a full size of 53.8$\times$49.5 ($\pm$0.1)\,mas at PA=--54.7\degr$\pm$0.6\degr, which corresponds to a brightness temperature of 2440\,K. This disk size and temperature are consistent, within 15\%, with those measured in long-baseline Band\,6 data and corresponding to phases 0.4--0.5 \citep{wong}. The large discrepancy in the derived brightness temperature, 4805 vs. 2440\,K, is mainly due to systematic errors and the unknown brightness distribution of the source. We assume here that a blackbody source of 2440\,K with a radius of $\langle r \rangle$=25.8\,mas=1.8\,\rstar\ underlies the circumstellar gas during the 2015 observations.  

\subsection{Gas in the extended atmosphere}
\begin{figure*}
\centering
\includegraphics[angle=270,width=0.8\textwidth]{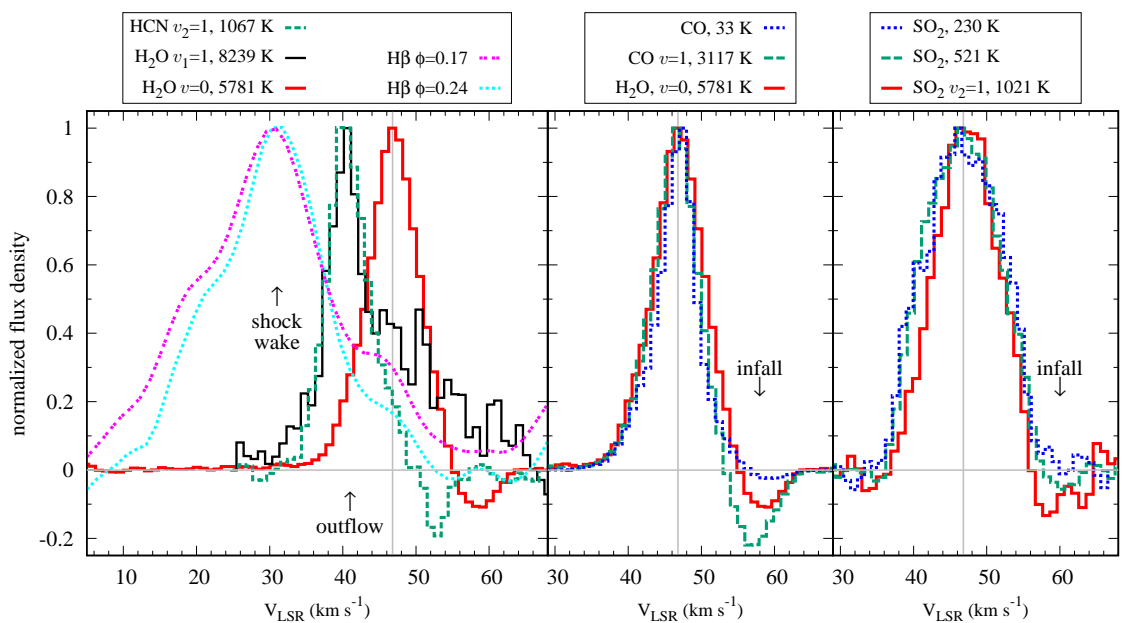}
\caption{Sample spectral profiles of transitions probing the atmosphere of Mira. The value of $E_u$ is for each transition in the legend. The lines were chosen to show features requiring a broad range of excitation temperatures. The center-of-mass velocity of Mira is marked with the vertical line. The panels cover different velocity ranges but are in scale. All molecular spectra were extracted within an aperture of about 0\farcs2 radius which almost fully encompasses the emission of TiO and TiO$_2$. 
{\bf Left:} High-excitation lines, H$_2$O 17$_{4,13} \to 16_{7,10}$, H$_2$O $12_{3,9}\to 11_{5,6}$ $\varv_3$=1$\to\varv_1$=1, and HCN $\varv_2$=1 $J$=4--3. The two latter lines are seen at a peculiar radial velocity of $\sim$40\,\kms\ and probe a region of the atmosphere which is affected by a shock. Also shown in this panel are two profiles of the H$\beta$ hydrogen recombination lines observed in 2010 at phases 0.17 and 0.24 bracketing the phase of the ALMA data, $\varphi$=0.21. The hydrogen lines probe the recombining wake of the strong shock. Their profiles are slightly affected by absorption in photospheric lines of TiO and their broad profile is a result of scattering. 
{\bf Center:} Transition from levels in a broad range of $E_u$ that show an absorption feature indicative of an infall. 
{\bf Right:} Lines of SO$_2$ of moderate and high excitation. The SO$_2$ profiles are wider (more parabolic) than of any other molecular species observed.}
\label{fig-sampleProfiles}
\end{figure*}

To gain a better understanding of the structure of the atmosphere and envelope of Mira during this phase, we investigated spectral features from molecules other than titanium oxides. The ALMA spectra cover over 140 transitions of several species (and their isotopologues): CO $\varv$=0 and 1; SiO $\varv$=0,1,2, and 3; SO $\varv$=0 and 1; SO$_2$ $\varv$=0 and $\varv_2$=1; AlO; SiS; HCN $\varv$=0 and $\varv_2$=1; and H$_2$O $\varv$=0 and $\varv_1$=1. (About a third of features remain unidentified but it is out of the scope of this paper to attempt a full identification.) The observed transitions cover a wide range of upper energy levels ($E_u$), from a few tens to thousands of K, probing parts of Mira's envelope with drastically different temperatures and densities. Example line profiles for species present in the innermost envelope are shown in Fig.\,\ref{fig-sampleProfiles}. 

Gas of low excitation is traced by pure emission lines which are spatially extended and centered at the stellar center-of-mass velocity, $V_{\rm LSR}$=46.8\,\kms. Generally, emission from gas with a temperature lower than 100\,K appears extended for our 145\,mas beam. The maps of emission arising in this low-excitation gas show the extended and complex wind that has been already described in detail \citep{ramstedt,wong,nhung}. We are more interested here in the inner regions of the envelope where the dust forms. These regions are traced by lines of moderate and high excitation which generally exhibit emission profiles centered at the stellar center-of-mass velocity (but see below). Additionally, low-excitation lines of very abundant species, such as CO $\varv$=0 $J$=3--2 with an upper energy level, $E_u$, of 33\,K and lines of SO$_2$ of moderate excitation \citep{SO2}, also trace these regions as they display inverse P-Cyg profiles indicative of material infalling onto the star (Fig.\,\ref{fig-sampleProfiles}, middle and right panels). The projected infall velocities are as high as 17\,\kms. While the emission components of the lines with low $E_u$ arise in a very extended envelope, the absorption must be produced very close to the star where the gas is densest and has a very high opacity \citep[cf.][]{pcyg}.


\paragraph{Absorption and infall in high-excitation gas}
The infalling part of the envelope is also seen in absorption in transitions requiring high excitation temperatures, for instance in CO $\varv$=1 $J$=3--2 \citep[$E_u$=3117\,K,][]{COvib} and H$_2$O $\varv$=0 17$_{4,13} \to 16_{7,10}$ \citep[$E_u$=5781\,K,][]{H2Ovib}. To produce absorption, this infalling gas has to have a temperature lower than the background continuum source. Our measured sizes of the line-absorbing regions from various species are smaller than or comparable to the continuum source, which agrees with our interpretation that the absorption originates from gas seen against the stellar disk. The physical radial extent of the gas cannot be directly measured at the angular resolution of our data. However, note that absorption in the two lines is only possible in a region where the rotational energy levels above a few thousand K are effectively populated. The radial extent of these regions is roughly the same as the angular size of the corresponding emission components. The beam-deconvolved sizes of 93$\times$77 ($\pm$3)\,mas of the CO emission and 82$\times$71 ($\pm$2)\,mas for H$_2$O set the upper limits on the size of the region dominated by infall motions. They correspond to a radius of about 5.5\,$R_{\star}$.  

\paragraph{Emission from the extended atmosphere}
When a region of about 0\farcs2 radius is analyzed\footnote{This size corresponds to the entire extended atmosphere and encompasses the emission regions of titanium oxides (Sect.\,\ref{sec-obs-submm}).}, most of the emission features arise from the {\it extended atmosphere}. Their emission profiles show a variety of shapes. For instance, the relative strengths of emission and absorption components produce slightly different net emission profiles, as illustrated in the middle panel of Fig.\,\ref{fig-sampleProfiles}. Some differences are related to the intrinsic difference in abundance distribution. This is most apparent in profiles of SO$_2$ (right panel of Fig.\,\ref{fig-sampleProfiles}) which are broader than all other molecular lines observed.    

\paragraph{Recombining shock-wake}
Our optical spectra allow us to locate the shocked material. We examined the optical spectra obtained in 2010 for phases 0.17 and 0.24 and presented in Paper\,I (Appendix\,C; see also \citealp{narval}). These spectra are the closest in phase to the ALMA observations ($\varphi$=0.21). They cover the H$\beta$ Balmer line ($\lambda_{\rm lab}$=4861.35\,\AA) blueshifted by 16.0\,\kms, i.e. to $V_{\rm LSR}$=30.8($\pm$1.0)\,\kms. The recombination lines trace the material cooling after the passage of the supersonic pulsation shock and of a temperature of the order of 10$^4$\,K \citep{fox}. The lines are shown in Fig.\,\ref{fig-sampleProfiles} (left panel). We do not find any molecular species in the ALMA data that would correspond to this velocity. This is not surprising as molecules are thought to be effectively dissociated by the shock \citep{cherchneff2006,gobrecht}. The recombination lines at phases close to 0.2 are a signature of the most recent pulsation shock that just left the optical/infrared photosphere. Because the shock emerges from below the optical photosphere at a visual phase of about --0.2 \citep[e.g][]{Hinkle-COII,richter2001}, it has had enough time to travel since then to a radius of $\sim$2\,R$_{\star}$ at the epoch of ALMA observations ($\varphi=0.2$) and therefore was above the submm radio photosphere ($\langle r \rangle$=1.8\,\rstar).



\paragraph{Shock signatures in molecular gas}
Although molecules are absent in the recombining shock wake, our data show that molecular gas is affected by the shock. We identified transitions of H$_2$O and HCN that require excitation of their vibrational modes \citep{H2Ovib,HCNvib} and whose emission is centered at a peculiar radial velocity $V_{\rm LSR} \approx 40.3$\,\kms, i.e. blueshifted by 6.5\,\kms\ with respect to the stellar velocity. This velocity is exactly the same as that of rovibrational $\Delta \varv$=3 absorption lines of CO at the same pulsation phase of Mira \citep{Hinkle-COII}. These absorption lines at 1.6\,$\mu$m require an excitation temperature of $\sim$3700\,K at $\varphi$=0.2 and are thought to represent molecular material cooling after the passage of the shock in the deepest observable layers of the star \citep{Hinkle-COI,Hinkle-COII}. The lines we observed with ALMA at this velocity are seen in emission and likely represent even hotter (>3700\,K) gas closer to the active shock. Still, the temperatures must be below those at which H$_2$O and HCN are effectively dissociated and characteristic to the recombination wake. Two strongest of those lines, {\it para}-H$_2$O $12_{3,9}\to 11_{5,6}$ $\varv_1$=1$\to\varv_3$=1 with $E_u$=8239\,K and HCN $\varv_2$=1 $J=4\to3$ with $E_u$=1067\,K, are shown in Fig.\,\ref{fig-sampleProfiles}. The former line is a rovibrational transition\footnote{Its spectroscopic line strength originates from the vibration-rotation interaction of the $\varv_1$=1 and $\varv_3$=1 states.} and has the highest $E_u$ among all lines we were able to identify in the ALMA spectrum. We interpret the emission at the peculiar velocity as arising in gas located in the expanding part of the atmosphere and excited directly by the new shock, likely in the post-shock layer where molecules start to reform \citep[cf.][]{gobrecht}. The small size of the region of the H$_2$O $\varv_1$=1$\to\varv_3$=1 emission, (89$\pm$11)$\times$(43$\pm$18)\,mas, and its location at exactly the continuum center strongly support our interpretation. This is the first time the direct pulsation-shock signatures have been identified in submm observations of a Mira. It is also the first strong observational evidence that carbon-bearing species like HCN are formed by shock chemistry in M-type stars, as advocated in multiple theoretical studies \citep[e.g.][]{duari,cherchneff2006,gobrecht}. 

For our further discussions it is sufficient that we identified the shock-excited molecular region and defined its kinematical location. We do not aim to explain the complex shock phenomena in Mira's atmosphere in this paper. In the next sections, we associate the signatures of titanium bearing species with the different molecular regions that have been defined above. 

\section{Submillimeter lines of titanium oxides}\label{sec-mm}
\subsection{Line lists}\label{sec-lab}
In order to investigate the presence of the isotopologues of the titanium oxides at mm to FIR wavelengths, we created line lists based on spectroscopic data from different sources. 


Titanium monoxide is a radical with two unpaired electrons having a regular $^3\Delta$ ground electronic state, thus $\Omega$=1 is the lowest spin component and $\Omega$=3 is the highest one. TiO has a large dipole moment of $3.34 \pm 0.01$\,D in its ground electronic state \citep{TiO_dip_2003}. Predictions of the rotational spectrum of $^{48}$TiO were taken from the Cologne Database for Molecular Spectroscopy \citep[CDMS;][]{CDMS_1,CDMS_2}. The entry is based on the report by \citet{TiO_rot_1998}. We prepared entries of the minor isotopic species based on measurements of $^{46}$TiO, $^{48}$TiO, and $^{50}$TiO between 250 and 345\,GHz of \citet{TiO_TiO2_OSU_2008} and unpublished ones (Kania, in prep.). 

Titanium dioxide is a bent molecule with a $\tilde{X}^1 A_1$ electronic ground state and a large dipole moment of $6.33 \pm 0.07$\,D \citep{TiO2_A-X_dip_2009}. Predictions of the rotational spectra of $^{48}$TiO$_2$, $^{46}$TiO$_2$, and $^{50}$TiO$_2$ were taken from the CDMS. The entries are based on the mm-wave measurements and analysis of \citet{TiO2_rot_2011}, which included additional data of these isotopologues from Fourier transform  microwave (FTMW) measurements by \citet{TiO2_rot_2008}. Predictions for $^{47}$TiO$_2$ and $^{49}$TiO$_2$ were produced based on FTMW measurements of additional isotopologues from \citet{TiO2_OSU_2007} and unpublished ones (in prep.). 

In addition to the rotational lines of the oxides and their rare isotopologues, it is interesting to search for signatures of other Ti-bearing species mentioned in Sect.\,\ref{sec-species}. Accurate frequencies of rotational transitions are known for TiN, TiS, and TiF \citep{TiO_rot_1998,TiS,TiF} or can be predicted, as for TiH, with the spectroscopic constants. However, our sensitive observations did not cover strong transitions of any of these species.

\subsection{Identification of TiO, TiO$_2$, and their isotopologues}
Prior to this work, pure rotational lines of TiO and TiO$_2$ have only been known in the mm-wave spectrum of the red supergiant VY\,CMa \citep{kami_tio,beck}. Because Mira is only the second known source with those circumstellar molecules, we describe in detail the identification of the mm/submm features of TiO and TiO$_2$. 

\subsubsection{Identification of TiO}
A list of all the TiO transitions covered by APEX, ALMA, and {\it Herschel}, dates of observations, noise levels, and fluxes are given in Table\,\ref{tab-tio}. In all, we detected five different transitions of $^{48}$TiO. 

\begin{table*}
\caption{Transitions of TiO covered by APEX, {\it Herschel}, and ALMA.}\label{tab-tio}
\small
\centering
\begin{tabular}{cc cc cc cc c}
\hline\hline
Isoto-&$J_{\rm up}\,\to\,J_{\rm low}$,& $\nu$& Err($\nu$)\tablefootmark{a}&$S\mu^2$&$E_u$ &Dates of   &Vis. & Flux (1$\sigma$)\\
pologue &multiplet $\Delta_{\Omega}$    &(MHz) &(\kms)                      &(Debye$^2$)&(K)&observation&phase& (Jy\,\kms)\\
\hline\hline
\multicolumn{9}{c}{APEX}\\
\hline
$^{48}$TiO&~7$\to$\,6,\,$\Delta_1$&~221580.45 & 0.05 & 153.0&  41.0& 3--20 Dec. 2013, 1--7 Aug. 2016 & 0.5/0.35 & 0.89 (0.21)\\
          &~7$\to$\,6,\,$\Delta_2$&~224138.74 & 0.03 & 143.4& 180.7& 3--20 Dec. 2013, 1--7 Aug. 2016 & 0.5/0.35 & <0.60\\
          &11$\to$10,\,$\Delta_3$ &~355623.35 & 0.04 & 227.2& 386.2& 30 Jun.--6 Jul. 2014& 0.0 & 3.14 (0.73) \\
          &13$\to$12,\,$\Delta_3$ &~420224.14 & 0.04 & 274.7& 425.0& 13--14 Aug. 2013    & 0.1 & 9.67 (1.81)\\
          &15$\to$14,\,$\Delta_3$ &~484795.99 & 0.06 & 321.4& 470.0& 1--8 Jul. 2014      & 0.0 & 8.66 (1.50)\\
\hline                                
\multicolumn{9}{c}{\it Herschel}\\    
\hline                                
$^{48}$TiO&18$\to$17,\,$\Delta_1$& ~569537.72& 0.13 & 400.3& 258.2 &19 Jul. 2010 & 0.7& <22.57\\
          &18$\to$17,\,$\Delta_2$& ~575979.23& 0.13 & 396.5& 400.4 &19 Feb. 2010 & 0.3& <21.97\\
          &20$\to$19,\,$\Delta_3$& ~646070.05& 0.26 & 436.2& 609.6 &20 Jul. 2010 & 0.7& <28.20\\
          &25$\to$24,\,$\Delta_3$& ~807067.03& 0.75 & 549.9& 787.7 & 4 Feb. 2010 & 0.2& <30.80\\[2pt]
          &29$\to$28,\,$\Delta_3$&~935619.31& 1.48 & 640.2& 958.1& 10 Aug. 2011,28 Jan. 2012\tablefootmark{b}&0.9/0.4& <222.89\\[2pt]
          &34$\to$33,\,$\Delta_3$& 1095946.06& 2.98 & 752.8&1205.7 &20 Jul. 2010& 0.7 & 65.25 (18.31)\\
\hline                                
\multicolumn{9}{c}{ALMA}\\    
\hline                                
$^{50}$TiO&11$\to$10,\,$\Delta_1$ &~344722.34 &$>$0.02 & 240.9&~~97.8 & 12--15 Jun. 2014  & 1.0 & 0.31 (0.02) \\
$^{48}$TiO&21$\to$20,\,$\Delta_3$ &~678293.80 & ~~0.33 & 459.0& 642.1 &     16 Jun. 2014  & 1.0 &16.36 (1.00) \\[2pt] 
$^{46}$TiO&11$\to$10,\,$\Delta_2$ &~355975.66 &$>$0.02 & 239.8& 240.2 & 21 Jul. 2015& 0.2 & 0.54 (0.02) \\
$^{47}$TiO&11$\to$10,\,$\Delta_3$ &~357526.69 &$>$0.04 & 228.4& 386.8 & 21 Jul. 2015& 0.2 &<0.33 (0.02) \\
$^{48}$TiO&11$\to$10,\,$\Delta_3$ &~355623.35 & ~~0.04 & 227.2& 386.2 & 21 Jul. 2015& 0.2 & 4.44 (0.03) \\
$^{50}$TiO&11$\to$10,\,$\Delta_1$ &~344722.34 &$>$0.02 & 240.9&~~97.8 & 21 Jul. 2015& 0.2 & 0.39 (0.02) \\
\hline
\end{tabular}
\tablefoot{Spectroscopic data for $e$ and $f$ parity components were combined. The flux is given with a 1$\sigma$ error in the parenthesis. Upper limits on flux have a $3\sigma$ significance.
\tablefoottext{a}{The uncertainty of transition frequency expressed in velocity units; from CDMS.}
\tablefoottext{b}{Mapping observations. The examined spectrum is an average from two dates and located at an offset of 5\farcs2 (or 0.24 the beam width) from the actual stellar position.}
}
\end{table*}

\begin{figure}
\centering
\includegraphics[angle=0,width=0.99\columnwidth]{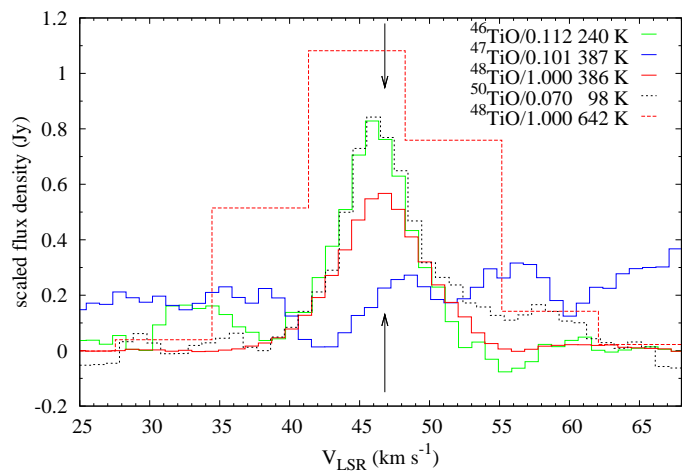}
\caption{Line profiles of TiO isotopologues observed with ALMA. One line of $^{50}$TiO observed in June 2014 is not shown because is covered by one resolution element only. The flux densities were scaled by their relative solar isotopic ratios. The arrows indicate the center-of-mass velocity of Mira. Spectra were extracted from regions encompassing the entire emission region.}\label{fig-TiOisoProfiles}
\end{figure}

In the APEX spectrum of Mira from 2013, we serendipitously detected a feature at 420.224\,GHz for which the most likely identification is the $\Delta_3$ $J$=13--12 line of TiO. As a verification of this identification, we next obtained with APEX a deep spectrum covering the $J$=7$\to$6 transition of $\Delta_1$ and $\Delta_2$ (221.580 and 224.139\,GHz). Both lines were observed earlier in VY\,CMa. The Mira observations were obtained in December 2013, i.e. two months after the observations of the feature at 420.224\,GHz. The deep integration resulted in a non-detection of the anticipated TiO $J$=7$\to$6 lines. Later in 2014, we obtained APEX observations covering two more lines, $\Delta_3$ $J$=11$\to$10 (at 355.623\,GHz) and $\Delta_3$ $J$=15$\to$14 (at 484.796\,GHz), which have $E_u$ values close to the $\Delta_3$ $J$=13$\to$12 line at 420.2\,GHz. Both were detected. An example APEX spectrum of TiO is shown in Fig.\,1 of Paper\,I. These discovery observations of TiO were later followed up with ALMA. Additionally, in 2015 we repeated observations of the two $J$=7$\to$6 lines with APEX and the $\Delta_2$ transition was detected at a 4.5$\sigma$ level in the combined 2013 and 2016 data. All the lines observed are very closely aligned in velocity and have a centroid position at 46.3$\pm$0.8\,\kms, which is consistent with the center-of-mass velocity of Mira.

The ALMA observations from 2015 were especially designed to observe the $\Delta_3$ $J$=11$\to$10 line of TiO and three additional transitions of the $^{46,47,50}$TiO isotopologues. The main reason to re-observe the $^{48}$TiO line (earlier covered by APEX) was to measure the size of the emission. All targeted lines were detected except the $\Delta_3$ $J$=11$\to$10 line of $^{47}$TiO. In addition, the ALMA observations in Bands 7 and 9 from June 2014 serendipitously covered one more transition of TiO, $\Delta_3$ $J$=21$\to$20 and one of $^{50}$TiO, $\Delta_1$ $J$=11$\to$10, which are observed at a high S/N but at a spectral resolution that is too poor to resolve the line profiles. The Band\,9 spectrum also covered one transition of $^{47}$TiO but it is buried in a strong line of SO\,$\varv$=1. The ALMA observations from Feb. and May 2014 also covered the $\Delta_1$ $J$=11$\to$10 transition of $^{50}$TiO but their combined sensitivity was not sufficient to detect the line. The spectrally-resolved lines observed with ALMA are shown in Fig.\,\ref{fig-TiOisoProfiles}. 

In searching for TiO emission at far-infrared wavelengths, we examined all available {\it Herschel}/HIFI scans. We found spectra covering six transitions of TiO but only one seems to be detected with a low confidence. Rest frequencies of TiO lines above $\sim$600\,GHz are more uncertain than at lower frequencies (cf. Table\,\ref{tab-ALMAdata}, column 4), making the identification in {\it Herschel} spectra more challenging. Nevertheless, we found a feature at the expected position of the 34$\to$33,\,$\Delta_3$ line whose flux is 3.5$\sigma$ above the local noise level (see Fig.\,1 in Paper\,I) and we consider it as tentatively detected.

Two transitions, 11$\to$10,\,$\Delta_3$ of TiO and 11$\to$10,\,$\Delta_1$ of $^{50}$TiO, were observed at multiple epochs and their fluxes remained unchanged within the observational uncertainties. The mm/submm data in hand do not indicate any substantial variability in the pure rotational lines of TiO.  


\subsubsection{Identification of TiO$_2$ emission}
Titanium dioxide has many rotational transitions with comparable intensity at submm wavelengths. Each individual line is rather weak and a sensitivity higher than that for TiO is required to observe TiO$_2$ emission \citep{kami_tio}. Nevertheless, we succeed in detecting the main isotopologue of TiO$_2$ and some of the rarer isotopic species.

The spectra of Mira collected with APEX covered several transitions of TiO$_2$. From these, only four lines are present at the expected frequencies, $J_{K_a,K_c}$=$19_{7,13} \to 19_{6,14}$, $15_{7,9} \to 15_{6,10}$, $38_{1,37} \to 38_{0,38}$, and $31_{5,27} \to 30_{4,26}$. The emission features are of a very modest S/N, 3--5. They are considered as tentatively detected and although not analyzed in detail here, were the stepping-stone to observing TiO$_2$ with ALMA.

\begin{figure*}
\centering
\includegraphics[angle=270,width=0.19\textwidth]{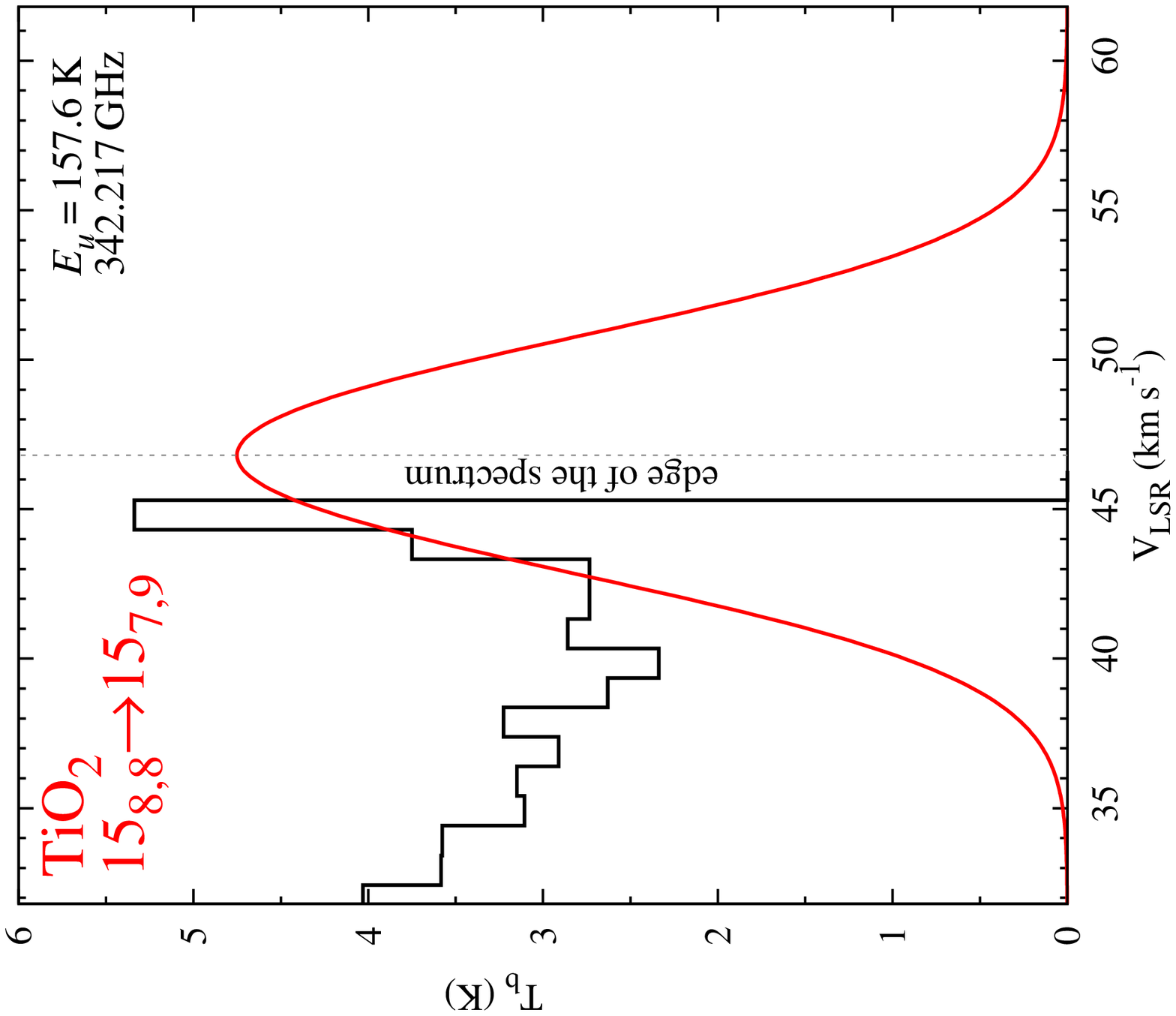}
\includegraphics[angle=270,width=0.19\textwidth]{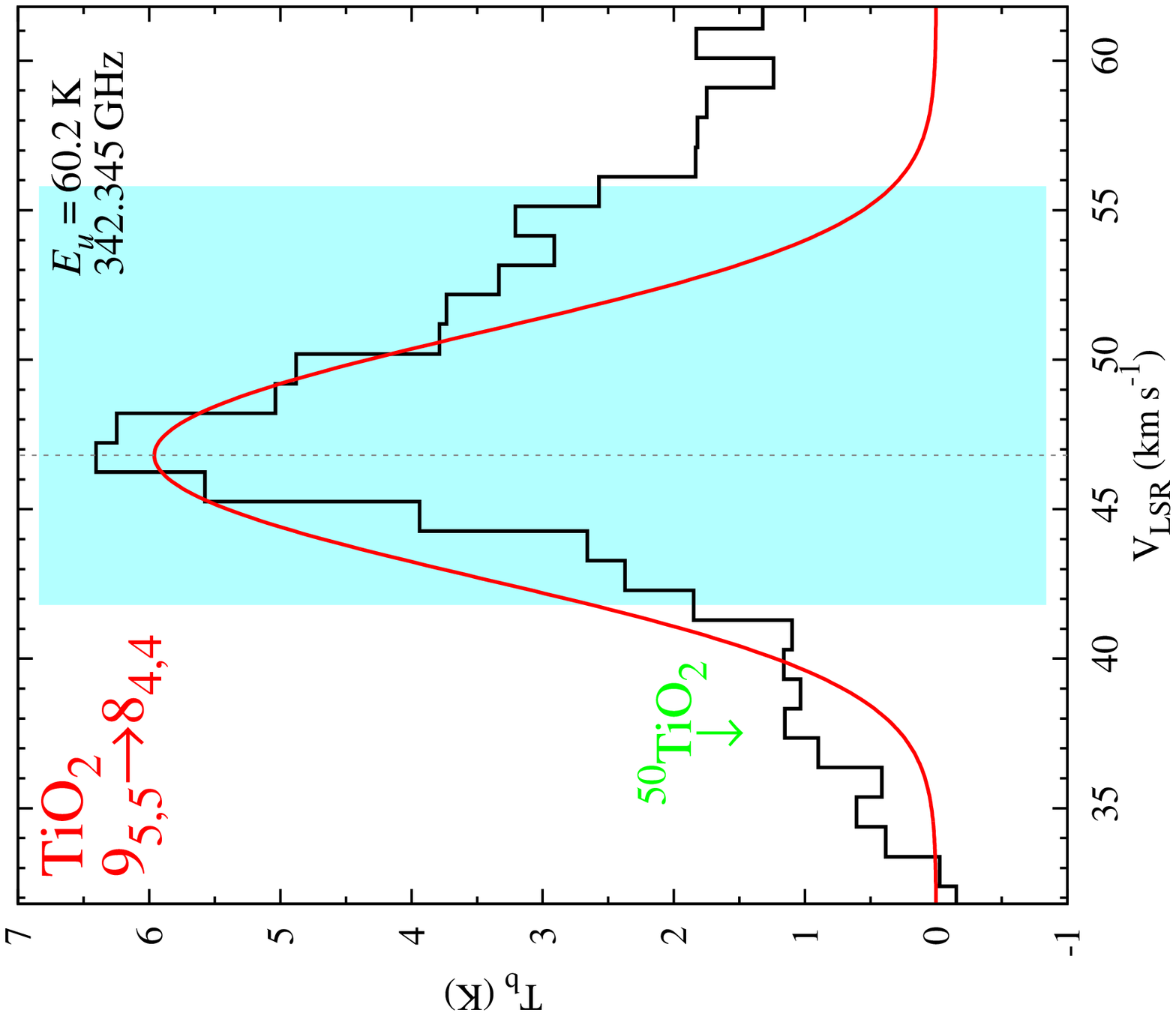}
\includegraphics[angle=270,width=0.19\textwidth]{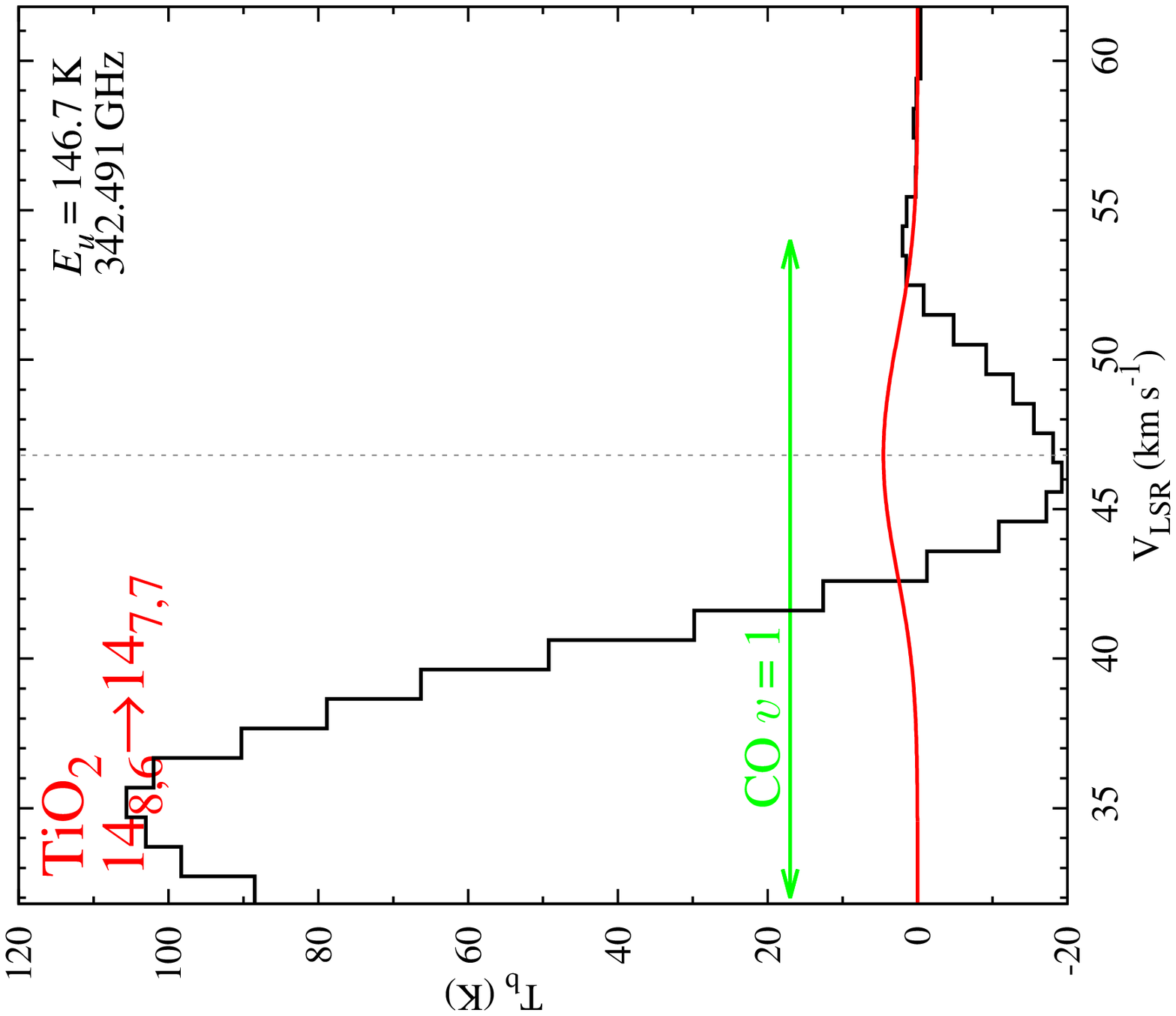}
\includegraphics[angle=270,width=0.19\textwidth]{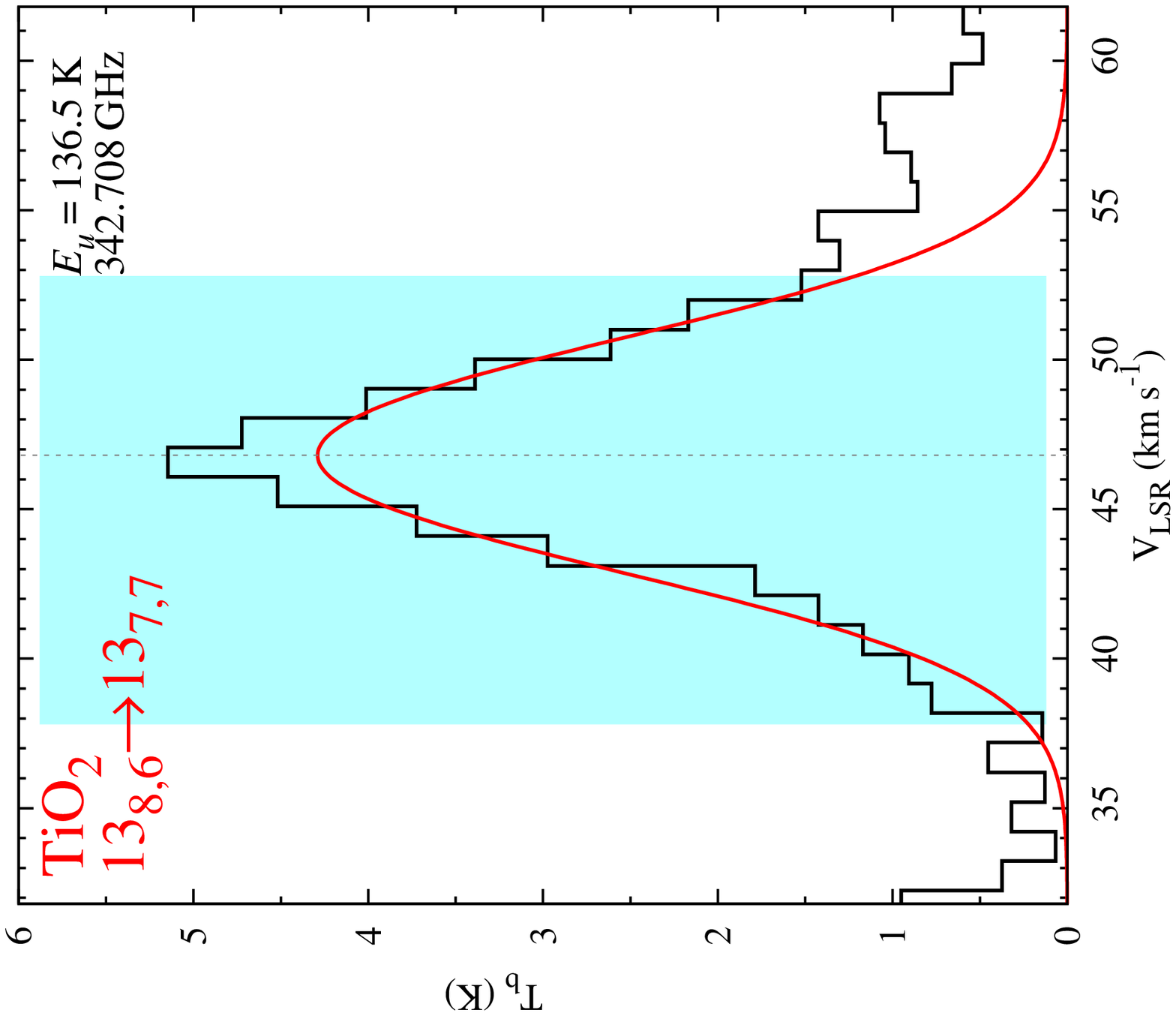}
\includegraphics[angle=270,width=0.19\textwidth]{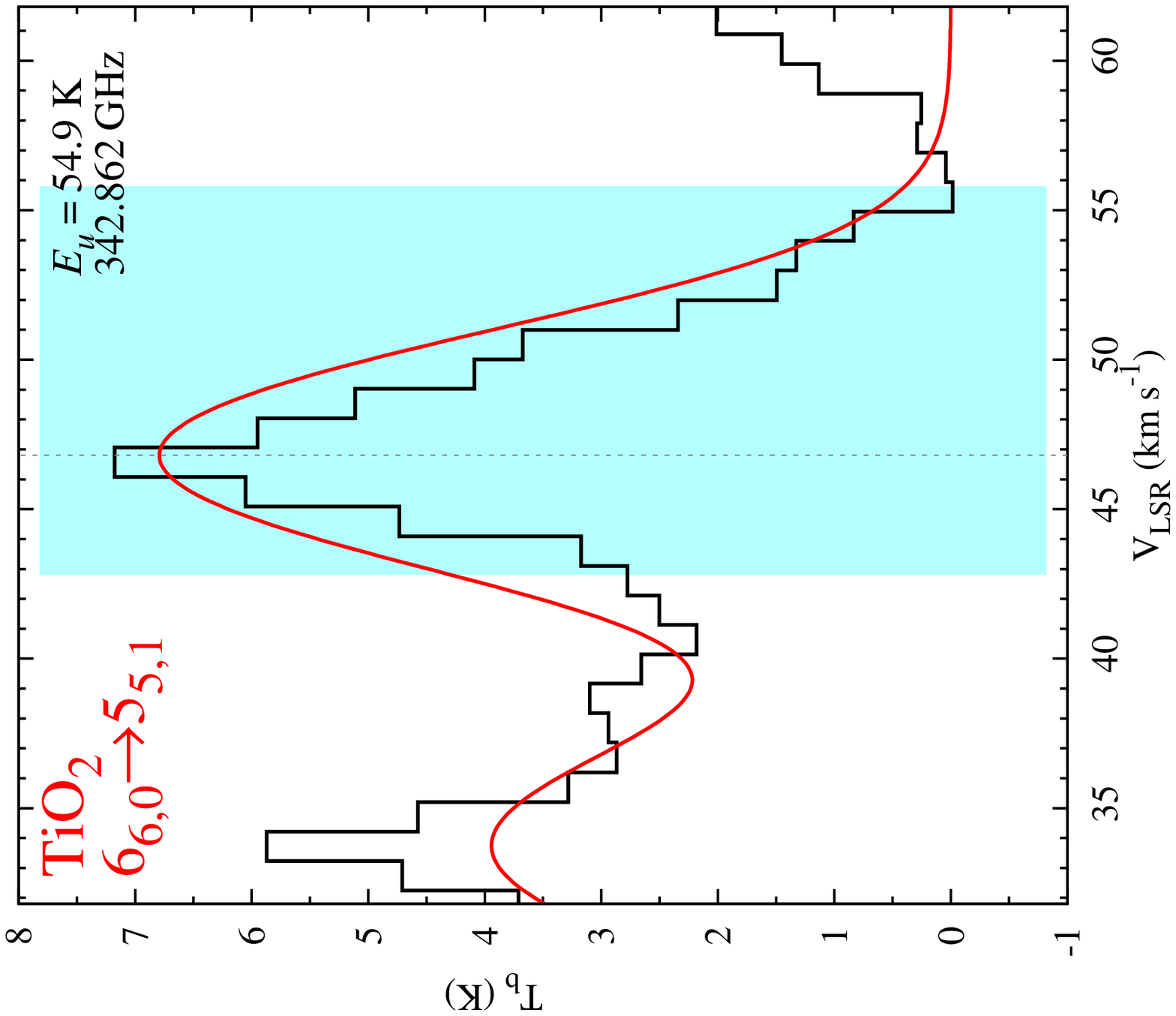}
\includegraphics[angle=270,width=0.19\textwidth]{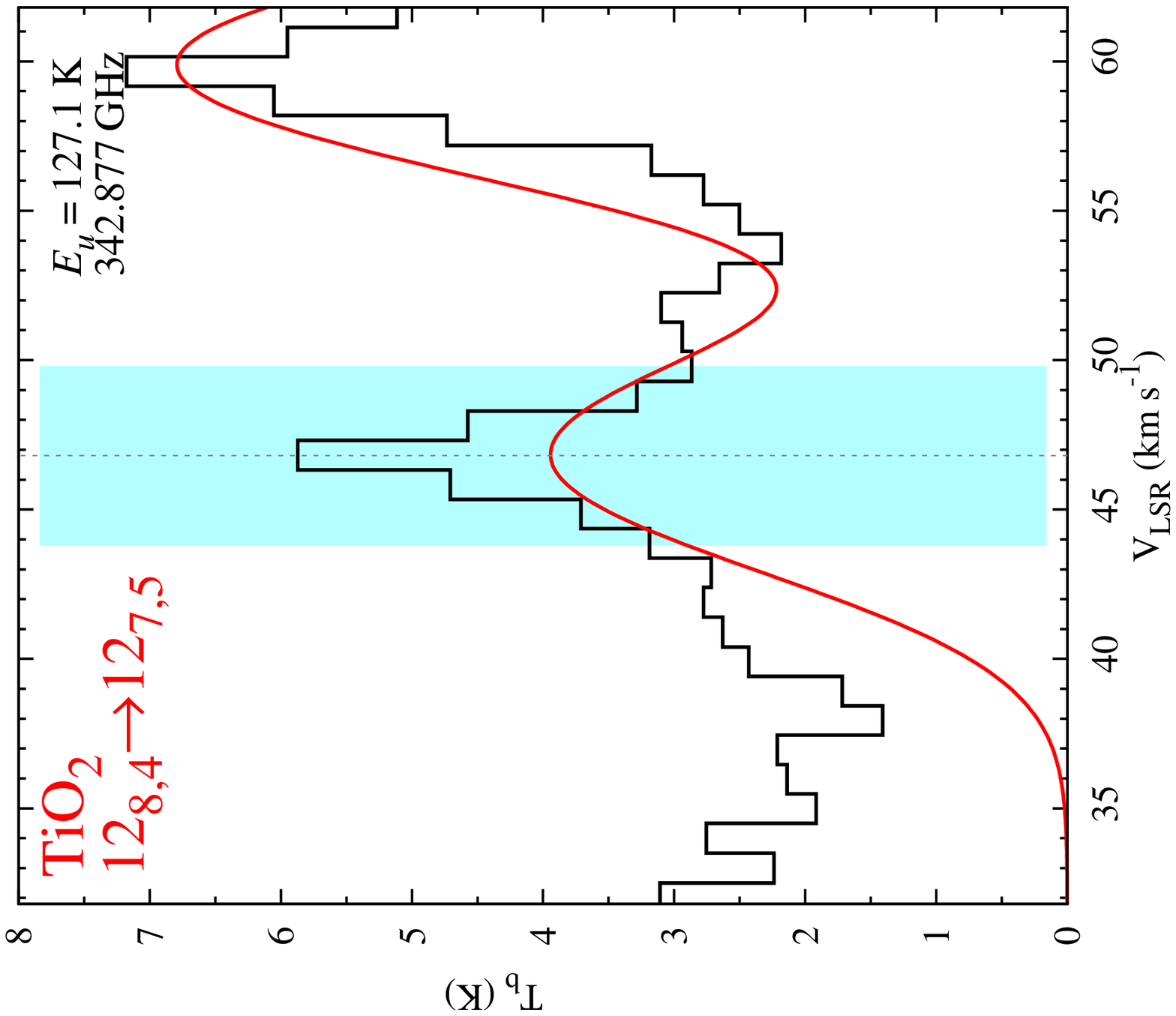}
\includegraphics[angle=270,width=0.19\textwidth]{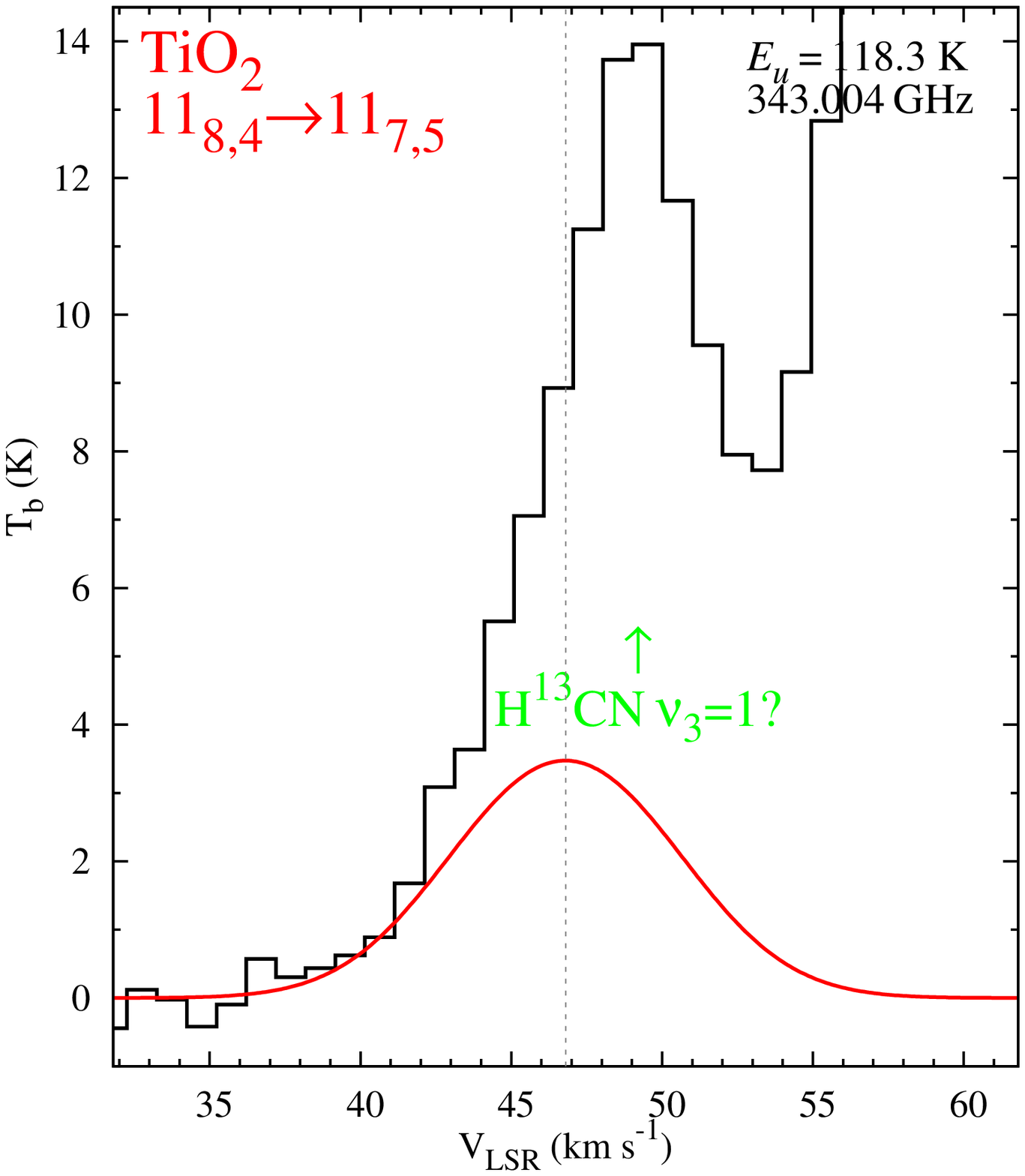}
\includegraphics[angle=270,width=0.19\textwidth]{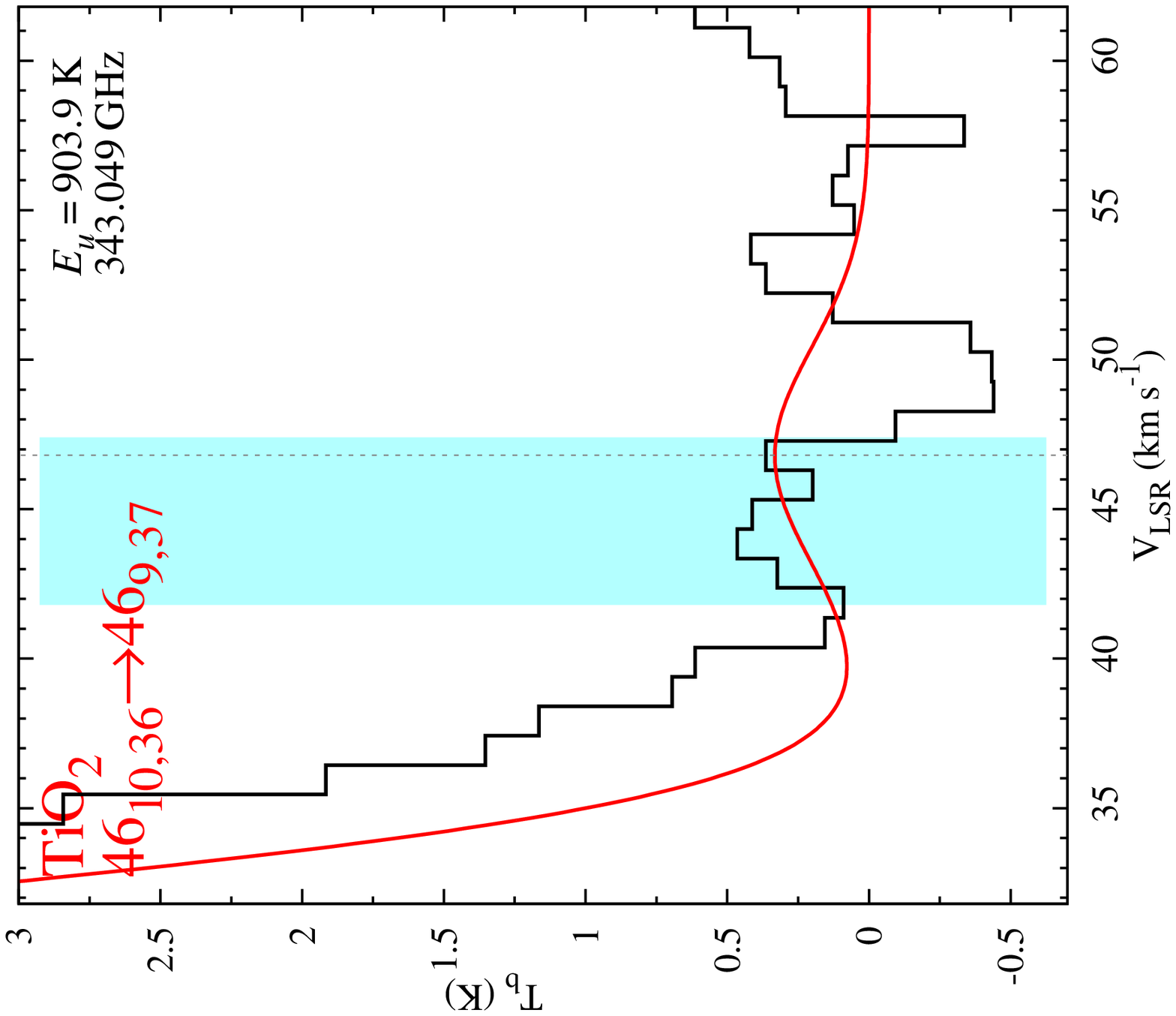}
\includegraphics[angle=270,width=0.19\textwidth]{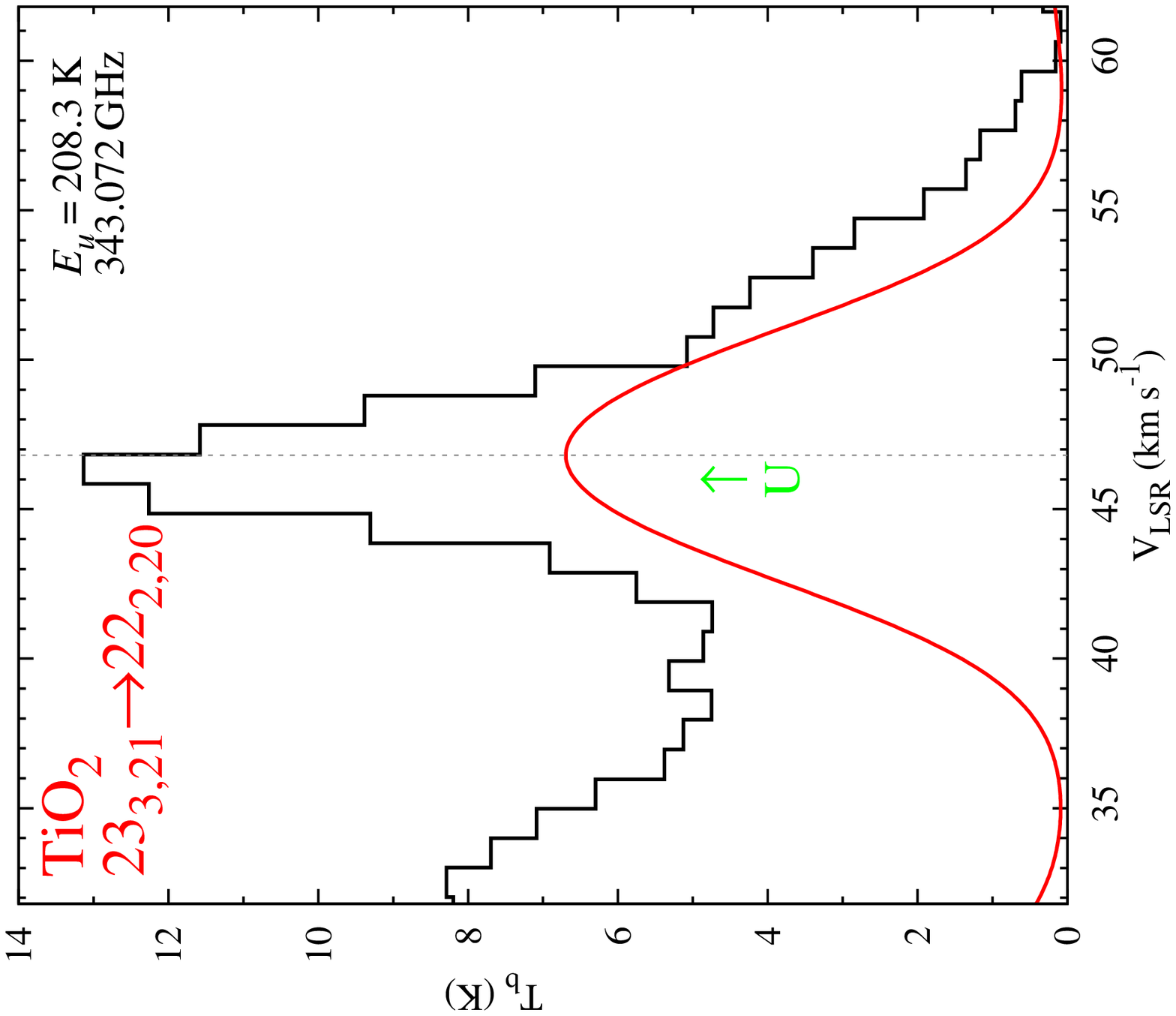}
\includegraphics[angle=270,width=0.19\textwidth]{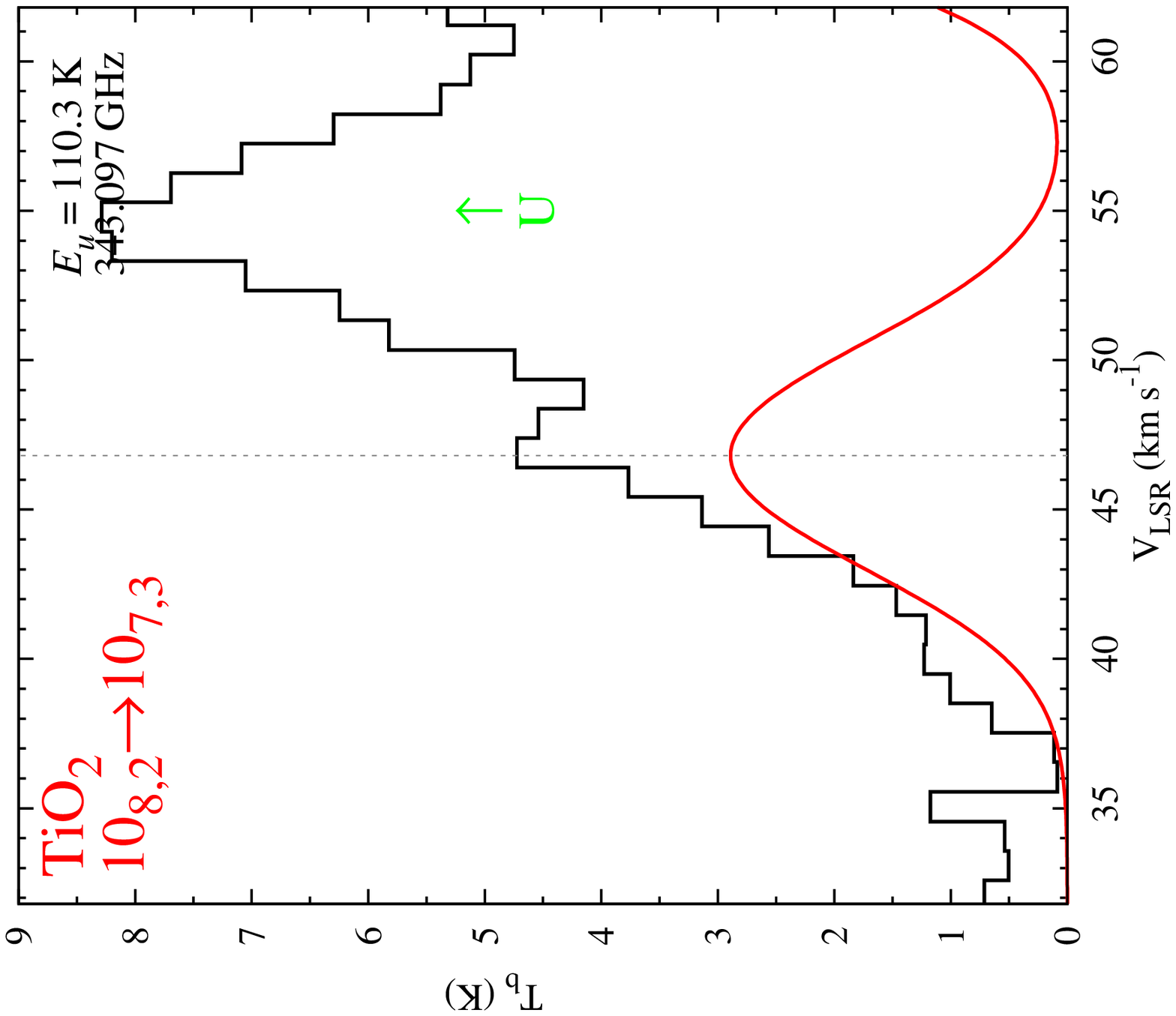}
\includegraphics[angle=270,width=0.19\textwidth]{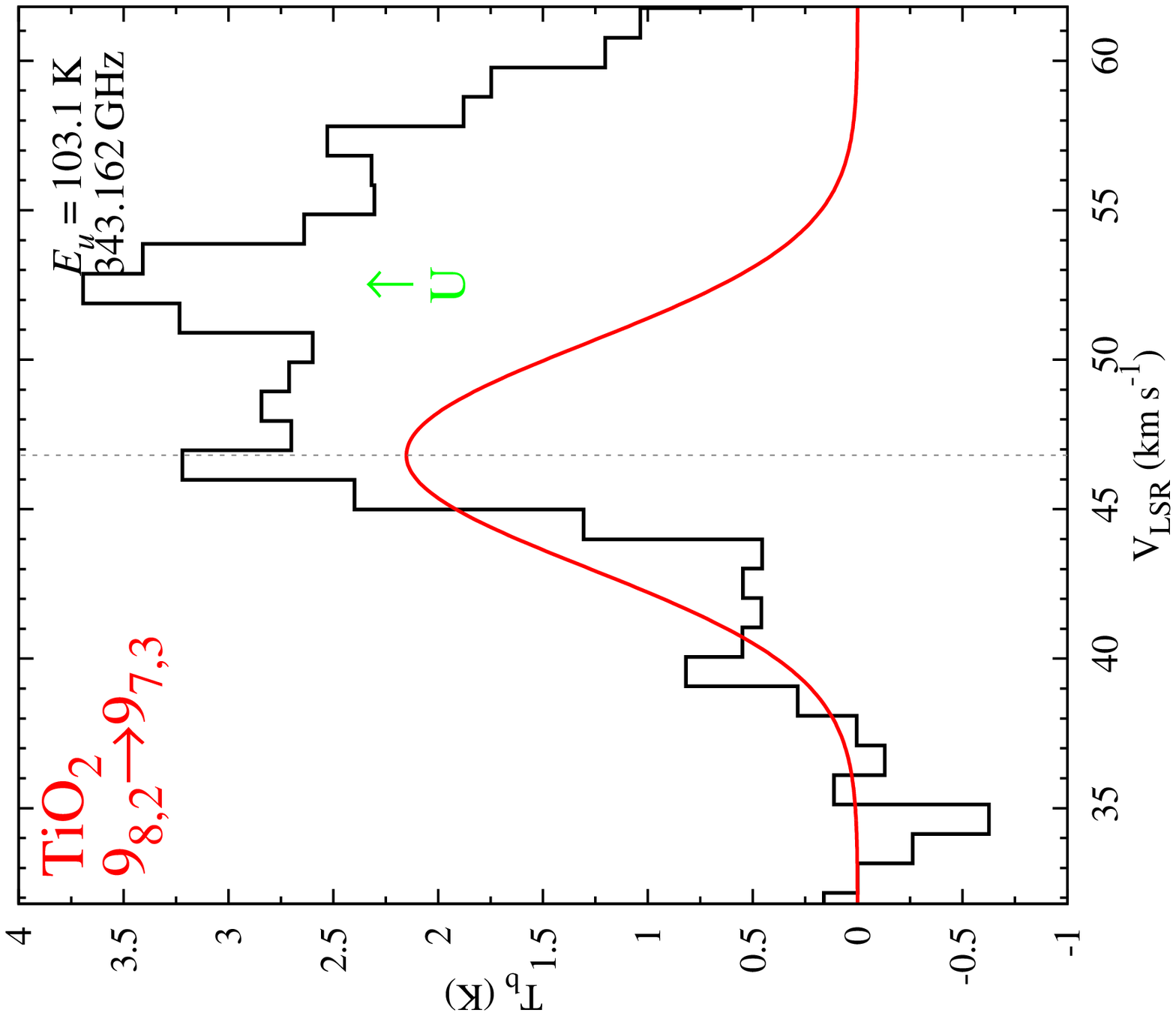}
\includegraphics[angle=270,width=0.19\textwidth]{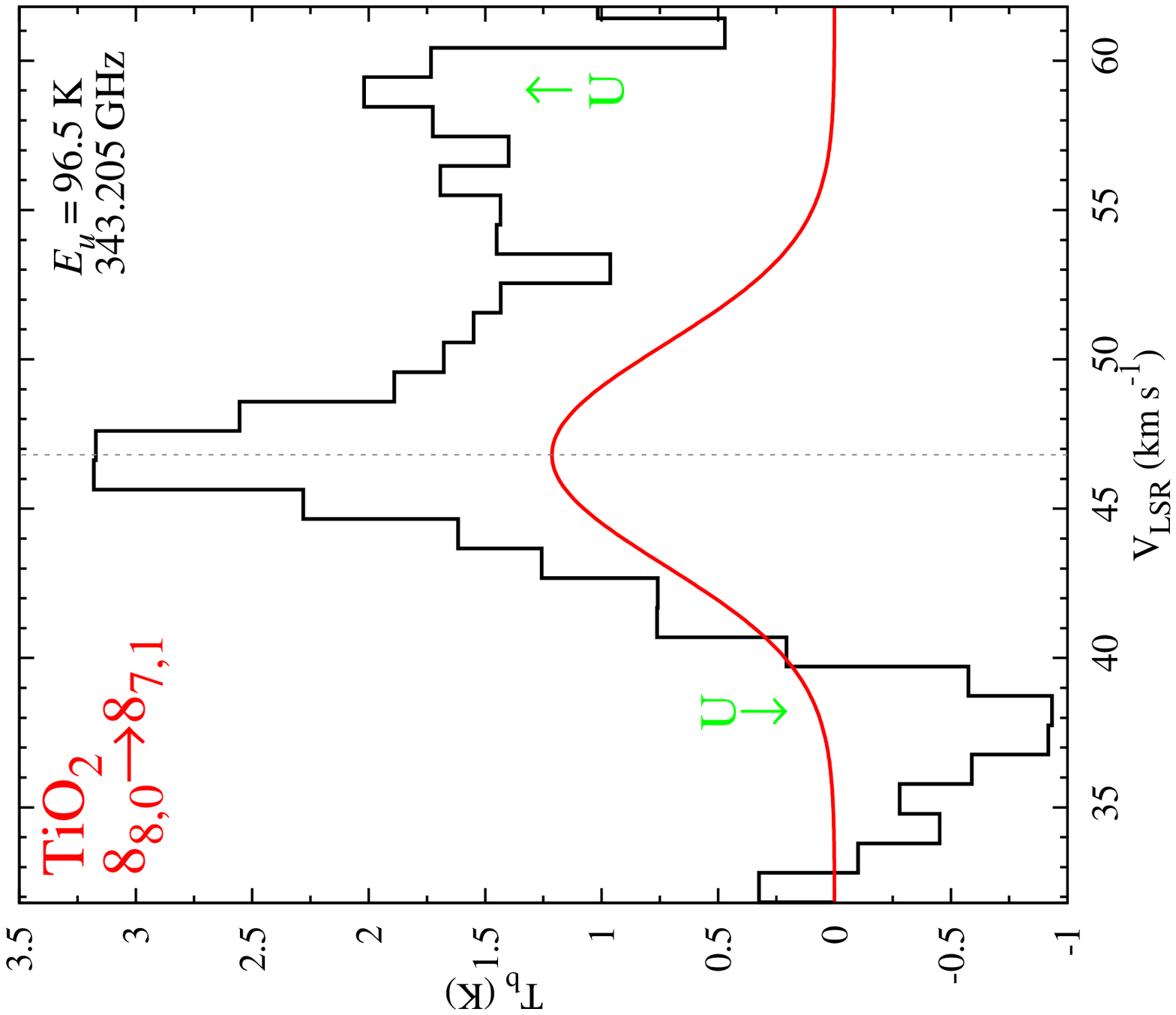}
\includegraphics[angle=270,width=0.19\textwidth]{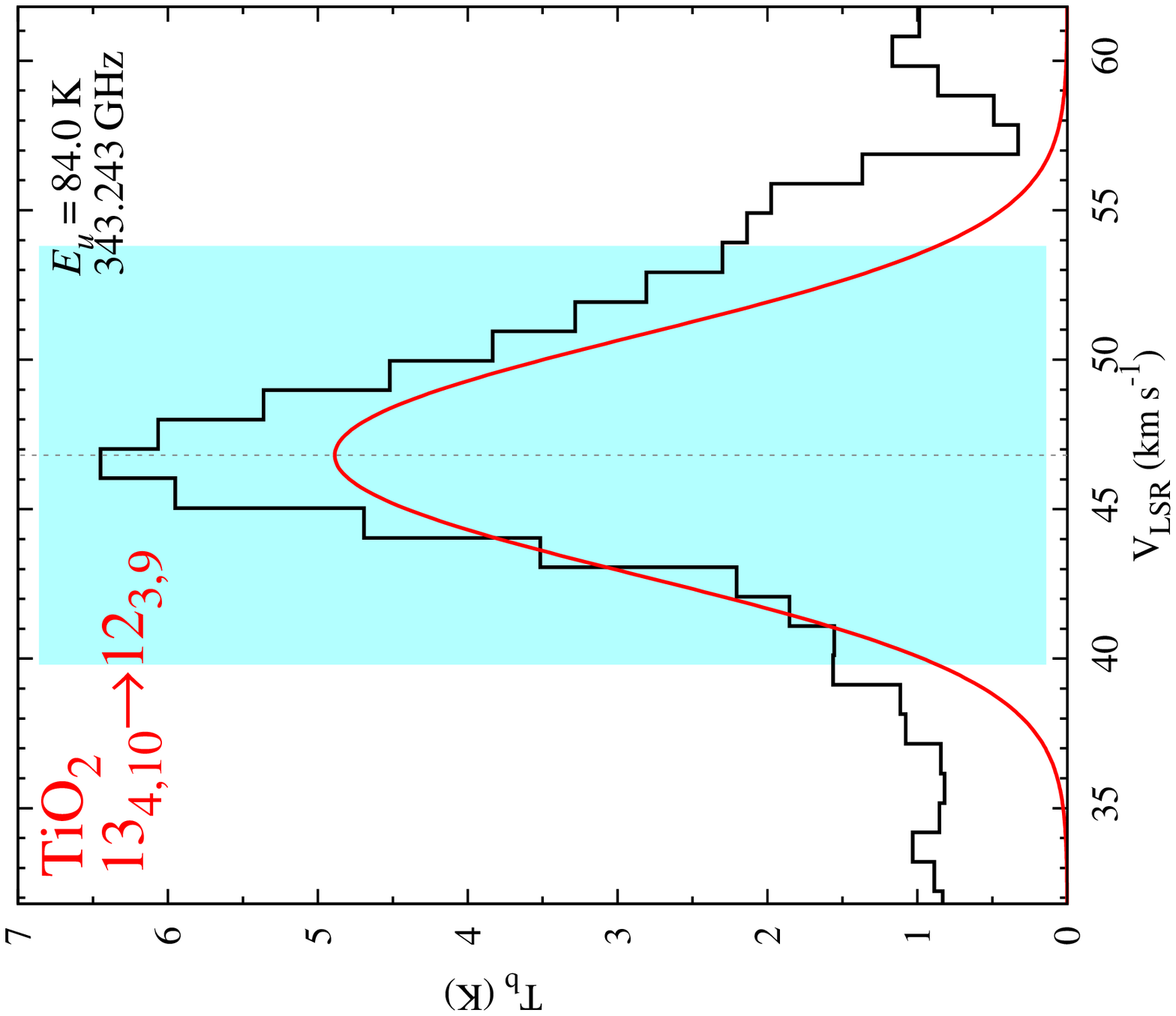}
\includegraphics[angle=270,width=0.19\textwidth]{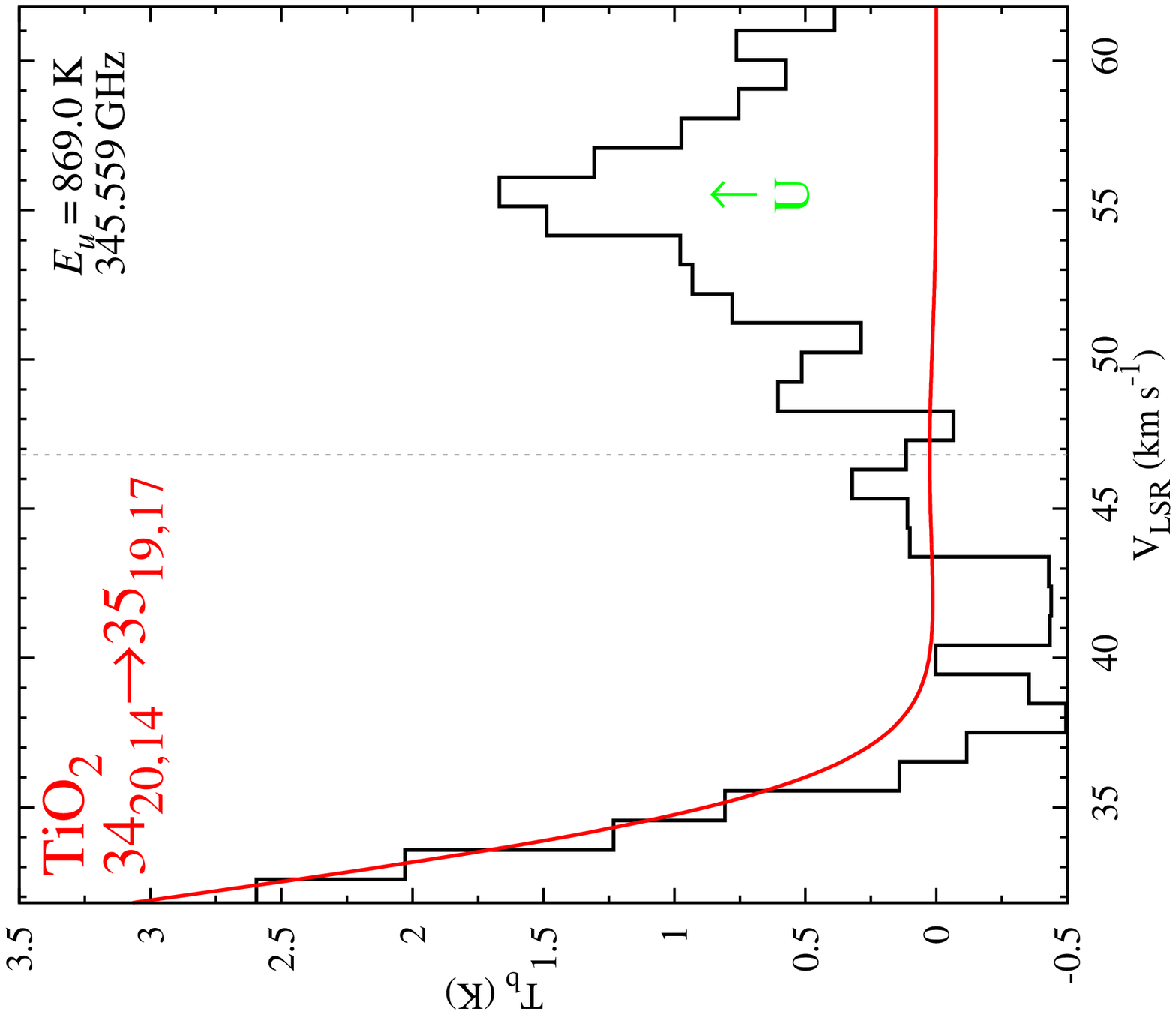}
\includegraphics[angle=270,width=0.19\textwidth]{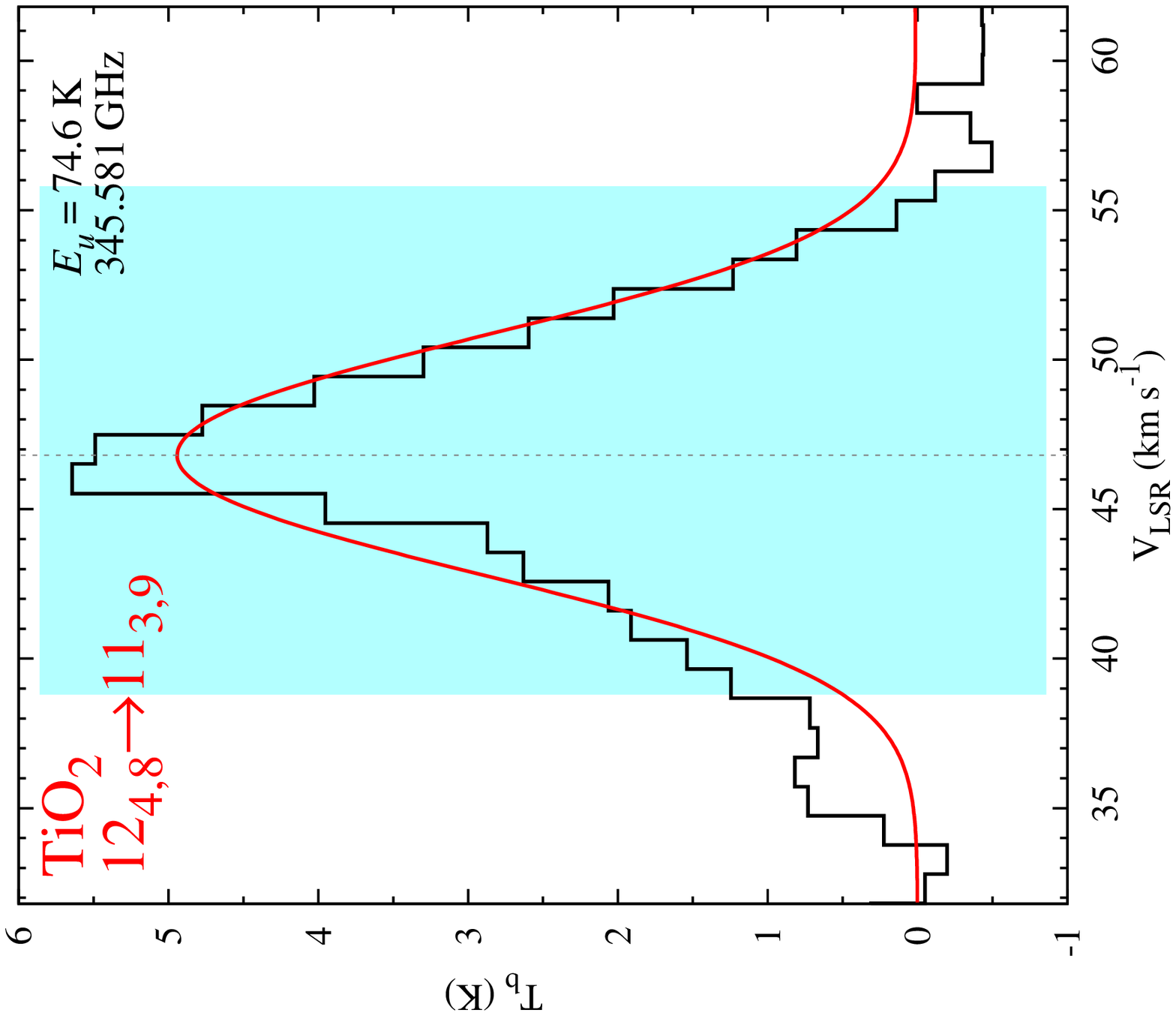}
\includegraphics[angle=270,width=0.19\textwidth]{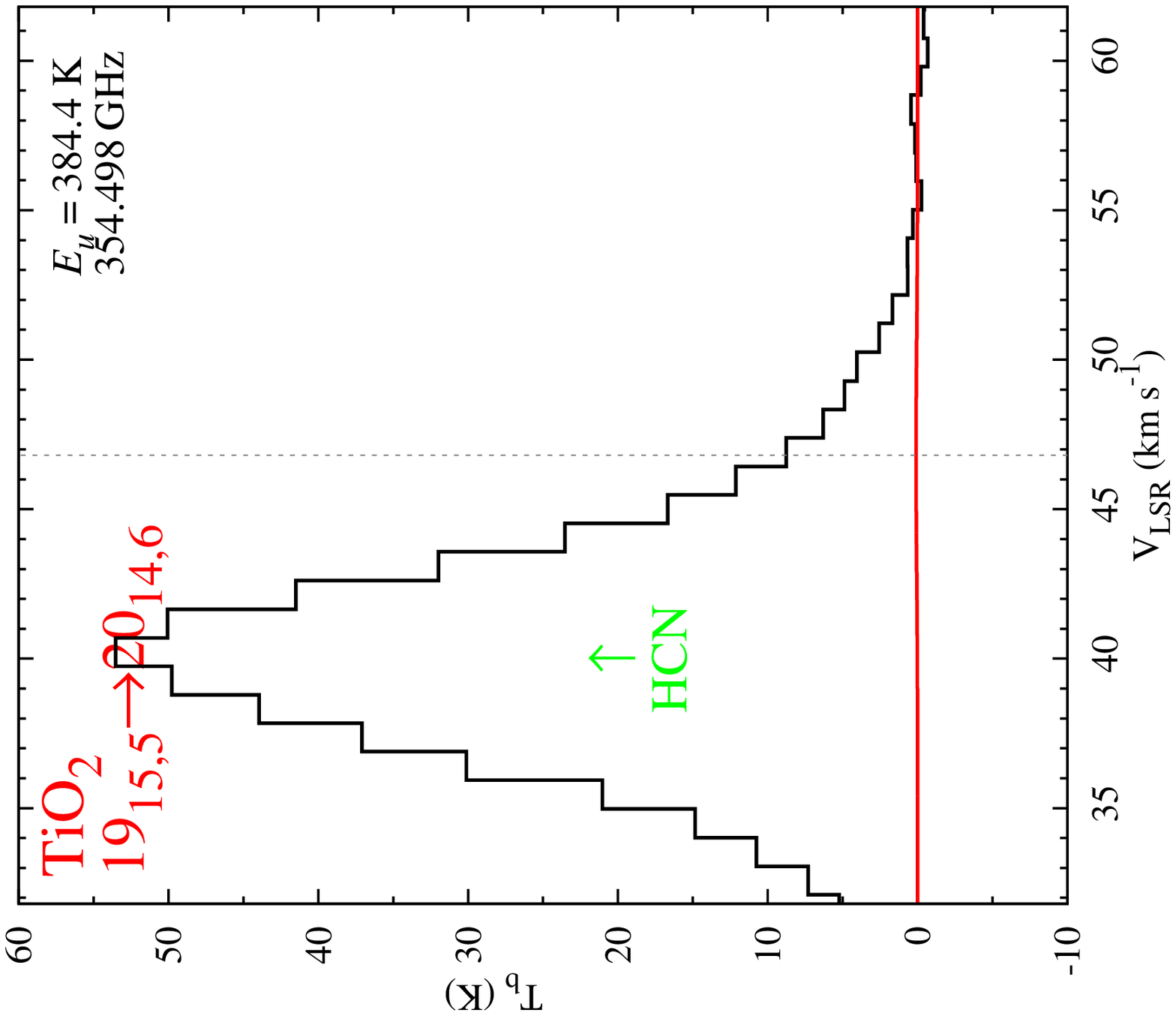}
\includegraphics[angle=270,width=0.19\textwidth]{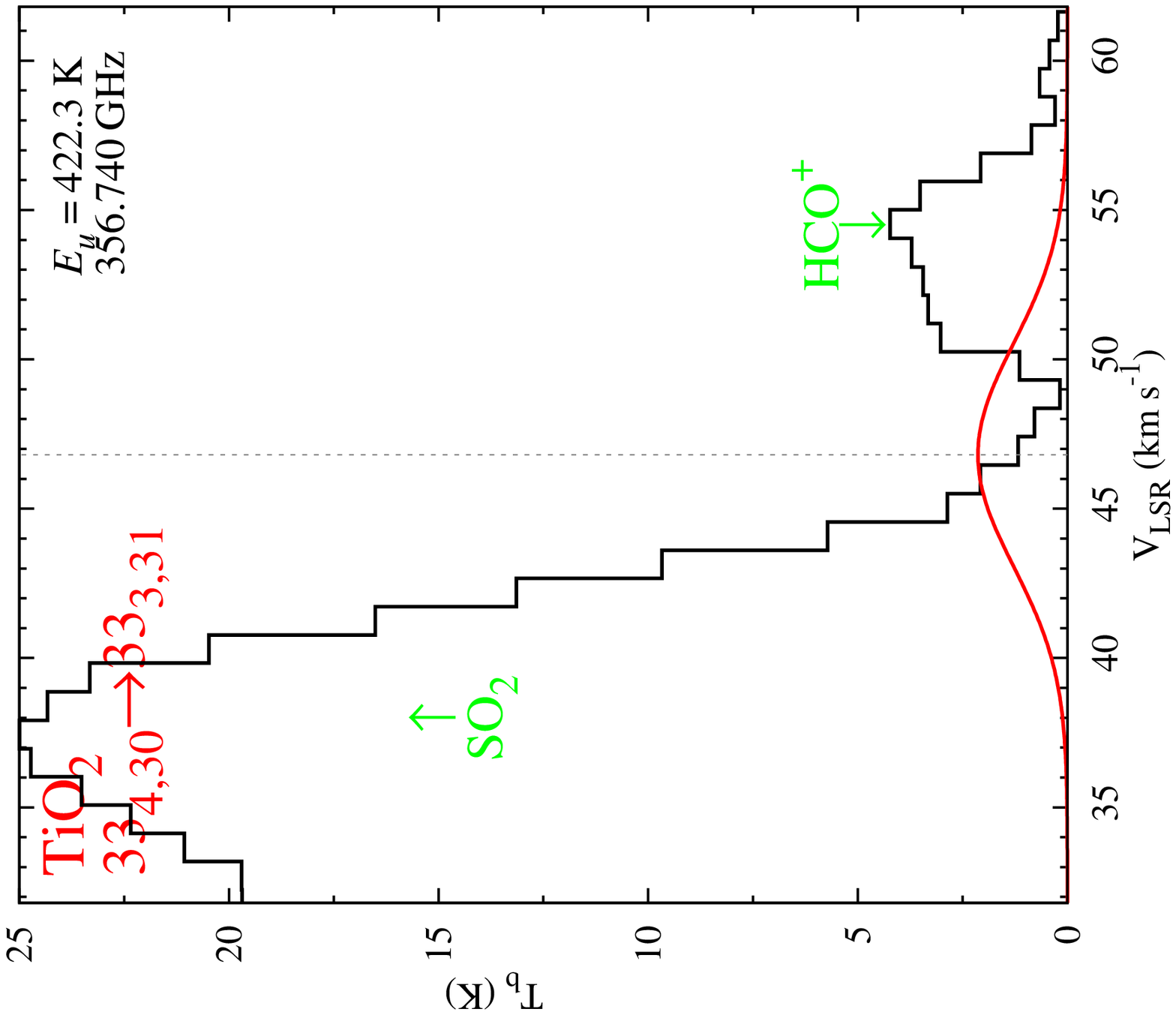}
\includegraphics[angle=270,width=0.19\textwidth]{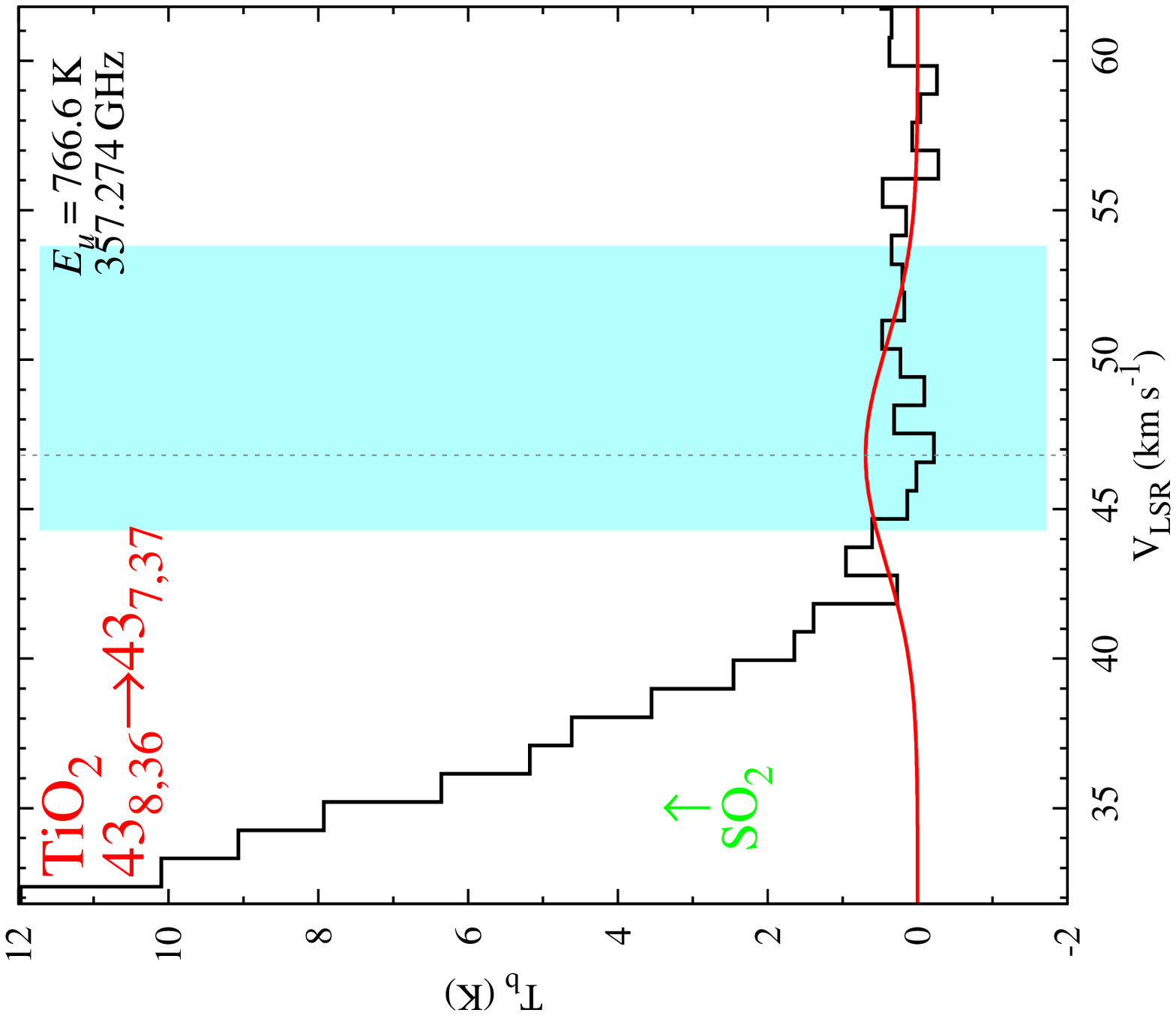}
\includegraphics[angle=270,width=0.19\textwidth]{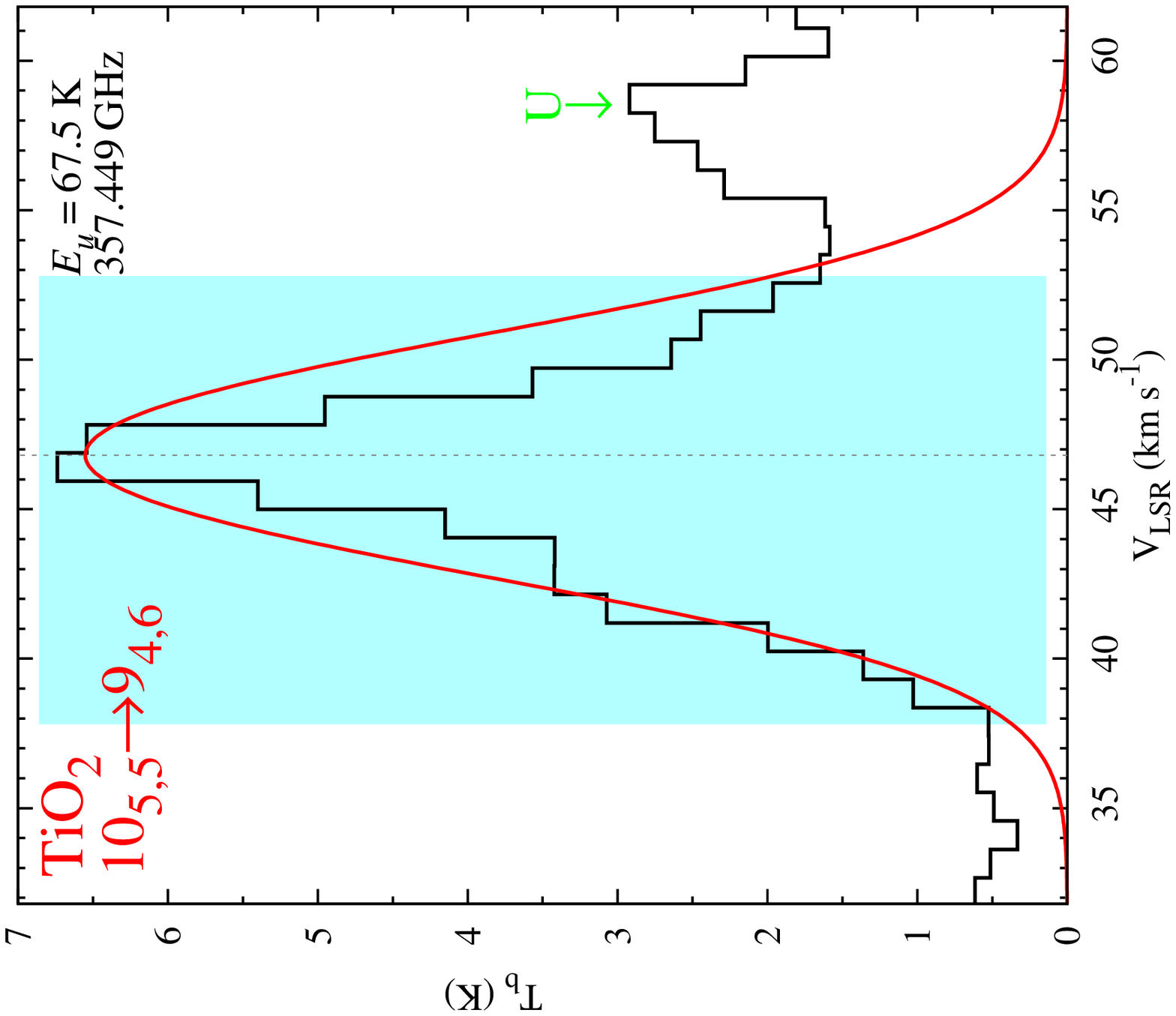}
\includegraphics[angle=270,width=0.19\textwidth]{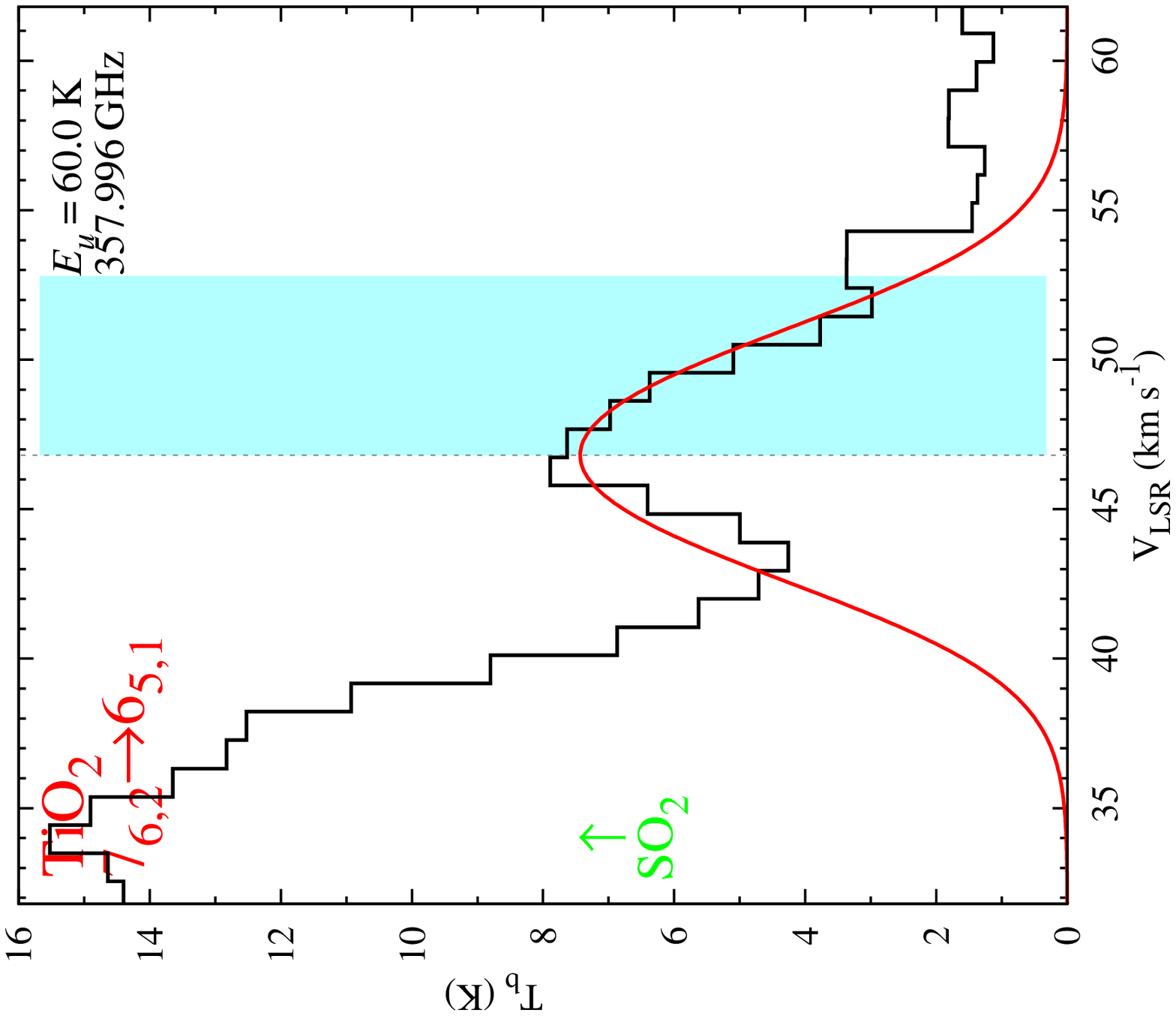}
\caption{All TiO$_2$ transitions covered by the ALMA 2015 data. The black histograms present the observations and the red line shows a simulation based on the best-fit gas-excitation model found with CASSIS. The highlighted parts of the profiles were used in our minimization routine. The vertical dashed line shows the central velocity of the TiO$_2$ features. Many of the lines shown are contaminated by emission of other species, strongest of which are labeled in green. Lines with $E_u$ higher than about 300\,K were not detected and their spectra are shown for completeness.}\label{fig-allTiO2}
\end{figure*}

Our most sensitive ALMA data, obtained in 2015, covered 20 transitions of TiO$_2$ with $E_u \leq 904$\,K. Five of them are not strongly contaminated by emission of other lines and have a high S/N. Five other lines blend tightly with other stronger features so it is difficult to assess if they are present. Lines with $E_u>300$\,K are not detected. The remaining lines are partially blended with other features. Although they are recognizable as TiO$_2$ features, their fluxes cannot be measured directly. ALMA observations in Band\,7 from earlier epochs (Table\,\ref{tab-ALMAdata}) were not sensitive enough to detect TiO$_2$ emission. All the transitions covered are shown in Fig.\,\ref{fig-allTiO2}. The figure labels provide basic information about the covered lines.
%

\begin{figure*}
\centering
\includegraphics[angle=270,width=0.19\textwidth]{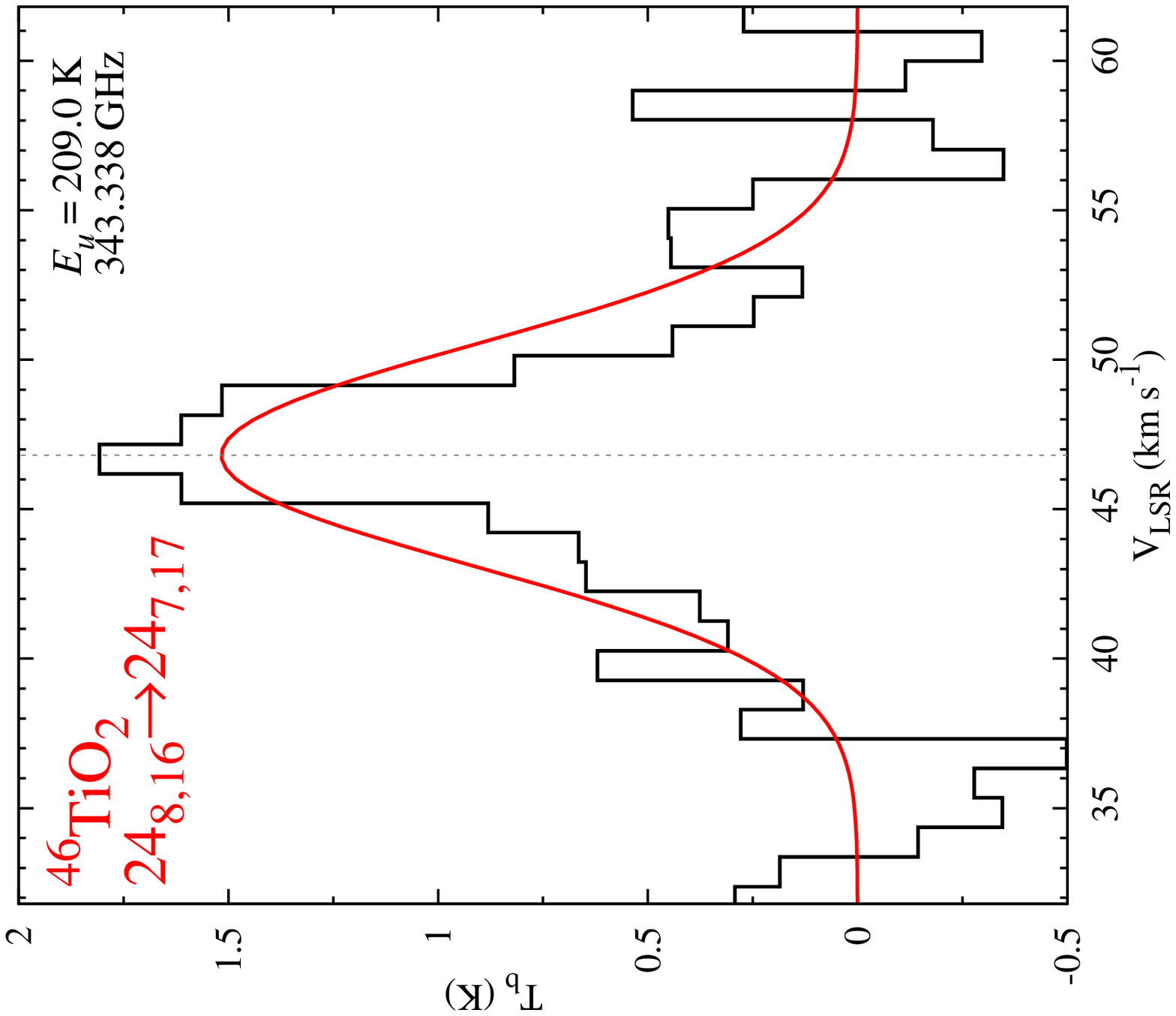}
\includegraphics[angle=270,width=0.19\textwidth]{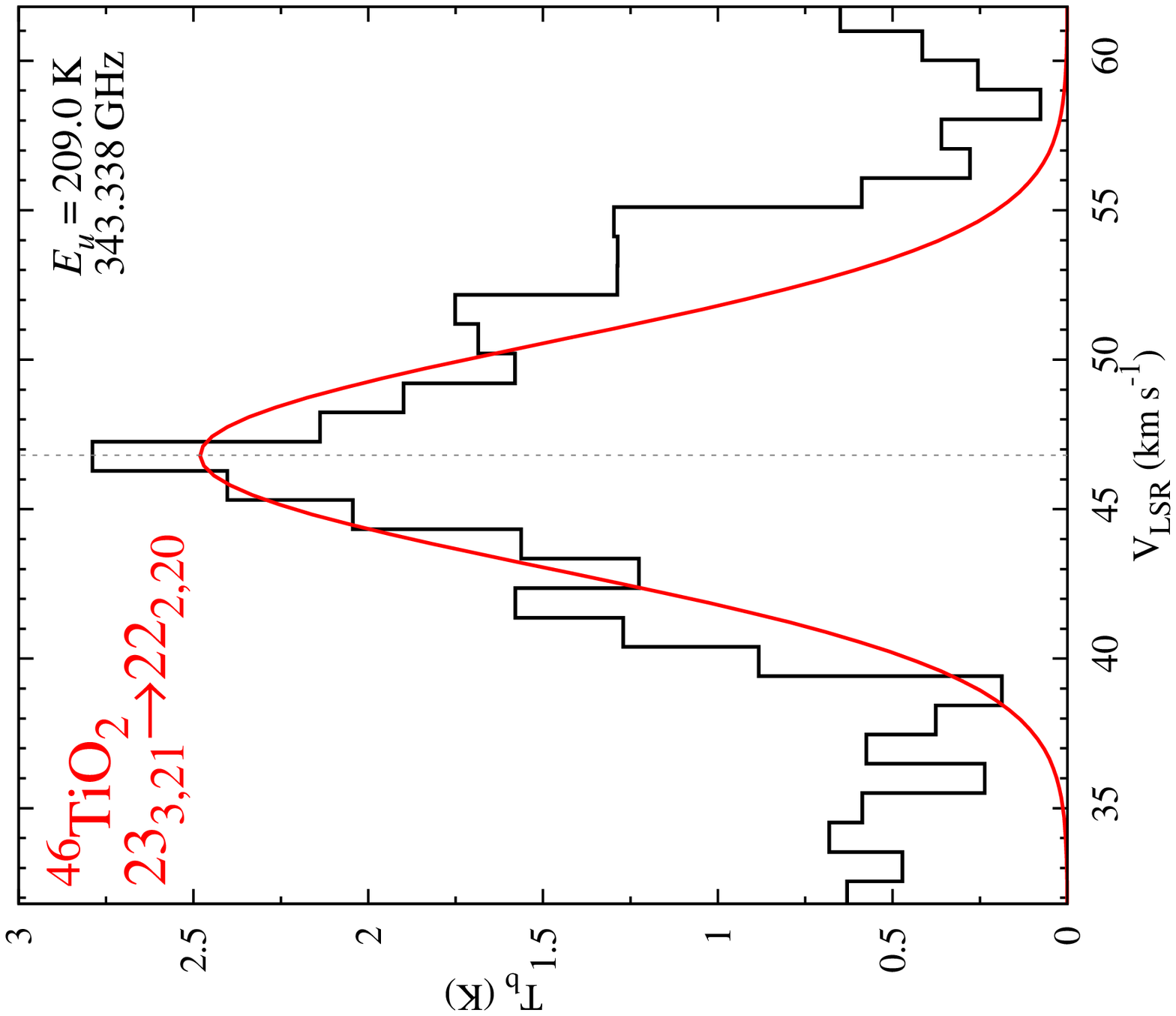}
\includegraphics[angle=270,width=0.19\textwidth]{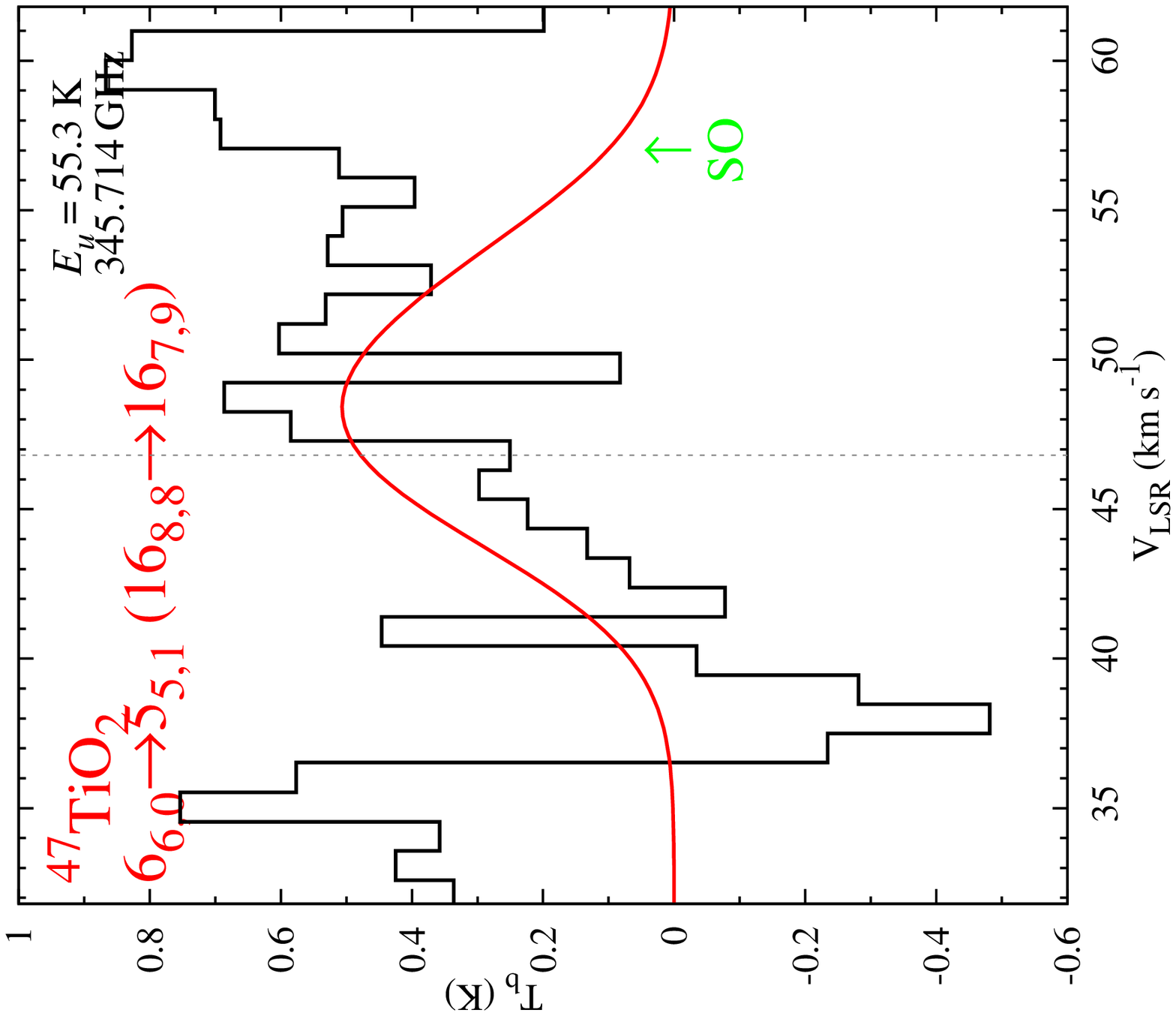}
\includegraphics[angle=270,width=0.19\textwidth]{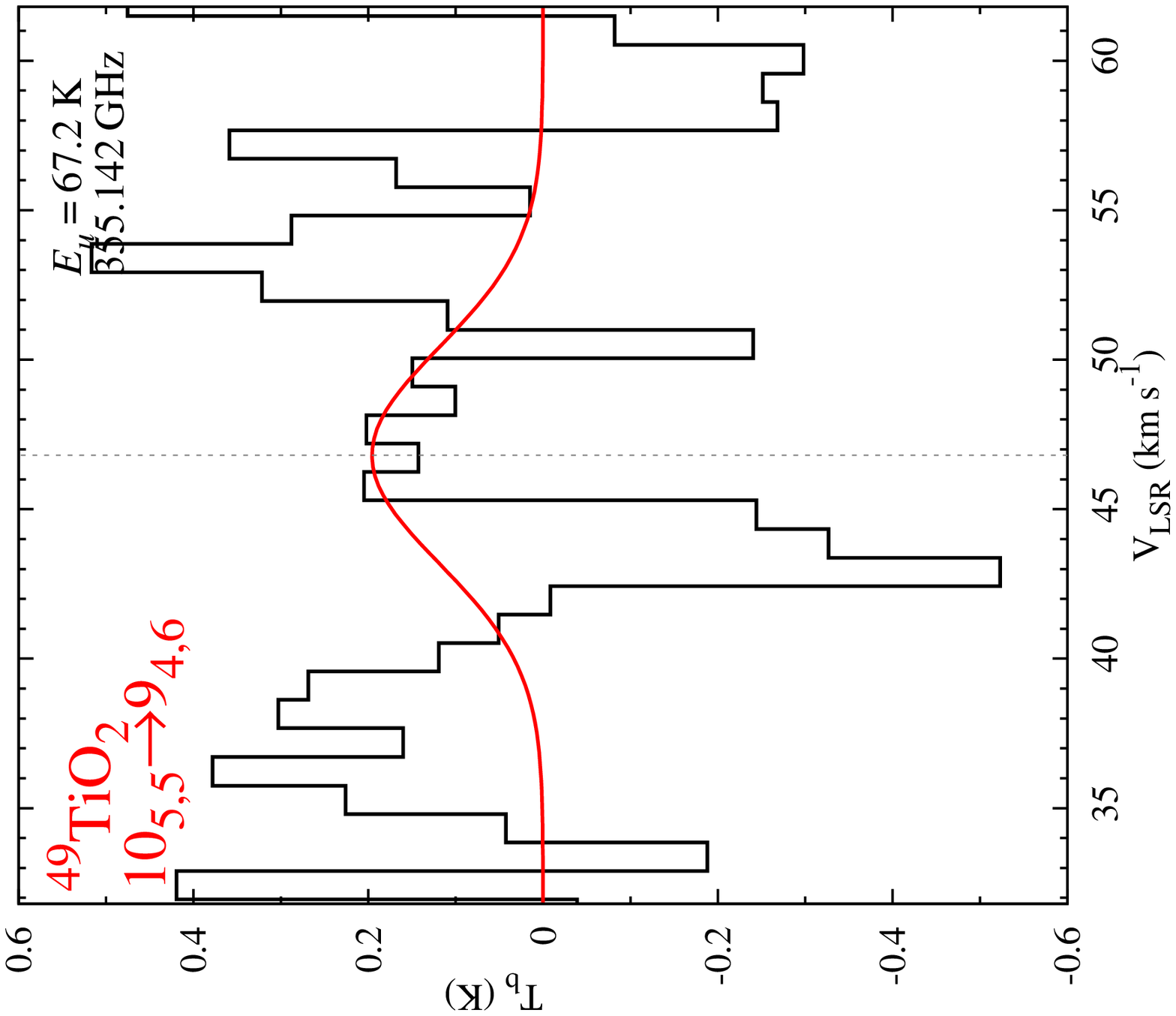}
\includegraphics[angle=270,width=0.19\textwidth]{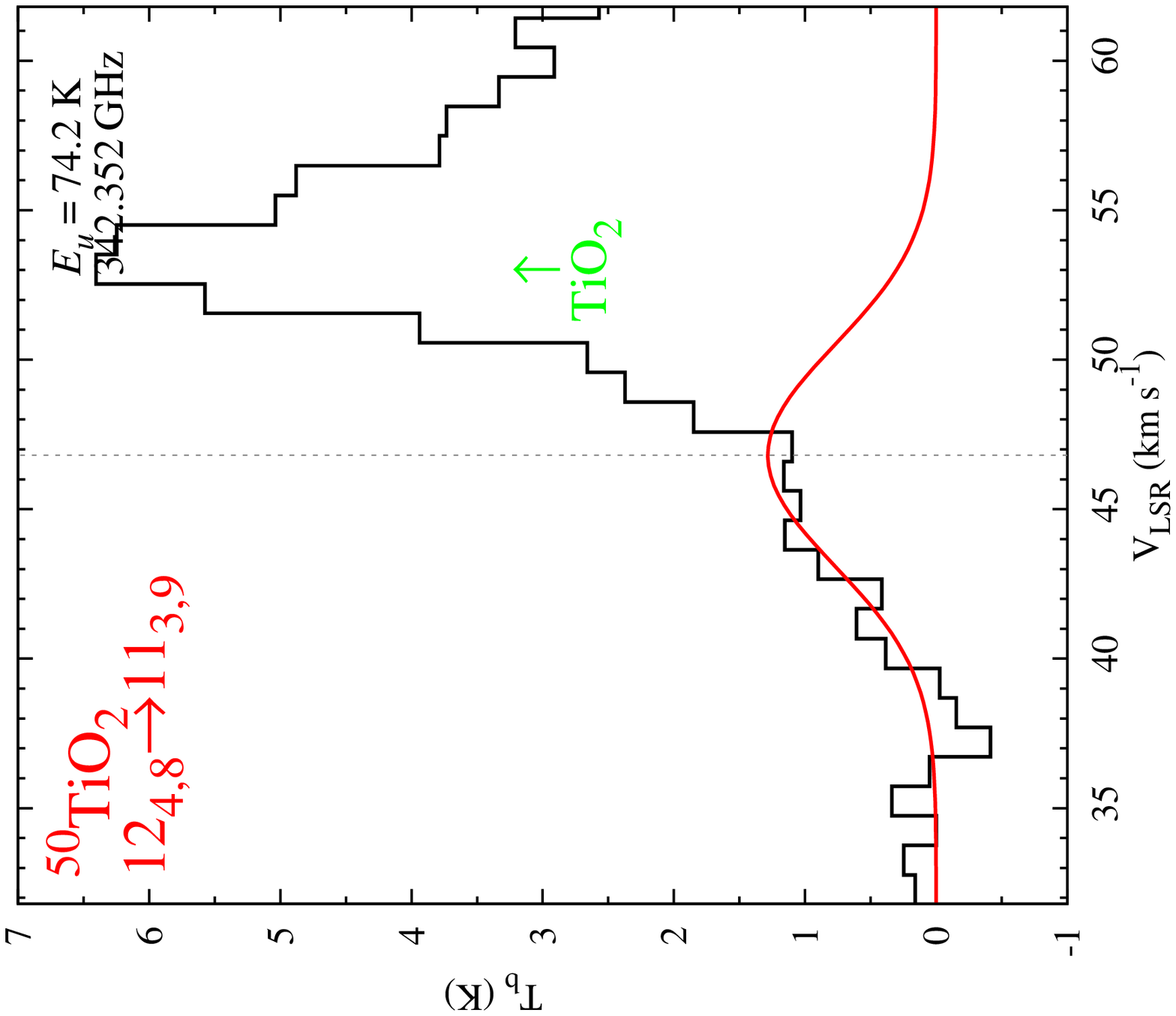}
\caption{Sample spectra of the rare isotopologues of TiO$_2$. Black histograms show observations and the red lines show CASSIS simulations corresponding to upper limits on the column densities of the rare isotopologues. Only features of $^{46}$TiO$_2$ and $^{50}$TiO$_2$ are considered as tentative detections.}\label{fig-allTiO2-iso}
\end{figure*}

After excitation conditions of TiO$_2$ gas have been constrained using the ALMA Band\,7 data from 2015 (Sect.\,\ref{sect-exc}), we examined the Band\,9 data for the presence of TiO$_2$. The identification was more challenging there because of the low spectral resolution and severe line blending. The strongest predicted emission feature is a close blend of two transitions of TiO$_2$, $13_{11,3} \to 12_{10,2}$ and $19_{9,11} \to 18_{8,10}$, near 675.9\,GHz and we found there an emission feature of the predicted intensity. Other lines are either too weak or are blended with much stronger emission features that have not yet been identified. Still, the observation of the 675.9\,GHz blend confirms our identification and excitation analysis.

Because the S/N of the strongest TiO$_2$ lines observed with ALMA is $\sim$93, we searched for the rare isotopologues of TiO$_2$ which are expected to have spectral features approximately ten times weaker than $^{48}$TiO$_2$.
\begin{itemize}
\item We found two unblended (out of 11 covered) lines of $^{46}$TiO$_2$ and their identification as the $24_{8,16} \to 24_{7,17}$ and $23_{3,21} \to 22_{2,20}$ transitions is firm (see Fig.\,\ref{fig-allTiO2-iso}).  
\item We covered 14 transitions of $^{47}$TiO$_2$. Three weak unidentified features were found close to the calculated positions of $^{47}$TiO$_2$, but the line centers are displaced by a few \kms\ and none can be confidently identified as $^{47}$TiO$_2$. 
\item Nine transitions of $^{49}$TiO$_2$ were covered and one unblended feature was identified as possibly $^{49}$TiO$_2$, i.e. $7_{6,2} \to 6_{5,1}$. Its S/N is too low, however, to consider this identification as firm.  
\item A feature near 342.30\,GHz is blended with a line of $^{48}$TiO$_2$ (Fig.\,\ref{fig-allTiO2}, first row, second column) and is most likely the $12_{4,8} \to 11_{3,9}$ transition of $^{50}$TiO$_2$. No other clean lines of this isotopologue are present in the spectrum. 
\end{itemize}
Overall, we are confident only about identification of the two lines of $^{46}$TiO$_2$ and one of $^{50}$TiO$_2$. Sample spectra are shown in Fig.\,\ref{fig-allTiO2-iso}.


\subsection{Spatio-kinematical characteristics of TiO-bearing gas}\label{sec-spkin-TiO}
\begin{figure}
\centering
\includegraphics[angle=270,width=0.8\columnwidth]{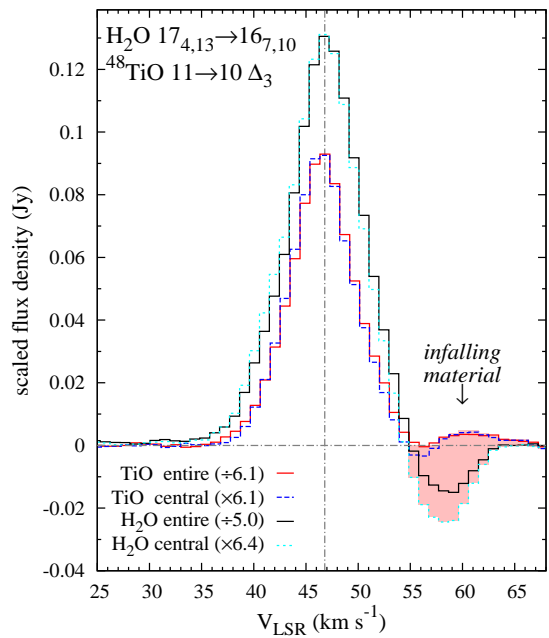}
\caption{Profiles of the TiO 11$\to$10,\,$\Delta_3$ transition are compared to these of H$_2$O $\varv$=0 $17_{4,13} \to 16_{7,10}$. Spectra were extracted for the central beam and within the entire emission region. The flux densities were arbitrarily scaled by the listed factors for a more readable comparison. Shaded in red  are signatures of infall in Mira.}\label{fig-TiOandH2O}
\end{figure}

\begin{figure*}
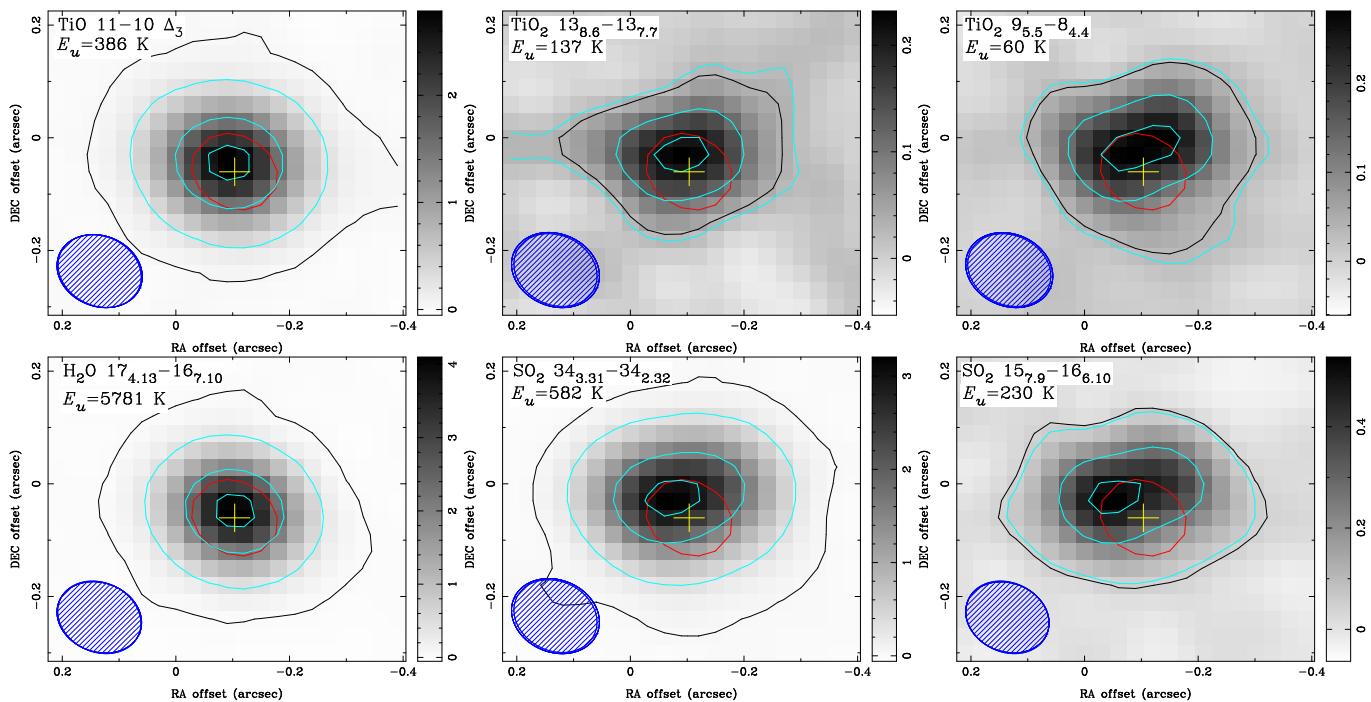

\centering
\includegraphics[angle=270,width=0.32\textwidth]{mirmap3rms_TiOmainem.ps}
\includegraphics[angle=270,width=0.32\textwidth]{mirmap3rms_TiO2_342708mainem.ps}
\includegraphics[angle=270,width=0.32\textwidth]{mirmap3rms_TiO2_342344mainem.ps}
\includegraphics[angle=270,width=0.32\textwidth]{mirmap3rms_H2Omainem.ps}
\includegraphics[angle=270,width=0.32\textwidth]{mirmap3rms_SO2_strong.ps}
\includegraphics[angle=270,width=0.32\textwidth]{mirmap3rms_SO2_230K.ps}
\caption{Maps of molecular emission in the extended atmosphere of Mira. The emission was integrated in the spectral range corresponding to the main emission component between about 35 and 55\,\kms\ (optimized for each transition). The emission is shown in gray in units of Jy\,\kms\ and with cyan contours at 10, 50, and 90\% of the maximum flux. Additionally, a black contour at the 3$\sigma$ noise level is shown. The location of continuum center of Mira\,A is marked with a plus symbol whose size corresponds to the physical size of the stellar disk. The red contour shows the continuum emission at 50\% of the peak value and represents the continuum extent smeared by the synthesized beam. The beams' FWHMs are shown with hatched ellipses.}\label{fig-mom0}
\end{figure*}

\begin{figure*}
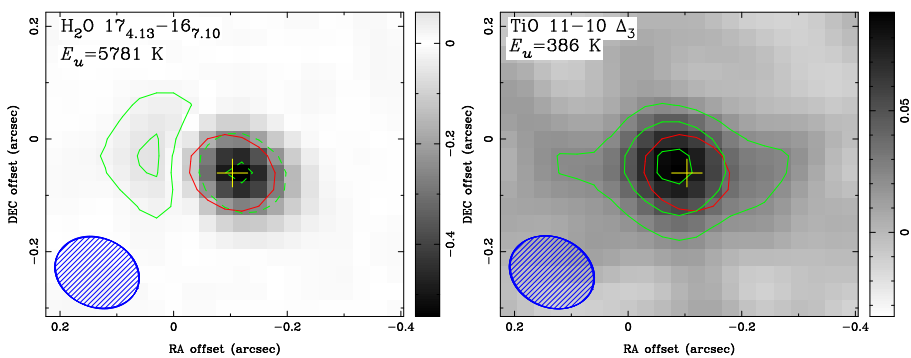

\includegraphics[angle=270,width=0.32\textwidth]{mirmap_H2Oinfall.ps}
\includegraphics[angle=270,width=0.32\textwidth]{mirmap_TiOinfall.ps}
\caption{Maps of the infall signatures. The maps were produced by integrating the flux density within 55--64\,\kms\ for H$_2$O $17_{4,13}\to 16_{7,10}$ and within 57--67\,\kms\ for TiO $J$=11$\to$10 $\Delta_3$ (cf. Fig.\,\ref{fig-TiOandH2O}). The water line at this range is dominated by absorption (negative flux), which is shown with green dashed contours, and residual emission outside the stellar disk (positive flux), which is shown with solid contours. The infall in the TiO line is seen mainly in emission (solid contours). The symbols have the same meaning as in Fig.\,\ref{fig-mom0}.}
\label{fig-infall}
\end{figure*}

The ALMA observations of the 11$\to$10,\,$\Delta_3$ transition have the best angular resolution (0\farcs154$\times$0\farcs125) and S/N ($\sim$248) among all the data collected for TiO and its isotopologues. We base our analysis mainly on this single dataset. The line profile is shown in Fig.\,\ref{fig-TiOandH2O} and maps are shown in Figs.\,\ref{fig-mom0} and \ref{fig-infall}. 

The line profile is dominated by emission centered at the stellar center-of-mass velocity and resembles profiles typical for species of moderate-to-high  excitation and limited extent, typical of the {\it extended atmosphere}. At a closer inspection and for the central resolution element (central beam), however, a weak absorption component is seen at 56.5\,\kms\ and the overall profile can be classified as an inverse P-Cyg type, as shown in Fig.\,\ref{fig-TiOandH2O}. In addition, in the spectral range where the comparison profile of H$_2$O in Fig.\,\ref{fig-TiOandH2O} is dominated by absorption, the TiO feature displays a weak emission component, at 57--67\,\kms. 


The spatial distribution of the \emph{main} emission component, within the 35--56\,\kms\ interval, is very well approximated by a Gaussian of a beam-deconvolved FWHM of 115.9$\times$93.3 ($\pm$2.7) mas and a position angle of the major axis (PA) of 106\degr$\pm$5\degr. This emission is much more extended than the submm photosphere. The centroid position is 9.3$\pm$0.2\,mas east and 16.3$\pm$0.4 mas north from the continuum peak (Fig.\,\ref{fig-mom0}), so the net location of TiO is not far from the stellar center. Most of the gas giving rise to TiO emission is therefore surrounding the star within the radius (FWHM/2) of about 4\,\rstar.

The TiO main emission region is only slightly larger than that of the AlO $N$=9--8 line of 123$\times$74\,mas observed at the same angular resolution (Paper\,I). The spatial offset of the centroid with respect to the continuum center is 12\,mas, much smaller than that for AlO (40\,mas). On the basis of the map in which the $N$=6--5 emission of AlO is well resolved, the apparent centroid of AlO emission is displaced from the stellar disk as a consequence of a highly anisotropic distribution of AlO around the star. Although details of TiO distribution require observations at a higher angular resolution, our maps suggest that while the extent of TiO emission is similar to that of AlO, the TiO emission is distributed much more isotropically around the star than AlO. 


The absorption of TiO seen at $V_{\rm LSR}$=56--57\,\kms\ is spatially compact and is at the center of the continuum source. It is therefore a signature of the portion of TiO-bearing gas seen in the line of sight toward the stellar disk. To appear in absorption, the temperature of TiO gas must be lower than that of the photosphere, $<$2440\,K. By comparing the emission-to-absorption intensity ratio in the profile of TiO to that of species requiring excitation temperatures of the order of 1000\,K, as in Fig.\,\ref{fig-TiOandH2O}, one can see that although TiO is excited in the hot infall, the bulk of TiO-bearing gas must reside in gas characterized by a much lower temperature (see  Sect.\,\ref{sect-exc}).    

The weak TiO emission in the 57--67\,\kms\ range is seen mostly east-south from the edge of the stellar disk and a similar feature is seen in maps of other species of moderate-to-high excitation, for instance in the H$_2$O $17_{4,13}\to 16_{7,10}$ line from $E_u$=5781\,K. The maps of the redshifted emission in TiO and H$_2$O are shown in Fig.\,\ref{fig-infall}. Because the absorption component is much stronger in the highly-excited line of H$_2$O, the net line intensity is negative close to the direction toward the star and pure emission is seen only east from the star, by as far as 0\farcs2 in the beam-smeared maps. The TiO emission is much less affected by the redshifted absorption and the full extent of the emission component can be seen directly in Fig.\,\ref{fig-infall}. The emission can be traced up to 0\farcs2 from the star toward east and west. The emission is a signature of warm gas falling on the star. Because we measure only the tangential motion and away from the direction of the kinematical center, the material seen in emission must be falling on the star with velocities in excess of the maximal redshift 17\,\kms. Being faint, the TiO emission must arise from a portion of gas that is only a small fraction of the bulk gas seen in the main emission component. If the infall was spherically symmetric around the star, the redshifted emission and absorption components seen above 56\,\kms\ should have a corresponding blueshifted component at velocities below about 36\,\kms. However, it is not observed, most likely because that material is obscured by the radio photosphere. 

%


We conclude that the submm features originate from gas located mainly in the extended atmosphere of Mira, at radii $\lesssim$4\,\rstar. The distribution of the bulk of TiO-bearing gas is more isotropic than that of AlO and is not directly affected by the pulsation shock that was active at the moment of the ALMA observations (Sect.\,\ref{sec-env}). A small fraction of TiO-bearing gas is also present in the infall. This infalling material may be warmer and more anisotropically distributed than the main mass of TiO. 

\subsection{Spatio-kinematical characteristics TiO$_2$-bearing gas}
\begin{figure}
\centering
\includegraphics[angle=270,width=0.8\columnwidth]{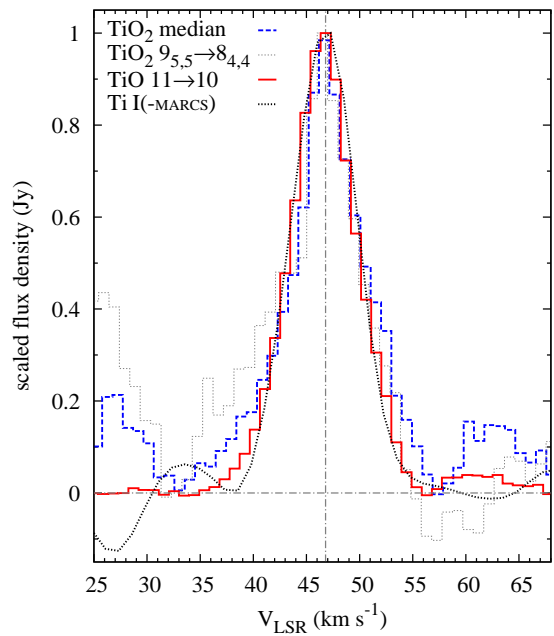}
\caption{Representative line profiles of Ti-bearing species in Mira. 
{\bf \ion{Ti}{I}:} An optical spectrum of \ion{Ti}{I} is shown with dotted black line. It is an average of two spectra acquired for phases 0.17 and 0.24 and represents the phase of ALMA observations ($\varphi$=0.21). To correct the average profile for the photospheric line of \ion{Ti}{I}, a synthetic spectrum based on MARCS model atmospheres was subtracted. The subtraction is not perfect and the profile of circumstellar \ion{Ti}{I} is distorted by photospheric features of TiO. 
{\bf TiO:} Our best ALMA observations of the pure rotational line of TiO is shown as a red histogram. 
{\bf TiO$_2$:} Two profiles are shown for TiO$_2$: one is a median of six normalized transitions and represents lines of different excitation potentials (blue dashed histogram); the other shows a profile of the 9$_{5,5}$--8$_{4,4}$ transition (gray dotted histogram).}\label{fig-profiles-TiTiOTiO2}
\end{figure}

The TiO$_2$ lines appear as pure emission features (Fig.\,\ref{fig-allTiO2}). However, because they are relatively weak, line blending is a problem in characterizing their profiles. To reduce the effect of line blending, we combined the six cleanest normalized lines into a median profile shown in Fig.\,\ref{fig-profiles-TiTiOTiO2}. Strikingly, it displays two extended wings. In the 9$_{5,5}$--8$_{4,4}$ transition alone, compared to the median profile in Fig.\,\ref{fig-profiles-TiTiOTiO2}, the red wing is completely missing and weak absorption or zero level is seen instead. It is possible that part of the TiO$_2$ gas is present in the infall but the corresponding absorption component is too weak to produce an unambiguous absorption feature. The TiO$_2$ profiles appear to be slightly broader for transitions from higher-lying energy levels but this trend may be mimicked by the heavy line crowding. 

The broad non-Gaussian wings make TiO$_2$ profiles distinctive from any profiles of other species observed. This is particularly apparent when the TiO$_2$ profile is compared to that of TiO, as in Fig.\,\ref{fig-profiles-TiTiOTiO2}. The typical FWHM of 8.5\,\kms\ of the emission profile is however only slightly larger than that of TiO (FWHM=7.5\,\kms). While the cores of the TiO$_2$ emission lines may originate from the same gas in the extended atmosphere, the material apparent in the wings should be characterized by an increased abundance of TiO$_2$ relative to TiO. 

The emission region of TiO$_2$ is larger than the ALMA beam in our Band\,7 observations and has a beam-deconvolved full size of about (159$\pm$10)$\times$(78$\pm$11)\,mas at PA=107\degr. These values were measured as a Gaussian fit to an image of seven combined TiO$_2$ transitions to obtain a better S/N. In case the emission is actually patchy or multiple components are present, our Gaussian fit would yield an inadequate measure of the size. Nevertheless, the Gaussian area of the emission region of TiO$_2$ is only 10\% larger than that of TiO and it appears more asymmetric, i.e. has a higher eccentricity. The orientation of the long axis of the Gaussian fits is the same for both oxides. The comparison of the maps of TiO and TiO$_2$ emission in Fig.\,\ref{fig-mom0} suggests that TiO$_2$ emission is indeed more extended even if the difference in excitation and S/N is taken into account. The centroid position of TiO$_2$ emission is 13($\pm$2)\,mas north and 30($\pm$2)\,mas east from the continuum center, which is 1.6 times farther away from the star than what we measured for TiO. This displacement is apparent in Fig.\,\ref{fig-mom0}. The location and distribution of the TiO$_2$ resembles more that of warm SO$_2$, which is representative of species present in the silicate formation zone and the wind.

The TiO$_2$ traced in submm lines is located mainly in the extended atmosphere of Mira, in layers where we observe TiO, but its spatial distribution is more anisotropic than that of TiO. With the radii (FWHM/2) of 2.7--5.5\,\rstar\ of the emission region, a smaller amount of TiO$_2$-bearing gas extends to farther distances from the star than the monoxide emission (<4\,\rstar). Overall, TiO$_2$ resides in the quiescent part of the envelope that is not affected directly by the supersonic shock.

\subsection{Excitation and column densities of TiO and TiO$_2$}\label{sect-exc}
\begin{figure}
\centering
\includegraphics[angle=270,width=0.99\columnwidth]{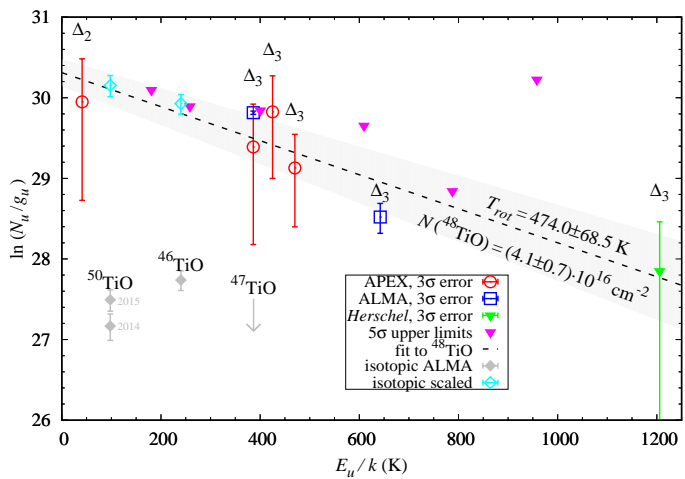}
\caption{Rotational diagram for TiO. Red, blue, and green points represent detected lines and the magenta triangles mark 5$\sigma$ upper limits. The linear fit which constrains the excitation temperature and column density of the main isotopologue, $^{48}$TiO, is shown with the dashed line and the shaded area corresponds to 1$\sigma$ uncertainty of the fit. Grey points and an arrow (lower left) correspond to measurements of $^{46}$TiO and $^{50}$TiO, and to an upper limit for $^{47}$TiO. The two 2015 measurements for $^{46}$TiO and $^{50}$TiO scaled by reciprocal of their solar abundances are shown with cyan diamonds. They were included in the final fit.}\label{fig-rt}
\end{figure}

We used the rotational-diagram method \citep{popdiagr} to derive the temperature and column density of TiO that is traced in the main emission component. The diagram is shown in Fig.\,\ref{fig-rt}. A fit to all lines, including the line tentatively detected with {\it Herschel}, yielded an excitation temperature $T_{\rm ex}$=491$\pm$96\,K and a column density $N$($^{48}$TiO)=(4.0$\pm$1.0)$\cdot$10$^{16}$\,cm$^{-2}$ for a source size of 116$\times$93\,mas. The errors quoted here represent only the quality of the fit. As discussed in Paper\,I, the values of $T_{\rm ex}$ and $N$ are derived under the assumption that all lines are optically thin, the level populations follow the Boltzmann distribution, the rotational temperatures in the three fine-structure components are equal, and the line fluxes do not change with time or phase so that data from different epochs and different spin components can be combined. The rotational lines of TiO are not formed in the immediate proximity of shocked regions (Sect.\,\ref{sec-spkin-TiO}) so LTE conditions are likely to apply. Some other assumptions might be violated in Mira, but our estimates here are the best constraints on $T_{\rm ex}$ and $N$ possible with the current data.

After we constrained the excitation temperature of the main TiO isotopologue, we used the measured fluxes of the rare TiO species to calculate their relative abundances. This is done under the assumptions that the different species are characterized by the same temperature and there is no isotopic fractionation. Both conditions are very likely fulfilled in the atmosphere of Mira. The ratios with 1$\sigma$ uncertainties are given in the first row of Table\,\ref{tab-iso}. In addition, we obtain $^{47}$Ti/$^{48}$Ti<0.08 from the flux ratio of the $\Delta_3$ $J$=11$\to$10 transition of $^{48}$TiO and the upper limit of $^{47}$TiO,   corrected for the intrinsic lines strengths.


The excitation analysis of TiO$_2$ is more difficult. Although many transitions in a broad range of $E_u$ were covered, the severe blending of lines leaves only four relatively clean lines whose fluxes could be measured over the entire profile. (The lines tentatively detected with APEX were not taken into account in our analysis here.) A simple rotational-diagram fit to those four measurements yielded $T_{\rm rot}=180\pm67$\,K and $N$(TiO$_2$)=$(9.5\pm1.8)\cdot 10^{15}$\,cm$^{-2}$ (for a fixed source solid angle of 0.0138\,mas$^2$). However, these results are not very satisfactory owing to the insufficient coverage of $E_u$, i.e. only 60--137\,K, and the poor S/N of the measurements. To make a better use of the data, we employed the LTE package of CASSIS\footnote{\url{http://cassis.irap.omp.eu}} and performed a $\chi^2$ minimization in which a grid of models was compared to ten observed line profiles. In the models, we approximated the profiles by a Gaussian with a FWHM of 8.5\,\kms\ and a central velocity of 46.8\,\kms. Only the uncontaminated parts of the profiles were included in the minimization procedure. In the CASSIS approach, we also used the information on lines that were not detected extending the analysis to transitions from energy levels as high as $E_u$=904\,K. Optical-depth effects were taken into account in the routine but the lines turned out to be optically thin (with $\tau_0 \leq 1.0$ for line centers). The parameters best characterizing the TiO$_2$ gas, $T_{\rm rot}=174\pm7$\,K and $N$(TiO$_2$)=$(1.15\pm0.12)\cdot 10^{16}$\,cm$^{-2}$ (1$\sigma$ errors), were used to simulate the spectrum shown in Fig.\,\ref{fig-allTiO2}. The single-component model reproduces the data relatively well but the observed line profiles are not exactly Gaussian. It is possible that our isothermic model is not adequate for the TiO$_2$ gas but the quality of data and their spread in parameters space do not warrant multi-component models. 

With the temperature constrained for TiO$_2$, we next used CASSIS to derive the column densities for the rare isotopologues. By analyzing the two lines of $^{46}$TiO$_2$, we derived $N$($^{46}$TiO$_2$)=(3.7$\pm$0.3)$\cdot 10^{15}$\,cm$^{-2}$ or the abundance ratio of $^{46}$TiO$_2$/$^{48}$TiO$_2$=0.32$\pm$0.04. This result is not fully satisfactory because our fit to $^{48}$TiO$_2$ profiles is particularly imperfect in the line center which dominates the profile of the low-S/N features of $^{46}$TiO$_2$ (cf. Figs.\,\ref{fig-allTiO2} and \ref{fig-allTiO2-iso}). From half a profile of the only feature of $^{50}$TiO$_2$ we obtain an upper limit $N$($^{50}$TiO$_2$)$\leq$2.9$\cdot 10^{15}$\,cm$^{-2}$ or $^{50}$TiO$_2$/$^{48}$TiO$_2 \leq$0.25. Upper limits on several transitions of $^{47}$TiO$_2$ imply $N$($^{47}$TiO$_2$)<6$\cdot 10^{14}$\,cm$^{-2}$ or $^{47}$TiO$_2$/$^{48}$TiO$_2$<0.05. Similarly, for $^{49}$TiO$_2$ we were able to derive an upper limit only, $N$($^{49}$TiO$_2$)<5$\cdot 10^{14}$\,cm$^{-2}$ or $^{49}$TiO$_2$/$^{48}$TiO$_2$<0.04. These CASSIS results, summarized in the third row of Table\,\ref{tab-iso} take into account 1$\sigma$ noise levels and do not suffer from the low-S/N problem or from saturation effects as in the rotation-diagram method. 

The isotopic ratios can also be derived by comparing the same transition of the rare isotopologue to that of $^{48}$TiO$_2$. Such a direct comparison is possible for: $12_{4,8} \to 11_{3,9}$ of $^{50}$TiO$_2$, $10_{5,5} \to 9_{4,6}$ of $^{49}$TiO$_2$, $9_{5,5} \to 8_{4,4}$ of $^{48}$TiO$_2$. (Although the transition $23_{3,21} \to 22_{2,20}$  of both $^{46}$TiO$_2$ and $^{48}$TiO$_2$ was observed, the direct flux ratio does not give a good measure of their relative abundances because the $^{48}$TiO$_2$ line is contaminated by emission of an unknown origin.) The results are given in the fourth row of Table\,\ref{tab-iso}. Note that the isotopic ratios derived this way are independent of the actual temperature but may be in principle underestimated owing to opacity effects. 

All the constraints on the isotopic ratios are summarized in Table\,\ref{tab-iso}. Within errors, the combined results are indicative of relative abundances that are not drastically different from solar values except for $^{47}$Ti which may be under-abundant by a few percent. We made advantage of this result to improve our temperature and column density determination for TiO. We scaled the 2015 ALMA flux measurements for $^{46}$TiO and $^{50}$TiO by their abundances relative to $^{48}$TiO and included them in the rotational diagram fit of TiO. This improved the quality of the fit and yielded $T_{\rm ex}$=474$\pm$69\,K and  $N$($^{48}$TiO)=(4.1$\pm$0.7)$\cdot$10$^{16}$\,cm$^{-2}$, which we treat as our best estimates for TiO.  

Useful for our further discussion are the abundances of TiO and TiO$_2$ with respect to hydrogen. As discussed in Paper\,I there is currently no reliable way to estimate the content of hydrogen in the probed regions. Nevertheless, hydrodynamical models of Mira of the CODEX grid \citep{ireland2011} give the hydrogen column density in the range $10^{22}-10^{24}$\,cm$^{-2}$ which indicates TiO abundance of $10^{-8}-10^{-6}$. Values higher than 10$^{-7}$ are unrealistic as would require the elemental abundance of titanium to be higher than the cosmic value (1.0$\times$10$^{-7}$). The source-averaged abundances of TiO and TiO$_2$ must be therefore of the order of $10^{-8}-10^{-7}$, but their uncertainties are uncomfortably large. 

No reliable excitation analysis was possible for the warm TiO component seen in the infall and the -- presumably cool -- gas manifested in the broad wings of TiO$_2$.

\begin{table}
\caption{Isotopic ratios.}\label{tab-iso}
\small
\centering
\begin{tabular}{c cc cc}
\hline\hline
Method & $^{46}$Ti/$^{48}$Ti & $^{47}$Ti/$^{48}$Ti & $^{49}$Ti/$^{48}$Ti & $^{50}$Ti/$^{48}$Ti\\
\hline\hline
TiO rot.diagr.     & 0.12$\pm$0.03 & <0.12 & & 0.08$\pm$0.02\\
TiO line ratio     &               & <0.08 & &              \\
TiO$_2$ CASSIS     & 0.32$\pm$0.04\tablefootmark{a} & <0.05 & <0.04 & $\lesssim$0.25\\
TiO$_2$ line ratio &               & <0.06 & <0.08  & <0.22\\[3pt]
Solar              & 0.11          & 0.10  & 0.07   & 0.07 \\ 
\hline
\end{tabular}
\tablefoot{The uncertainties are 1$\sigma$ random errors. 
\tablefoottext{a}{An uncertain result with high systematic errors.}
}
\end{table}

%

%
%
\section{Optical spectroscopy of Ti- bearing species}\label{opt}

\subsection{Neutral atomic titanium}\label{sec-Ti}
Several electronic multiplets of \ion{Ti}{I} are observed in absorption in the optical spectra of Mira, including some resonance transitions. Examples are presented   in Figs.\,C.1--C.5 of Paper\,I. They all are photospheric lines. One striking exception of a line that does not appear as a pure absorption feature is the resonance line $a^3F_2-z^3F_3$. Its upper energy level at $E_u$=27\,800\,K \citep{labTiI}, is the lowest one among all covered transitions of \ion{Ti}{I}. The line is dominated by emission in a great majority of our multi-epoch spectra. With a laboratory wavelength of 5173.74\,\AA, the line is located within the deep absorption band $\alpha$(0,0) of TiO. The molecular lines are saturated and tightly overlap with each other making the pseudo-continuum relatively flat in this spectral range. Owing to this location, \ion{Ti}{I} $\lambda$5173 is the most conspicuous line in the $a^3F-z^3F$ multiplet. 

At close inspection, the feature has an inverse P-Cyg type profile. Although such a profile usually indicates infall, in this case the feature is formed mainly by an emission profile overlapping with a photospheric absorption line of \ion{Ti}{I}. The photospheric origin of the absorption is supported by: (1) the presence of pure absorption profiles in the majority of \ion{Ti}{I} multiplets observed; (2) the presence of an absorption line of nearly the same intensity and width in stationary model atmospheres of MARCS \citep{MARCS} and for stellar parameters adequate for Mira; and (3) the central velocity of the absorption feature, corrected for the overlapping emission, corresponds well to the velocity of the optical photosphere. Inverse P-Cyg profiles of non-resonant ($E_u$=21\,760 and 25\,100\,K) lines of \ion{Ti}{I} were also observed in the near-infrared \citet{TiI} but were incorrectly interpreted as direct signatures of infall. 
%

A profile of the \ion{Ti}{I} emission line is compared to pure rotational features of TiO and TiO$_2$ in Fig.\,\ref{fig-profiles-TiTiOTiO2}. It is an average of two \ion{Ti}{I} spectra taken at phases 0.17 and 0.24 in 2010 from which we subtracted a photospheric spectrum. The latter was a MARCS synthetic spectrum shifted to a velocity of $V_{\rm LSR}$=47.6\,\kms. At this velocity the simulated electronic structure of the TiO $\alpha$(0,0) band matches best the observations.\footnote{Note that optical photosphere is not expected to be stationary and therefore its velocity is not necessarily the same as that of the center of mass.} The center and width of the \ion{Ti}{I} emission at phase $\sim$0.2 is the same as for the majority of molecular emission features arising in the extended atmosphere and traced with ALMA. In particular, the profile is almost the same as that of the pure rotational line of TiO. 
 
The \ion{Ti}{I} $\lambda$5173 emission is seen in every pulsation cycle and therefore is not an episodic phenomenon as some Al-bearing emission features reported in Paper\,I. When analyzed in the multi-epoch spectra, the emission intensity relative to that of continuum shows considerable variations. The relative intensity is lowest at visual maxima and systematically increases with phase reaching a maximum of 130\% near visual minima. Those changes are modulated by the continuum flux variations by huge factors of 158--1580 (5.5--8.0 mag), depending on the cycle (Paper\,I). The \ion{Ti}{I} emission follows these dramatic changes displaying also intrinsic line variations at a level of 30\%. Absolute intensities of the emission component cannot be measured directly in our spectra owing to unreliable flux calibration and an uncertain contribution of the photospheric component in each phase.

The apparent peak of the \ion{Ti}{I} emission profile changes its position over time between $V_{\rm LSR}$=43.8 and 47.3\,\kms\ with no obvious correlation with phase. These variations are most likely caused by the overlap of the absorption and emission components which both vary in time. Near maximum visual light, when the emission component is weak with respect to  the continuum and with respect to the photospheric line, the absorption component dominates the entire profile shifting the net emission peak toward longer wavelengths. The intrinsic changes in the velocity of the pure emission component must thus be considerably smaller than 3.5\,\kms\ and, in fact, we believe that the line is stationary. It is most apparent in the photosphere-corrected profile shown in Fig.\,\ref{fig-profiles-TiTiOTiO2} which is centered at the stellar center-of-mass velocity. In profiles dominated by the emission component, the FWHM is of 5.5$\pm$0.5\,\kms\ and the full base width is of about 11\,\kms, which are characteristic for the material in the extended atmosphere seen with ALMA data at radii $\lesssim$4\,\rstar.  


To produce the \ion{Ti}{I} $\lambda$5173 transition in emission, it may seem necessary to populate its upper energy level, which is 27\,800\,K above the ground. Direct collisional excitation of this state can be entirely excluded. Although the pulsation shock is capable of producing such high-excitation conditions, any shock-triggered emission would show dramatic changes in intensity and velocity which are not observed in \ion{Ti}{I}. For the same reasons we can exclude influence of the radiation of the shock, whether it is a direct excitation or fluorescence \citep[cf.][]{willson1976}.   

An alternative mechanism that can explain the \ion{Ti}{I} emission is resonant scattering. The random shifts in the centroid position of \ion{Ti}{I} with a low amplitude of <3.5\,\kms, if present, could be partially explained by an inhomogeneous and variable distribution of the scattering medium. Such asymmetries have been observed close to the photosphere of Mira in the distribution of gas species (Paper\,I) and of hot dust \citep{lopez}. Additionally, for resonant scattering one would expect the scattered flux to be tightly correlated with the intensity of the continuum and, in consequence, the continuum-normalized line intensity should be roughly constant with phase. Such a correlation is observed in $\lambda$5173 but is not strict. Considering the enormous visual light variations of Mira with flux changing by a factor of 250 for the epochs covered by our optical spectra, the reported variability of 30\% in the continuum-normalized line flux can be considered insignificant and may be partially explained by line-saturation effects. Resonant scattering is therefore a reasonable explanation of the \ion{Ti}{I} emission. 

The resonantly scattered emission of \ion{Ti}{I} does not require high electron densities and temperatures but requires sufficient population of the ground level. This requirement suggests low-excitation conditions for the \ion{Ti}{I}-bearing gas. The two strongest lines of the multiplet to which $\lambda$5173 belongs have their lower levels at 245\,K and 557\,K. We did not analyze these lines here because they are severely distorted by overlapping TiO features. It is evident, however, that they produce strong emission features similar to that of $\lambda$5173. At the same time, \ion{Ti}{I} multiplets with higher-lying levels are seen only in absorption. Although we are unable to calculate the excitation temperature because of the heavy line blending, we can conclude that the \ion{Ti}{I} emission must arise from circumstellar material of a temperature of a few hundred K. It is also important to note that the sole fact that we see the \ion{Ti}{I} emission within the saturated $\alpha$ band of TiO implies that the Ti-bearing gas must be located above the absorbing layers of TiO. This electronic molecular band is opaque enough to absorb any radiation from layers lying beneath its carrier. The most likely location of \ion{Ti}{I} is the extended atmosphere of Mira.


\subsection{Titanium ion, \ion{Ti}{II}}\label{sec-TiII} 
We were unable to identify any spectral features indicative of significant amounts of \ion{Ti}{II} in the envelope of Mira. Our optical spectra cover only one ground-state transition of \ion{Ti}{II} which however is a forbidden transition and is not useful for probing the dense shock-excited photospheric gas where this ion is expected to reside \citep{willson1976}. We found a weak emission feature at the position of the [\ion{Ti}{II}] line (7916.2\,\AA\ corrected to the center-of-mass velocity) but were unable to establish whether it is the atomic line or a rotational line of overlapping molecular bands. \citet{fox} postulated the presence of the \ion{Ti}{II} line at 4307.9\,\AA\ which should be blended with a line of \ion{Fe}{I}. We found a strong emission feature in some of our spectra at this position but other lines of the same \ion{Ti}{II} multiplet are missing. This raises questions about this identification. Archival ultraviolet spectra of Mira, e.g. those of \citet{HSTspec}, cover a strong resonance line of \ion{Ti}{II} at 3383.8\,\AA, but do not show anything conspicuous at this position, possibly owing to the lack of spectral resolution and sensitivity.

We conclude that the observations in hand are not well suited to conclusively investigate the presence of titanium ions in Mira but it is very likely that their abundances are negligible.  

\begin{figure*}
\sidecaption 
\includegraphics[angle=0,width=12cm]{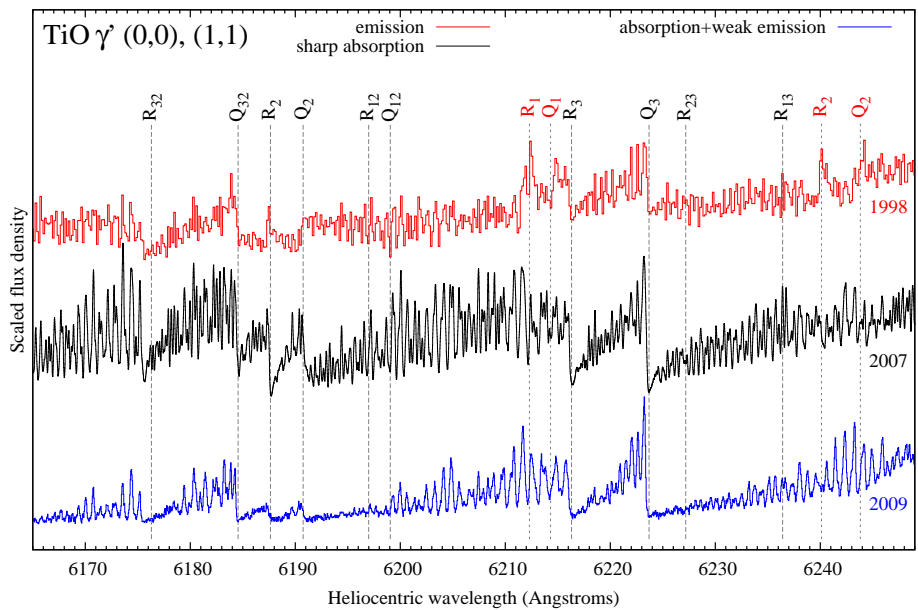}
\caption{Sample spectra of the TiO $\gamma^{\prime}$ (0,0) and (1,1) bands illustrating: the emission-dominated bands (top, red; 7 Nov. 1998); sharp absorption band-heads (middle, black; 20 Jan. 2007); and absorption bands partially filled with emission (bottom, blue; 8 Aug. 2009). The full collection of spectra is shown in Fig.\,\ref{fig-TiOgammaPrime}. Main TiO features are identified with vertical dashed lines, with black and red labels corresponding to the (0,0) and (1,1) bands, respectively.}
\label{fig-TiOoptExamples}
\end{figure*}

\subsection{TiO emission}\label{sec-opt-TiO}
The absorption spectrum of TiO has been recognized in $o$\,Ceti a century ago \citep{Fowler1909}. These variable bands are formed in the photosphere and are responsible for the dramatic changes in the visual light of Mira over its 333-day period \citep{reid02}. It is difficult to identify any \emph{circumstellar} features in the absorption spectra because of a very complex structure of the numerous and overlapping photospheric bands and because of incomplete laboratory measurements of TiO line positions. 

Circumstellar material is manifested in electronic bands when they are seen in emission. In Paper\,I, we reported emission in optical bands of AlO which is very episodic and may be related to shock phenomena. From the plethora of electronic bands of TiO covering the entire optical region, we noticed emission features only in the $\gamma^{\prime}$($\equiv B-X$) (0,0) and (1,1) bands between 6170 and 6250\,\AA. All spectra covering this range are shown in Fig.\,\ref{fig-TiOgammaPrime}. In the 35 spectra, strong emission was observed only once. There are however hints that weak emission appears in the same bands much more often but is almost completely consumed by the overlapping photospheric absorption. Examples of TiO spectra affected by emission and free of it are shown in Fig.\,\ref{fig-TiOoptExamples}.  

Strong and obvious emission of $\gamma^{\prime}$ was observed on 7 November 1998 near a visual phase of 0.9. Such TiO flaring must be extremely rare. On the same date, the AlO bands were also seen in emission (Paper\,I). The TiO emission is most apparent in the (1,1) transitions, in particular involving the $R_1$ and $R_2$ branches (at $\sim$6212 and $\sim$6240\,\AA\ in Fig.\,\ref{fig-TiOoptExamples}). It is also evident near the $R_2$ band-head of (0,0) ($\sim$6188\,\AA). In addition, many of the $\gamma^{\prime}$ (0,0) absorption band-heads appear much shallower than in a typical spectrum of Mira at the same phase because they are filled with emission components, as illustrated in Fig.\,\ref{fig-TiOoptExamples}. The S/N of the emission-dominated features is too low to constrain the temperature and kinematics of the emitting gas. However, from the appearance of the (1,1) emission we can infer that the emitting gas must have a temperature lower than about 1000\,K because otherwise emission would arise in a broader range of rotational transitions.    

Other identified emission features are much less conspicuous. The TiO bands usually show sharp triangular absorption band-heads that develop toward longer wavelengths but starting from 29 October 2008 the absorption heads became blunt and remained so till at least 2012. The last spectrum in our collection, from 2014, shows the sharp absorption heads back again. Two example spectra of sharp and blunt heads are compared in Fig.\,\ref{fig-TiOoptExamples} (black and blue, respectively). The difference is most pronounced in $R_2$ (6188\,\AA) and $Q_2$ (6224\,\AA) and band-heads of (0,0). When the spectra affected by weak emission are analyzed one by one in the chronological order (Fig.\,\ref{fig-TiOgammaPrime}), some variability in the emission component can be seen.


The upper electronic level of $\gamma^{\prime}$ at $E_u\!\approx$22\,800\,K is below that of the optical AlO band ($E_u\!\approx$28\,800\,K) which was identified in emission in Paper\,I. Our spectra cover electronic systems of TiO that have their upper levels lying above (e.g. $a-f$ and $\alpha$) and below (e.g. $\gamma$ and $\epsilon$) that of $\gamma^{\prime}$ but none shows signs of emission. It is unclear why the emission episode was limited to one electronic system of TiO. The emission cannot be associated with a regular pulsation shock because then it would appear in every cycle. On the other hand, there are no other potential sources that could produce emission from such high-lying energy levels by collisional excitation only. Taking the episodic nature of the strong emission event and its similarity to what was observed in electronic bands of AlO, we propose that the emission is triggered by an unusually strong shock that excites material in the parts of the atmosphere that are quiescent in a regular cycle \citep[cf. Paper\,I,][]{richter2003}. 

The origin of the weaker emission in the $\gamma^{\prime}$ band of TiO, which appears in the recorded spectra much more often, may be different. Resonantly scattered emission seems a reasonable mechanism to explain it. The inhomogenities present in the closest vicinity of the star and the ever-changing distribution of gas around the stellar photosphere can be responsible for the variable nature of this weak emission component. One can further speculate that the emission episode from 6 Nov. 1998 was an extreme case of this behavior and geometrical asymmetries produced a large volume of TiO-bearing gas surrounding the stellar disk that by far suppressed the absorption. However, this scenario does not explain why the emission appeared in only one electronic system.  

Mira was observed close to its 1990 maximum using techniques of aperture-masking interferometry by \citet{haniff}. The object was imaged at a resolution of $\sim$30\,mas in the pseudocontinuum at 7007$\pm$4\,\AA\ and within a spectral region 7099$\pm$4\,\AA\ which is dominated by the TiO $\gamma$ (0,0) band. The source imaged within the TiO band appeared $\sim$1.5 times larger than in pseudocontinuum. That led Haniff et al. to a conclusion that there is a halo of resonantly scattered emission of TiO around the stellar photosphere. The $\gamma$ system has the upper electronic level 20\,220\,K above the ground, only slightly below that of $\gamma^{\prime}$(0,0) discussed above. It is unlikely that the imaging observation caught the star during an episode of strong TiO emission, like the one seen in 1998 in $\gamma^{\prime}$. Rather, it was emission analogous to the $\gamma^{\prime}$ features that we observed in 2008--2012. These interferometric observations support our interpretation of the weak TiO emission as resonantly-scattered continuum. We do not see the emission component of $\gamma$ bands in our spectra owing to insufficient angular resolution. The interferometric data also locate the scattering medium at a radius of $\sim$3\,\rstar\ and demonstrate that the distribution of TiO has some degree of anisotropy (their Fig.\,3d). The circumstellar emission of TiO seen at optical wavelengths is therefore at least partially collocated with that traced in the pure rotational lines at submm wavelengths in Mira's extended atmosphere (cf. Sect\,\ref{sec-spkin-TiO}).

\paragraph{TiO isotopologues:} Using our optical spectra, we attempted to constrain the isotopic ratios of the different TiO isotopologues in a similar fashion as in \citet{Wyckoff1972} and \citet{Chavez}. Even in the spectra of highest dispersion, however, the isotopic effects are too subtle to reliably measure contributions of the different isotopic species.  In a procedure described in Appendix\,\ref{app2}, we found that the overall enhancement of the four rare isotopologues must be less than about 25\%.

\subsection{TiO$_2$ absorption} 
One known electronic system of TiO$_2$, $\tilde{A}^1 B_2-\tilde{X}^1 A_1$, is located between about 5360 and 5450\,\AA\ \citep{TiO2-vis,TiO2_A-X_dip_2009}. It overlaps with part of the $\Delta \varv$=--2 progression of AlO $B-X$ (very weak in Mira) and is located within the $\gamma^{\prime}$(3,0) band of TiO. Presumably, the dominant source of opacity at these wavelengths is TiO, making the identification of any TiO$_2$ features very challenging. We performed a simulation of the TiO$_2$ bands in {\tt pgopher} \citep{pgopher} using spectroscopic constants from the literature. (Although, the spectroscopic parameters presented in \citet{TiO2_A-X_dip_2009} do not reproduce the full spectrum of TiO$_2$ generated in laboratory conditions.) The simulation shows a profile whose general shape corresponds well to what is seen in the spectra of Mira at different phases of its cycle. However, discrete features of TiO$_2$ are not conspicuous in the observed spectrum, suggesting that TiO$_2$ bands are not present. A full model-atmosphere spectrum with a complete line list of TiO is necessary to verify the presence of TiO$_2$ in this spectral range. It should be noted that the published spectroscopic parameters we used represent only a small part of the electronic spectrum of TiO$_2$ and more spectroscopic work is necessary to verify if bands of TiO$_2$ are present in optical spectra of stars.  

\subsection{TiH absorption} 
To investigate the presence of TiH in the spectra of Mira we simulated the optical bands in {\tt pgopher}. We used the line list\footnote{\url{http://www.exomol.com/data/molecules/TiH/48Ti-1H/Yueqi/}} compiled by Z. Yueqi Na for the ExoMol project \citep{exomol} and based on the spectroscopic data of \citet{TiH}. In a broad range of temperatures characteristic of the circumstellar gas and of the photosphere, the strongest optical bands of TiH are the $B^4\Gamma-X^4\Phi$ and $A^4\Phi-X^4\Phi$ systems at about 5100--5385\,\AA\ and 9380--9700\,\AA. A direct comparison between the simulations and the observed spectra at different phases did not reveal any conspicuous features of TiH. These spectral ranges are dominated by the photospheric bands of TiO $\gamma^{\prime}$, AlO $B$--$X$, and H$_2$O (including the telluric bands). A comprehensive test for the presence of TiH would require removing the contribution of the other species to the observed spectra but this is currently not feasible  because of the same reasons as mentioned for TiO$_2$. Our analysis shows, however, that TiH cannot be a major Ti carrier in Mira and its envelope.

For the other potential molecular carriers of Ti mentioned in Sect.\,\ref{sec-species}, we could not find appropriate spectroscopic data (TiN, TiF, TiF$_2$) or the spectral signatures are located in ranges that were not covered by our spectra (TiS).

\section{Discussion}\label{discussion}
\subsection{Tracing titanium chemistry in O-type AGB stars} 
Our multi-wavelength observations of Mira allow us to trace the Ti-bearing species in the different phases of the photospheric and circumstellar gas. We compare the gas characteristics deduced from the observations with chemical models to better constrain the titanium chemistry in an O-rich star like Mira.

Chemical abundances in the {\it photosphere} are thought to be well reproduced by predictions of equilibrium chemistry \citep{tsuji,cherchneff2006}. At the photospheric temperature of Mira of $\sim$3000\,K, titanium should be locked in TiO and in nearly equally-abundant atomic Ti \citep[e.g.][]{tsuji,sharp}, with a negligible contribution of TiO$_2$. For solar composition, the abundance of TiO is expected to be 10$^{-7}$ with respect to H$_2$. The atomic Ti and TiO in the photosphere produce very conspicuous and well known spectral features. Although we did not attempt to derive the photospheric abundances, all our observations are consistent with the dominant presence of TiO and \ion{Ti}{I} in the photosphere.  


The \emph{circumstellar} chemistry of titanium is more complex. In first attempts to understand circumstellar chemistry, it has often been assumed that there is a smooth transition in abundances from the photosphere to the extended atmosphere and to the wind, and that all the abundances are given by the thermal-equilibrium (TE) conditions \citep[e.g.][]{tsuji,sharp,GS98}. However, this approach was refuted on observational and theoretical grounds in pulsating Mira stars, whose atmospheres are strongly affected by the periodic passage of shocks and which effectively produce molecules that are predicted to be absent in TE \citep[][and references therein]{cherchneff2006,cherchneff12}. Several models based on chemical kinetics aimed to describe and reproduce the molecular abundances of the shocked regions in O-rich AGB stars, including \citet{duari}, \citet{cherchneff2006} and \citet{gobrecht}. The latter study presents a model of the Mira star, IK\,Tau with an effective temperature of 2200\,K, which is slightly lower than that of Mira, 2200--3200\,K (cf. Paper\,I). We assume here that this model  is representative of the chemistry of the Mira's envelope and  compare the model predictions with our observational constraints.


Photospheric TiO is expected to be destroyed by the supersonic shock and reformed in the cooling wake of the shock. We do not observe any signatures of shock-excited TiO molecules. Although our submm observations reveal the presence of the hot molecular gas from the post-shock region, through transitions of HCN and H$_2$O (Sect.\,\ref{sec-env}), we do not observe TiO from there. However, our observations are essentially limited to a single phase and strongest TiO emission may occur at earlier phases when the shock is penetrating deeper layers of the star which are obscured by the radio photosphere. We reported in Sect.\,\ref{sec-opt-TiO} one instance of optical observation of TiO bands suggesting strong influence of shocks but this must have been a manifestation of a phenomenon occurring only in some cycles, possibly only those characterized by exceptionally strong shocks. Also, these emission episodes likely affect the gas of the extended atmosphere where shocks are too weak to directly destroy strongly-bound molecules such as TiO.   

If TiO is fully destroyed by the supersonic shock, it reforms in the post-shock material in the photosphere. The production of TiO from \ion{Ti}{I} continues in the extended atmosphere whose lower temperature allows for oxidation of titanium. The formation of TiO decreases the content of \ion{Ti}{I}. In general, the lower the temperature the higher the consumption of \ion{Ti}{I} in the extended atmosphere. Because the temperature changes with phase, the abundance of atomic Ti is expected to undergo cyclical changes and its highest abundance is expected close to a visual maximum. Our optical observations reveal that \ion{Ti}{I} is present in the extended atmosphere (Sect.\,\ref{sec-Ti}) but we could not quantify how much of Ti remains in the atomic gas. We can state with confidence, however, that the oxidation in the circumstellar medium is not 100\% efficient. In addition, we observe cyclical variability in the emission features of \ion{Ti}{I}, but we are unable to ascribe them to the potential changes in the abundance. The model of \citet{gobrecht} predicts variations of the \ion{Ti}{I} abundance in the extended atmosphere of two orders of magnitude (Cherchneff, priv. comm.), which is comparable to the order of the optical light variations. The high-amplitude flux changes mask the intrinsic variability of the circumstellar features of \ion{Ti}{I} observed by us. Overall, the model predictions are in good quantitative agreement with our observations of \ion{Ti}{I}.

In cool circumstellar gas, the abundance of TiO should increase with radius at the expense of \ion{Ti}{I}. On the other hand, at low temperatures TiO is converted to TiO$_2$. For reference, in TE the two oxides have comparable abundances at 1380\,K \citep{GS98}. In the non-equilibrium model of \citet{gobrecht}, both conversion processes are insignificant for the overall content of TiO and its abundance remains nearly constant throughout the extended atmosphere and does not vary significantly with phase. For an abundance of the order of 10$^{-7}$ -- i.e. close to the TE abundance -- TiO is expected to be the dominant molecular carrier of Ti and we observe considerable amounts of TiO in the atmosphere of Mira at submm and optical wavelengths. Our abundance estimate, although highly uncertain, is of the same order of magnitude as predicted in the model. The bulk of TiO-bearing gas has an excitation temperature of 475\,K, indicative of a location within the extended atmosphere. The ALMA observations allow us to trace the beam-smeared emission out to a radius of 240\,mas or 17\,\rstar\ (as measured for an isophote corresponding to the 3$\sigma$ noise level) so some amount of gaseous TiO may be present in the silicate-formation zone and in the wind. Beyond that region, the TiO-bearing gas is not excited sufficiently to produce conspicuous emission or it is depleted. Our observations do not indicate significant variability in the rotational lines of TiO which confirms the phase-invariant abundance predicted by \citet{gobrecht}.

The model of \citet{gobrecht} does not predict significant production of TiO$_2$ in the extended atmosphere. The small quantities that are produced are related to shock chemistry. At 6\,\rstar, Gobrecht et  al. predict a TiO$_2$ abundance that is two orders of magnitude lower than that of TiO. Although the relative contribution of TiO$_2$ increases with distance from the star, it is a minor carrier of titanium in the model. Our observations indicate that although the TiO$_2$ content is smaller than that of TiO, the difference is not as extreme as in the simulations -- the observed source-averaged column-density ratio is TiO/TiO$_2$=3.4$\pm$0.9. The main emission components of both oxides have almost the same spatial extent and kinematics indicating that they largely coexist is the same region of the extended atmosphere. However, the TiO$_2$ emission region appears slightly more extended in some directions, its spectral features are slightly wider in velocity, and its excitation temperature of 174\,K is significantly lower than that of TiO, indicating that there are parcels of gas where the TiO$_2$-to-TiO ratio is increased. These TiO$_2$-enhanced regions must be located farther from the star, in agreement with TiO$_2$ forming directly from TiO and with the oxidation being more efficient farther from the star where the temperatures are lower \citep[cf.][]{GS98}. The model is in general agreement with our observational constraints, except that TiO$_2$ is produced more efficiently in Mira. An additional exploration of the model (Cherchneff, priv. comm.) shows that the TiO$_2$ abundance is expected to change with phase by orders of magnitude -- TiO$_2$ is destroyed by the shocks and it takes about 0.2 of the period to reform the molecules in significant quantities. This behavior cannot be verified because our TiO$_2$ data are limited to a single epoch. 


We conclude that the non-equilibrium model of \citet{gobrecht} is in good quantitative agreement with observational constraints on the gas-phase titanium species in Mira. It is worth noting that the observations are in stark contrast to the older models that assumed chemical equilibrium. The most apparent difference is the almost complete oxidation of TiO to TiO$_2$ at temperatures above 1000\,K in the models \citep{GS98,sharp} whereas observations indicate that both species are abundant at much lower temperatures.

\subsection{Implications for seed formation}
On the assumption that TiO$_2$ provides the first condensation nuclei in the heterogeneous formation of inorganic dust, we would expect that the first solid titanium compounds form at very high temperatures, 1000--1500\,K. Also, taking the low elemental abundance of Ti, the nucleation would have to be very efficient to provide enough condensation cores for formation of silicates at the observed rates. In this section, we confront these expectations with the observed characteristics of gas-phase Ti-bearing species in Mira.

It is often assumed that titanium seeds would consist of solid TiO$_2$ (titania) or/and its clusters and to a lesser degree\footnote{At TE, solid TiO does not form \citep{GS98}.} of solid TiO. Solid TiO$_2$ forms mainly from gas-phase TiO and solid TiO forms from gaseous TiO$_2$ \citep{GS98}. Formation of both types of solids requires therefore an efficient formation of the simple gas-phase oxides. Our observations indicate that most of the Ti-bearing molecular gas resides in regions of the extended atmosphere of temperatures of 170 and 500\,K (Sect.\,\ref{sect-exc}), i.e. regions where we could expect most efficient formation of the titanium solids owing to this high density of gas-phase oxides. If this condensation indeed took place therein, the dust temperatures would be low, definitely much below 1000\,K. At these temperatures, however, silicates form easily \citep{jeong,goumans,gobrecht} and there is no need to introduce titanium oxides to initiate silicate nucleation \citep{GS98,jeong}. Because we find no traces of hot titanium oxides in the circumstellar environment of Mira, the formation of hot titanium condensates is very unlikely.    

Our observations show that nearly all titanium is present in the gas phase, in atomic and oxidized forms. Because the derived abundances for TiO and TiO$_2$ are uncertain by more than one order of magnitude, there is a chance that some amount of Ti is locked into solid compounds but we think it is unlikely. The TiO$_2$ to TiO abundance ratio of 30\% indicates that the  production of gaseous TiO$_2$ is inefficient. One can expect that the transition to solids has also a limited efficiency and in consequence any solids, if formed, would have an abundance that is a small fraction of the elemental abundance of Ti. Because the cosmic abundance of Ti is very low compared to other metals, a high nucleation degree would be necessary to consider it a key player in dust formation. It is therefore apparent that the amount of solid titanium oxides is inadequate to create sufficient nucleation seeds.

Although we have shown that solid TiO$_2$ does not form close to the star at high temperatures, let us consider a formation of homo-molecular condensates of TiO$_2$ at larger distances from the star, where temperatures are a few hundred K. The low-temperature environment would pose strong constraints on the forms of solid TiO$_2$ that can form. The crystalline form, rutile, can form from anatase and brookite after annealing in temperatures higher than 1188\,K and 1023\,K, respectively \citep{BB92}. Therefore, one does not expect to observe rutile in O-rich AGB stars. This can be verified in future through infrared spectroscopy, although solid titanium compounds have not been conclusively identified in spectra of AGB stars (Sect.\,\ref{sec-intro}). 

Even if titanium oxides are not dominant in the formation of first hot mono-molecular condensates, titanium can still play a positive role in dust formation in evolved stars. The low abundance of gas-phase Ti in the interstellar medium is the strongest argument that Ti is depleted into dust \citep{churchwell}. Titanium can mix with more common oxides, helping to make the dust-formation process more rapid and efficient. \citet{goumans} proposed, for instance, an association reaction of TiO$_2$ and SiO resulting in TiSiO$_3$ that could be another potential source of first condensates. This and similar reactions may eventually lead to incorporation of Ti into dust. Another argument in support of the positive role of Ti in dust formation comes from studies of presolar grains. Titanium was found in a corundum (crystalline Al$_2$O$_3$) grain originating from an AGB star and it was proposed that, similarly to what is seen in synthetic Al$_2$O$_3$, trace amounts of Ti can stabilize the crystalline structure of corundum \citep{stroud}.   

A plausible alternative channel of heterogeneous formation of inorganic dust might start with alumina instead of titanium seeds. An attempt to test the role of alumina in dust formation was presented in Paper\,I. The direct precursor of alumina dust, Al$_2$O$_3$, is gaseous AlO \citep{gobrecht}. Its location within the extended atmosphere ($\lesssim$4\,\rstar),  excitation temperature (330\,K), and abundance relative to hydrogen (10$^{-9}$--10$^{-7}$) are very similar to those derived for TiO. The source-averaged column density ratio of TiO to AlO is $\sim$8. Although this ratio may change considerably within the inhomogeneous atmosphere, it is surprising that Al gas whose elemental abundance is 35 times higher than that of Ti produces a much smaller abundance of monoxide molecules. Because AlO was observed at five times better angular resolution than TiO, part of the discrepancy may arise from an inaccurate determination of the size of the TiO emission region. A more tempting interpretation of this high ratio is that the column density of AlO is low owing to its efficient depletion into dust. All observations presented in Paper\,I are in favor of this, but qualitative determination of the depletion rates have not been possible. Considering all the material and discussions in this series of papers, Al$_2$O$_3$  appears so far to be the most likely condensate that initiates the dust formation sequence in O-rich stars. 

An important question is whether our findings on the lack of Ti seeds in Mira can be generalized and applied to the entire family of O-rich AGB stars or even all O-rich evolved stars of different masses. Although $o$\,Ceti is one of the closest AGB stars and in many aspects is prototypical, it was chosen to be the main target of our study owing to relatively easily accessible observations covering several wavelength regimes and multiple epochs. The circumstellar chemistry leading to dust formation can be strongly influenced by parameters such as the shock strength, the pulsation period, the mass-loss rate, effective temperature and metallicity of the star. As mentioned in Paper\,I (Sect.\,1.3 therein), different types of O-rich AGB stars were identified through their mid-infrared spectra, including a group rich in silicates (of which Mira belongs) and those that appear to be dominated by alumina dust \citep{lorenz,little}. One can therefore expect different channels for formation, growth, and processing of grains around these stars and a large sample of objects should be studied to fully explore the various possibilities. Nevertheless, the insufficient formation of Ti seeds concluded here is in accord with findings of a similar study of the red supergiant VY\,CMa \citep{kami_tio,beck}. The characteristics of Ti-bearing gas-phases species in Mira and VY\,CMa are very similar although many parameters of the stars and their envelopes are different, e.g. the mass-loss rate of Mira is three orders of magnitude lower than of VY\,CMa. The combined results of these studies suggest that titanium oxides are an unlikely source of the first hot condensates in a broad range of circumstellar environments.          


\subsection{Conclusions}
The gas-phase chemistry of titanium in the photosphere and atmosphere of Mira is dominated by \ion{Ti}{I}, TiO, and TiO$_2$. The observed characteristics of the gas-phase Ti-bearing species rule out chemical models of the circumstellar gas that is in thermal equilibrium but they support the more realistic models that take into account shocks in the atmosphere. We do not find traces of hot circumstellar Ti-bearing gas which makes the presence of hot titanium solids very unlikely. Also, the abundance of the gas-phase species is too high for substantial amounts of titanium locked in solids. Although titanium oxides are unlikely to produce the seeds that initiate dust nucleation in warm envelopes of O-rich AGB stars, titanium might still participate and support production of silicate dust in these stars. In conjunction with the results presented in Paper\,I, we consider alumina dust to be a better candidate for the seed nuclei.

\begin{acknowledgements}
We thank Hans von Winckel for making available to us the Mercator spectra of Mira.
Based on observations made with the Mercator Telescope, operated on the island of La Palma by the Flemmish Community, at the Spanish Observatorio del Roque de los Muchachos of the Instituto de Astrofísica de Canarias. Based on observations obtained with the HERMES spectrograph, which is supported by the Research Foundation - Flanders (FWO), Belgium, the Research Council of KU Leuven, Belgium, the Fonds National de la Recherche Scientifique (F.R.S.-FNRS), Belgium, the Royal Observatory of Belgium, the Observatoire de Genève, Switzerland and the Th\"uringer Landessternwarte Tautenburg, Germany. 
We acknowledge with thanks the variable star observations from the AAVSO International Database contributed by observers worldwide and used in this research.
Based on data obtained from the ESO Science Archive Facility and made with ESO Telescopes at APEX and the La Silla and Paranal Observatories under programme IDs 074.D-0114(A), 089.D-0383(A), 097.D-0414(A), and 097.D-0414(B).
Based on observations obtained at the Canada-France-Hawaii Telescope (CFHT) which is operated by the National Research Council of Canada, the Institut National des Sciences de l'Univers of the Centre National de la Recherche Scientique of France, and the University of Hawaii. 
%
%
{\it Herschel} is an ESA space observatory with science instruments provided by European-led Principal Investigator consortia and with important participation from NASA.
%
%
Based on analysis carried out with the CASSIS software developed by IRAP-UPS/CNRS and CDMS and JPL spectroscopic databases. 
This paper makes use of the following ALMA data: ADS/JAO.ALMA\#2011.0.00014.SV, 2013.1.00047.S, 2012.1.00524.S, 2013.1.00156.S. ALMA is a partnership of ESO (representing its member states), NSF (USA) and NINS (Japan), together with NRC (Canada), NSC and ASIAA (Taiwan), and KASI (Republic of Korea), in cooperation with the Republic of Chile. The Joint ALMA Observatory is operated by ESO, AUI/NRAO and NAOJ.
\end{acknowledgements}

\begin{appendix}
\section{Mercator/HERMES spectra}
Table\,\ref{tab-mercator} presents details of the archival observations obtained with HERMES at the Mercator telescope.
\begin{table}
\caption{A log of Mercator/HERMES spectral observations of Mira.}\label{tab-mercator}
\small
\centering
\begin{tabular}{c llll}
\hline\hline
Date & JD & Exposure & Vis.  & Vis \\
     &    & time (s) & phase & mag \\
\hline\hline
2009-07-21&	2455033.7&	600	&0.63&	9.2\\
2009-08-08&	2455051.7&	900	&0.68&	9.1\\
2009-08-15&	2455058.7&	600	&0.70&	9.2\\
2009-09-06&	2455080.7&	600	&0.77&	8.9\\
2009-09-15&	2455089.6&	200	&0.79&	8.4\\
2009-10-02&	2455106.7&	100	&0.85&	8.0\\
2009-10-20&	2455124.6&	130	&0.90&	5.4\\
2009-10-26&	2455131.5&	100	&0.92&	4.6\\
2009-11-23\tablefootmark{a}&	2455159.5&	100	&0.00&	3.9\\
2010-02-07&	2455235.4&	 60	&0.23&	5.8\\
2011-01-04&	2455566.3&	  6	&0.23&	5.1\\
2014-08-31&	2456900.7&	 20	&0.23&	5.8\\
\hline
\end{tabular}
\tablefoot{
\tablefoottext{a}{Instrumental artifacts above 8000\,\AA.}
}
\end{table}

\section{Isotopic composition from optical bands of TiO}\label{app2}
We attempted to measure contributions of the five isotopologues of TiO to the optical absorption spectrum of Mira. Following \citet{Wyckoff1972} and \citet{Chavez}, we analyzed mainly the (0,0) band of the $\gamma$ system. For the spectral synthesis, we used a list of lines of \citet{plez} where we replaced original wavelengths of transitions in the main branches by those measured in laboratory by \citet{Ram}; for satellite branches, the wavelengths were computed from the spectroscopic constants provided by \citet{Ram}, while isotopic shifts were computed using Dunham's expansions of these constants. This modified line list was used to generate a synthetic spectrum from the MARCS model photosphere \citep{MARCS} with physical parameters best representing $o$\,Ceti (solar chemical composition, $T_{\rm eff}$=3200\,K, $log g$=0, microturbulence of 2\,\kms). The observations were represented by a spectrum acquired close to the maximum visual light, on 3 Oct. 2009. Spectra in maximum correspond to the most compact phase of the photosphere and are relatively well represented by the static MARCS model. Even with this careful simulation procedure, the synthetic spectrum reproduces the TiO band very roughly. Owing to this imperfect simulation, we analyzed the isotopic composition by scaling the cumulative abundance of all the four rare titanium isotopologues rather than just one. The observed spectrum is consistent with the solar isotopic composition of Ti and any enhancement of the rare species would have to be smaller than 25\%. Upper limits on enhancements of individual isotopologues would be even higher. 

\section{Optical spectra of Mira}
\clearpage
\begin{figure*} [tbh]
\centering
\includegraphics[angle=270,width=0.85\textwidth]{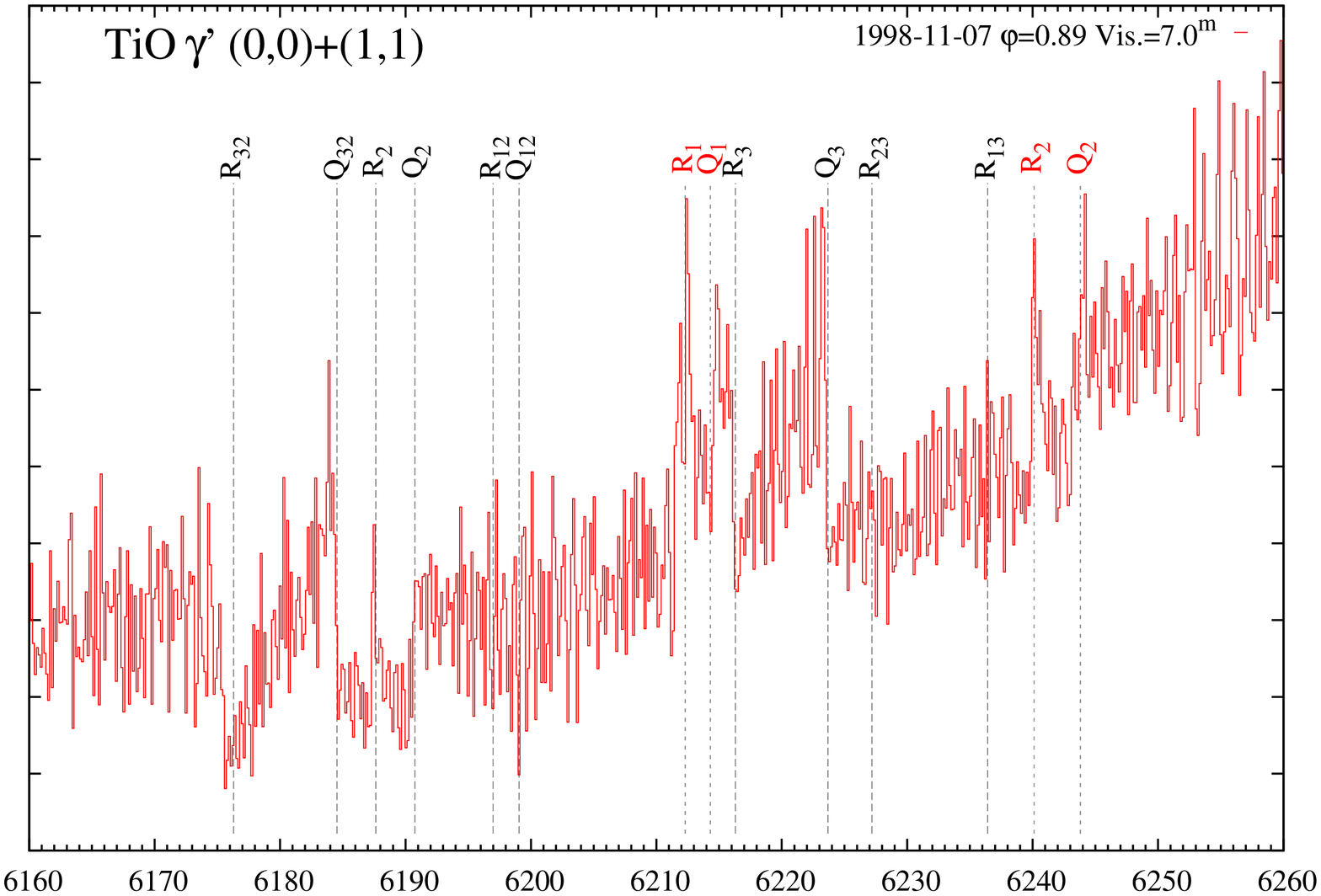}
\includegraphics[angle=270,width=0.85\textwidth]{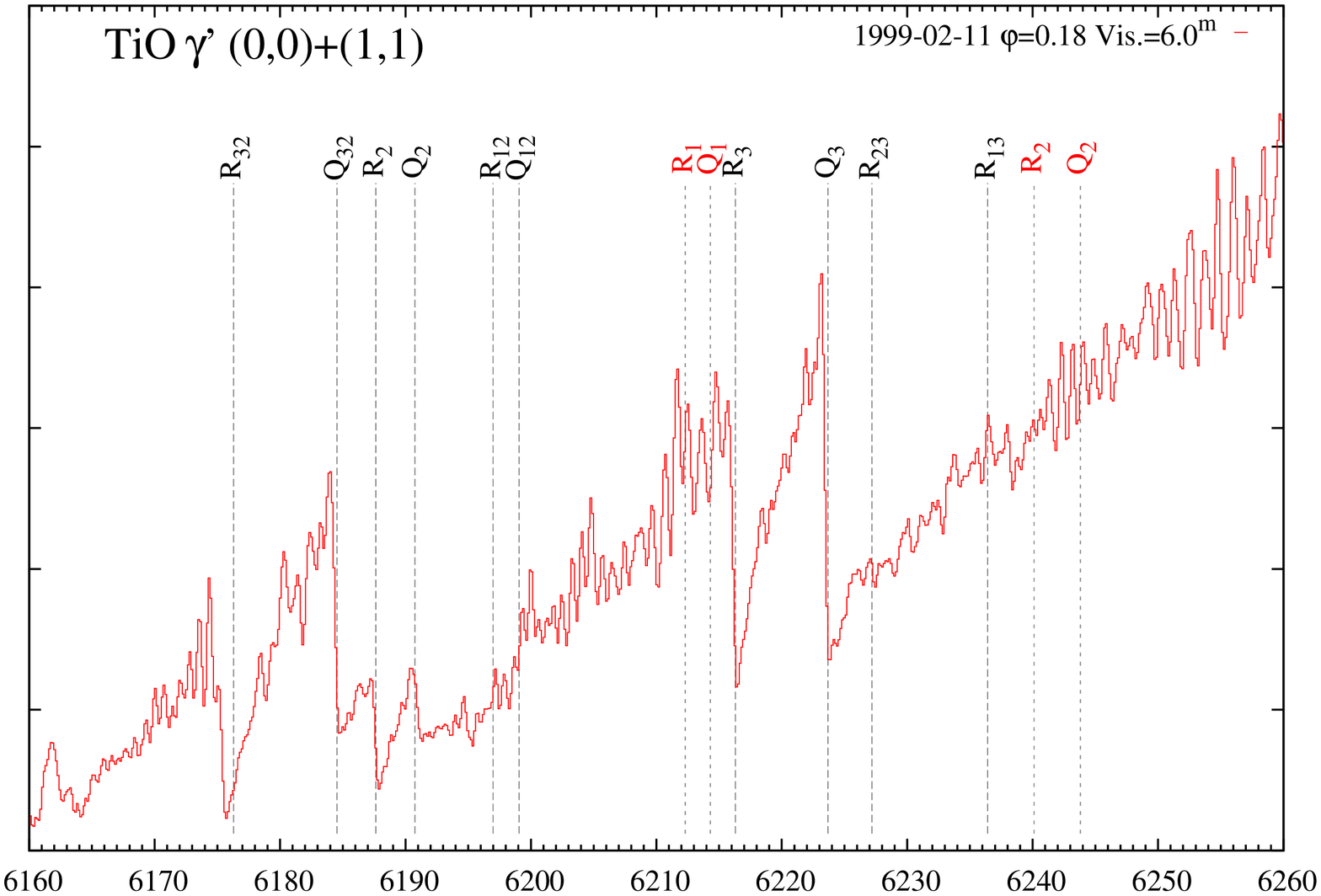}
\caption{Spectra of Mira covering the part of the $\gamma^{\prime}$ system of TiO that occasionally displays emission features. The date and the corresponding visual phase and magnitude are indicated in the top right corner of each panel. The spectra are shown in the chronological order. Branches forming main electronic features are indicated in black and red for the (0,0) and (1,1) bands, respectively. The wavelength scale in \AA\ is in the heliocentric rest frame.}\label{fig-TiOgammaPrime}
\end{figure*}

  \setcounter{figure}{0}%

\begin{figure*} [tbh]
\centering
\includegraphics[angle=270,width=0.85\textwidth]{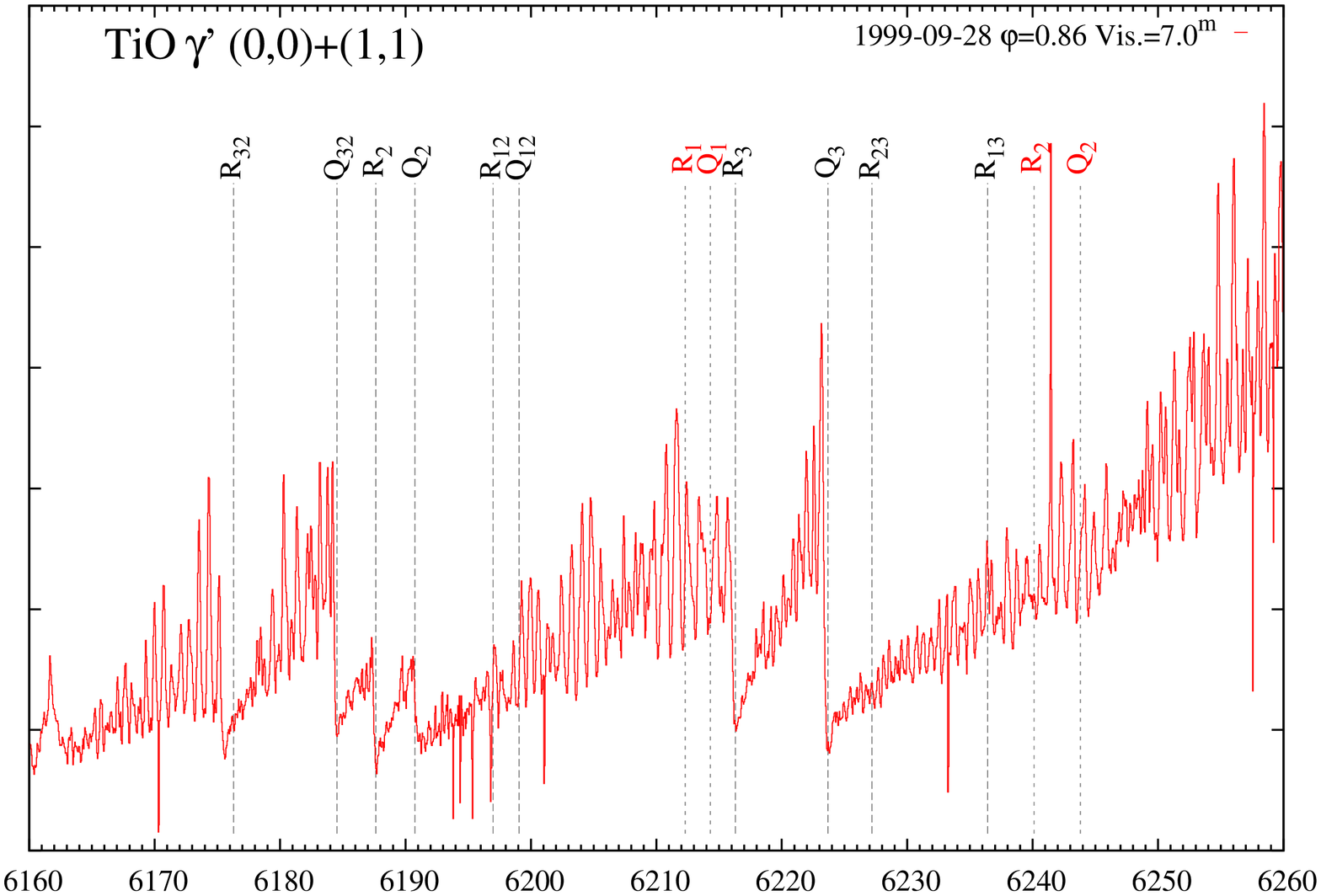}
\includegraphics[angle=270,width=0.85\textwidth]{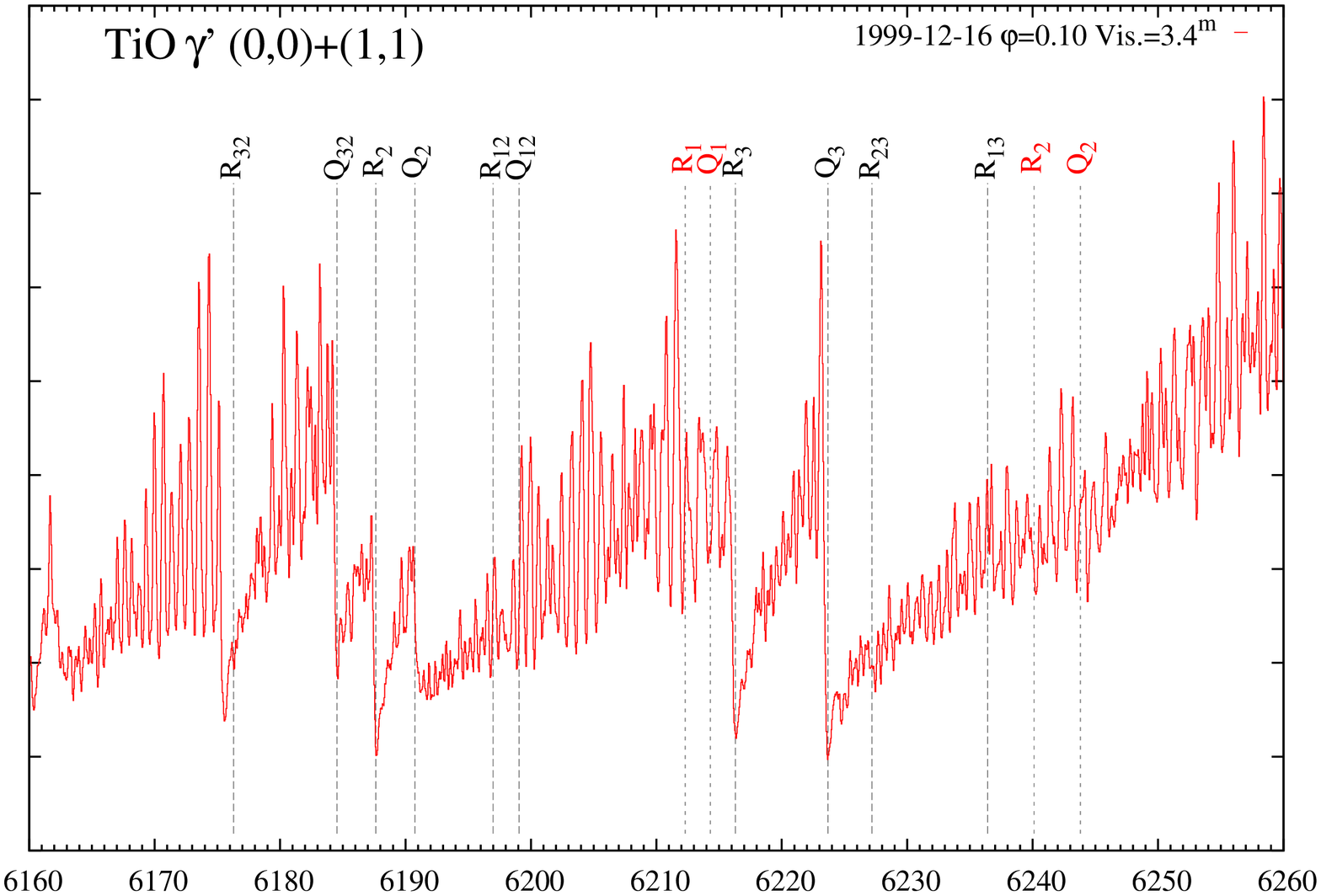}
\caption{Continued.}
\end{figure*}

  \setcounter{figure}{0}%

\begin{figure*} [tbh]
\centering
\includegraphics[angle=270,width=0.85\textwidth]{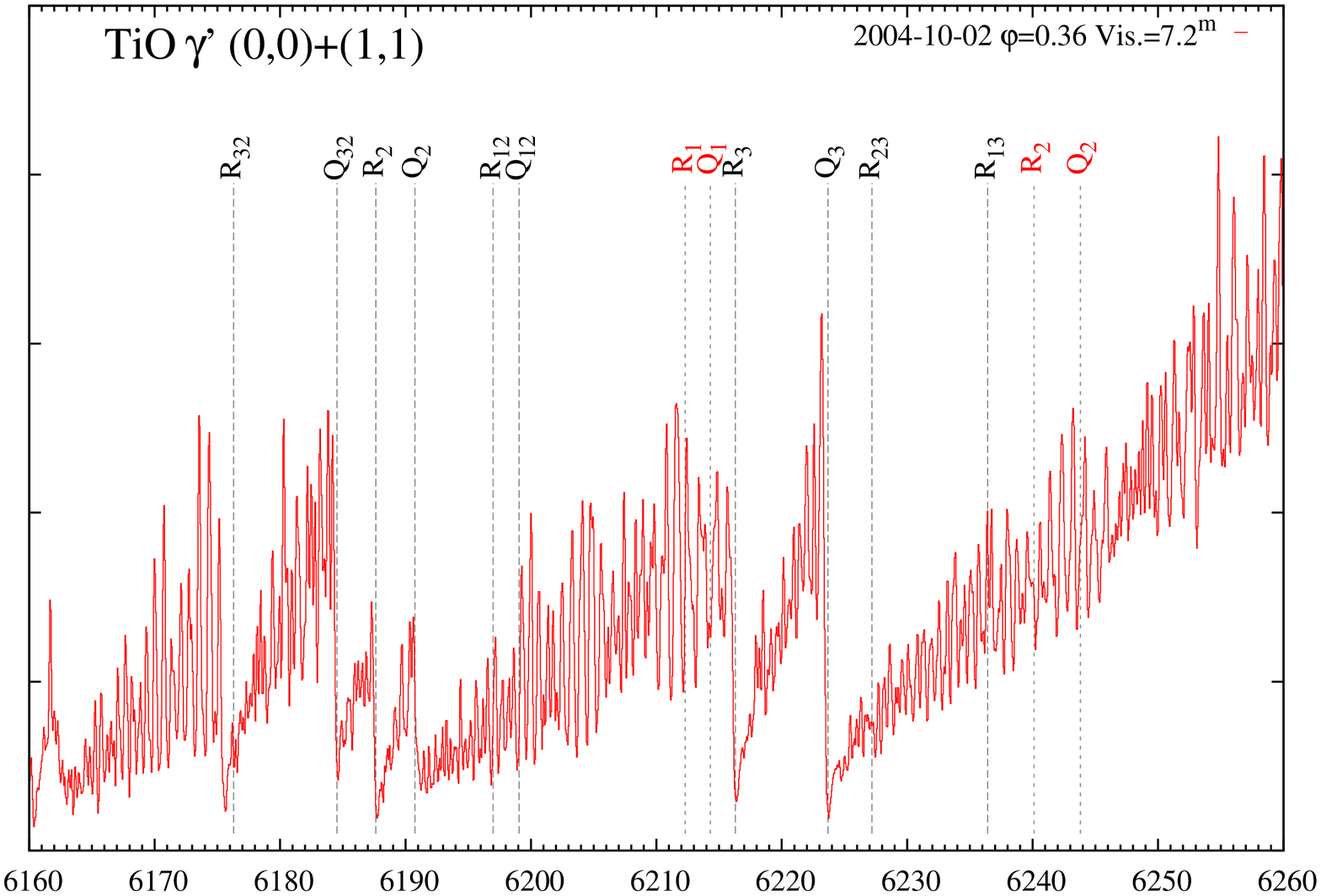}
\includegraphics[angle=270,width=0.85\textwidth]{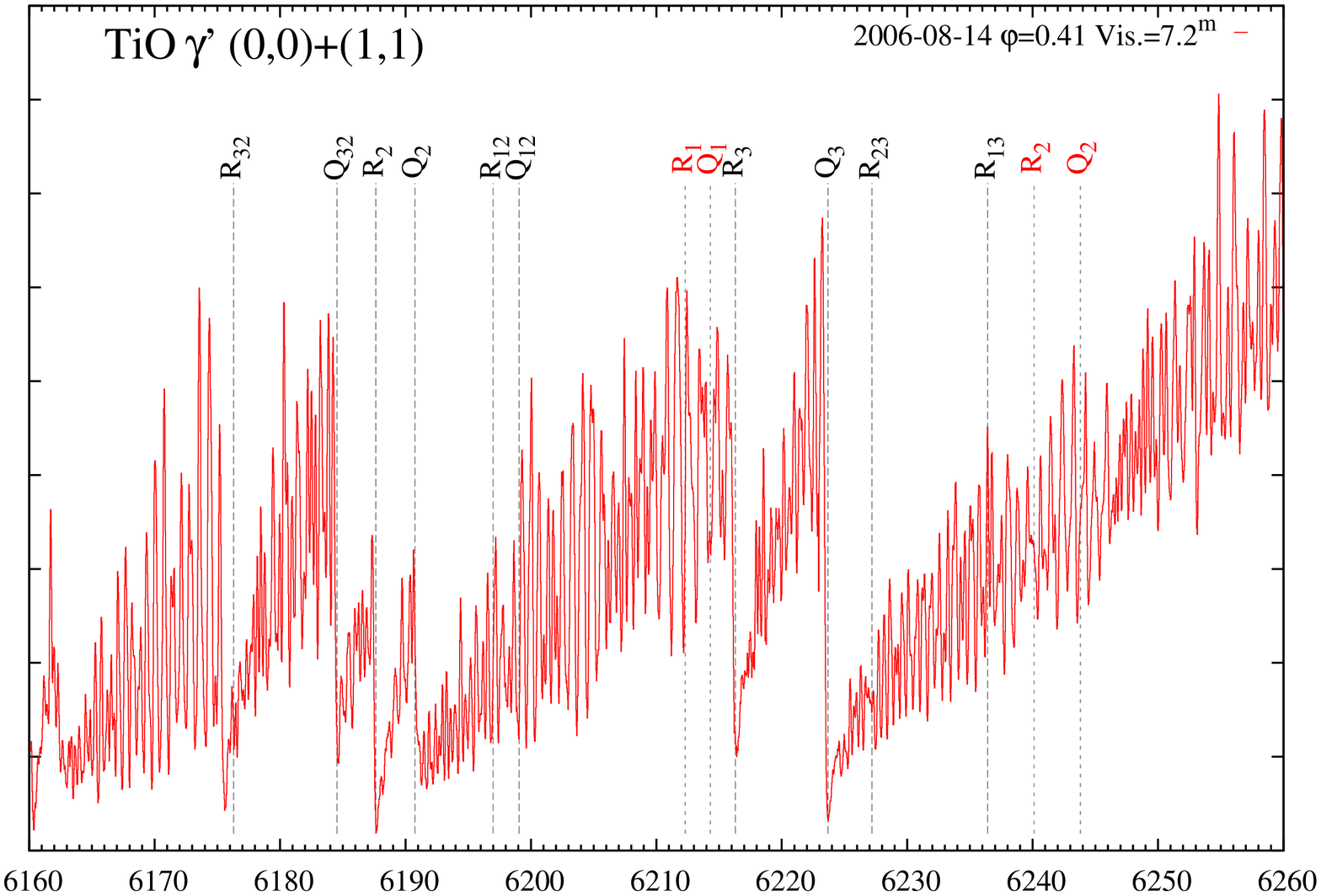}
\caption{Continued.}
\end{figure*}

  \setcounter{figure}{0}%

\begin{figure*} [tbh]
\centering
\includegraphics[angle=270,width=0.85\textwidth]{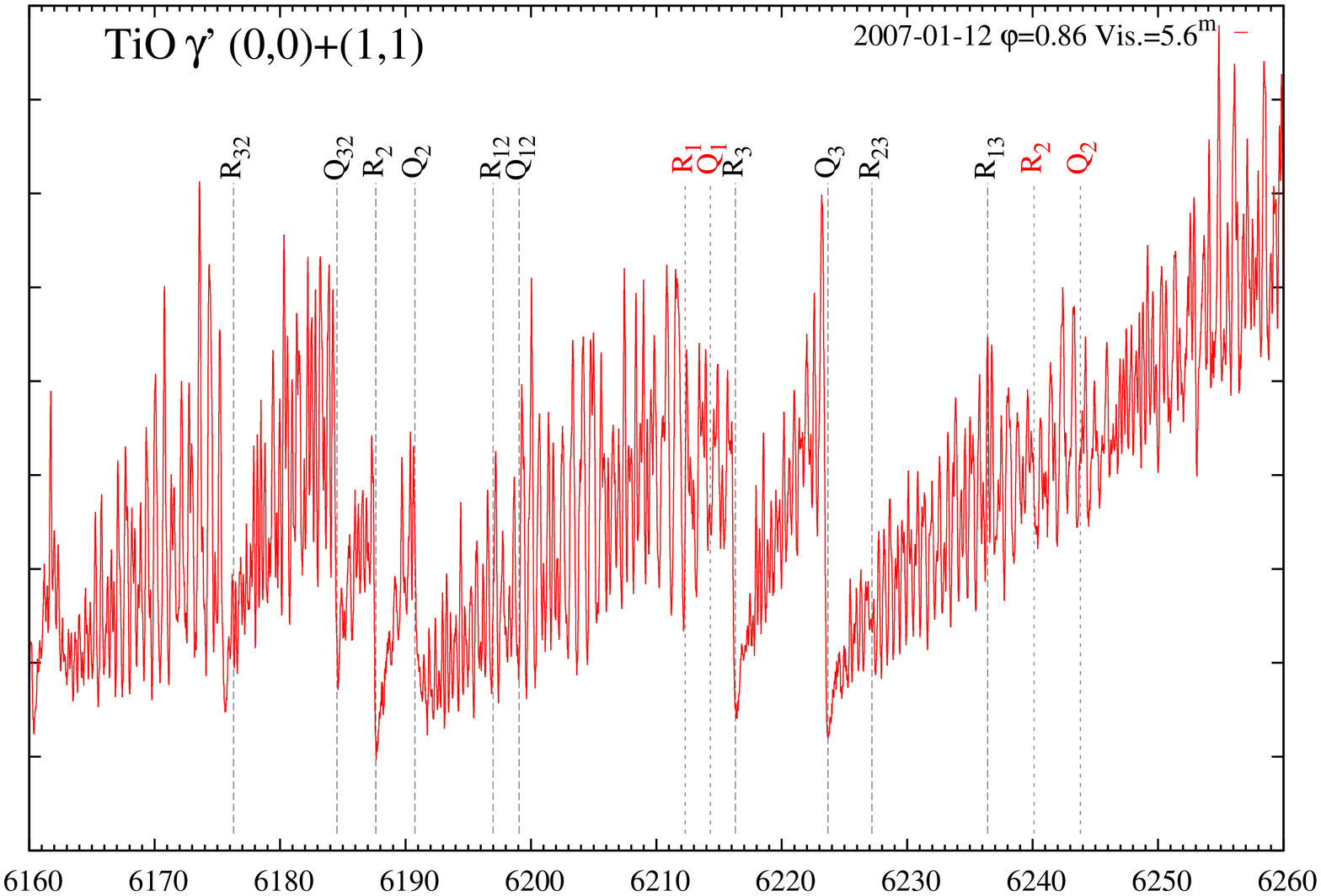}
\includegraphics[angle=270,width=0.85\textwidth]{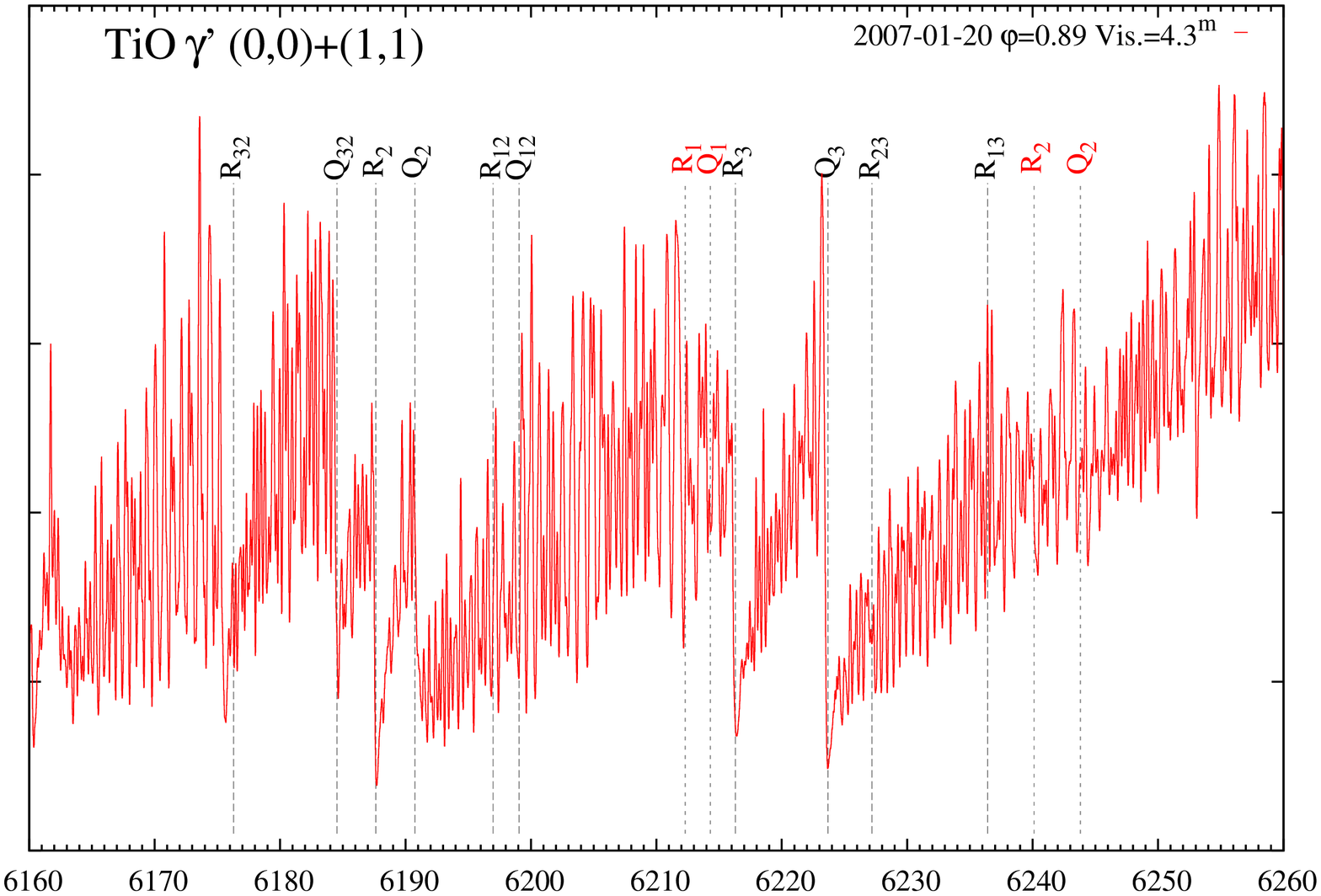}
\caption{Continued.}
\end{figure*}

  \setcounter{figure}{0}%

\begin{figure*} [tbh]
\centering
\includegraphics[angle=270,width=0.85\textwidth]{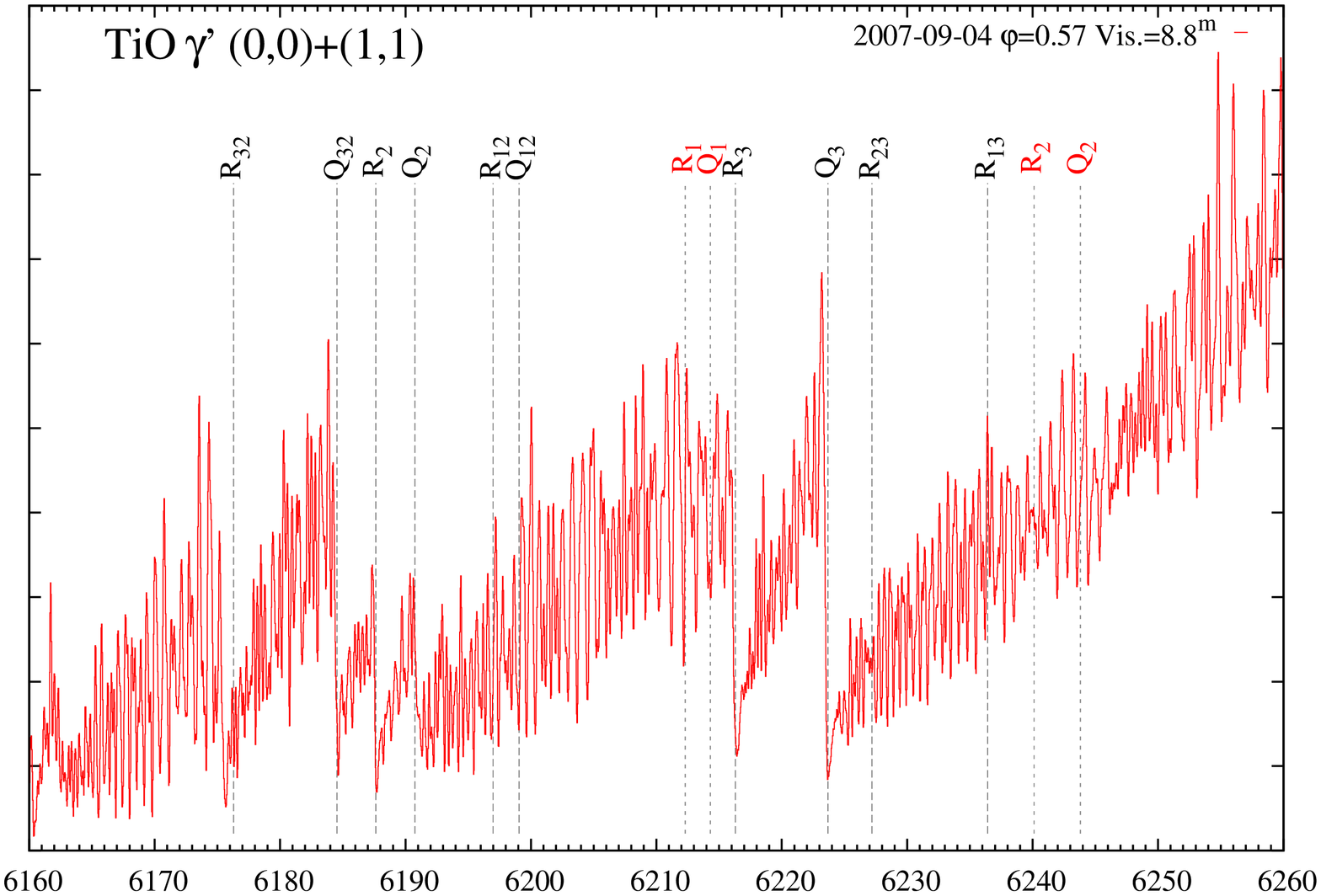}
\includegraphics[angle=270,width=0.85\textwidth]{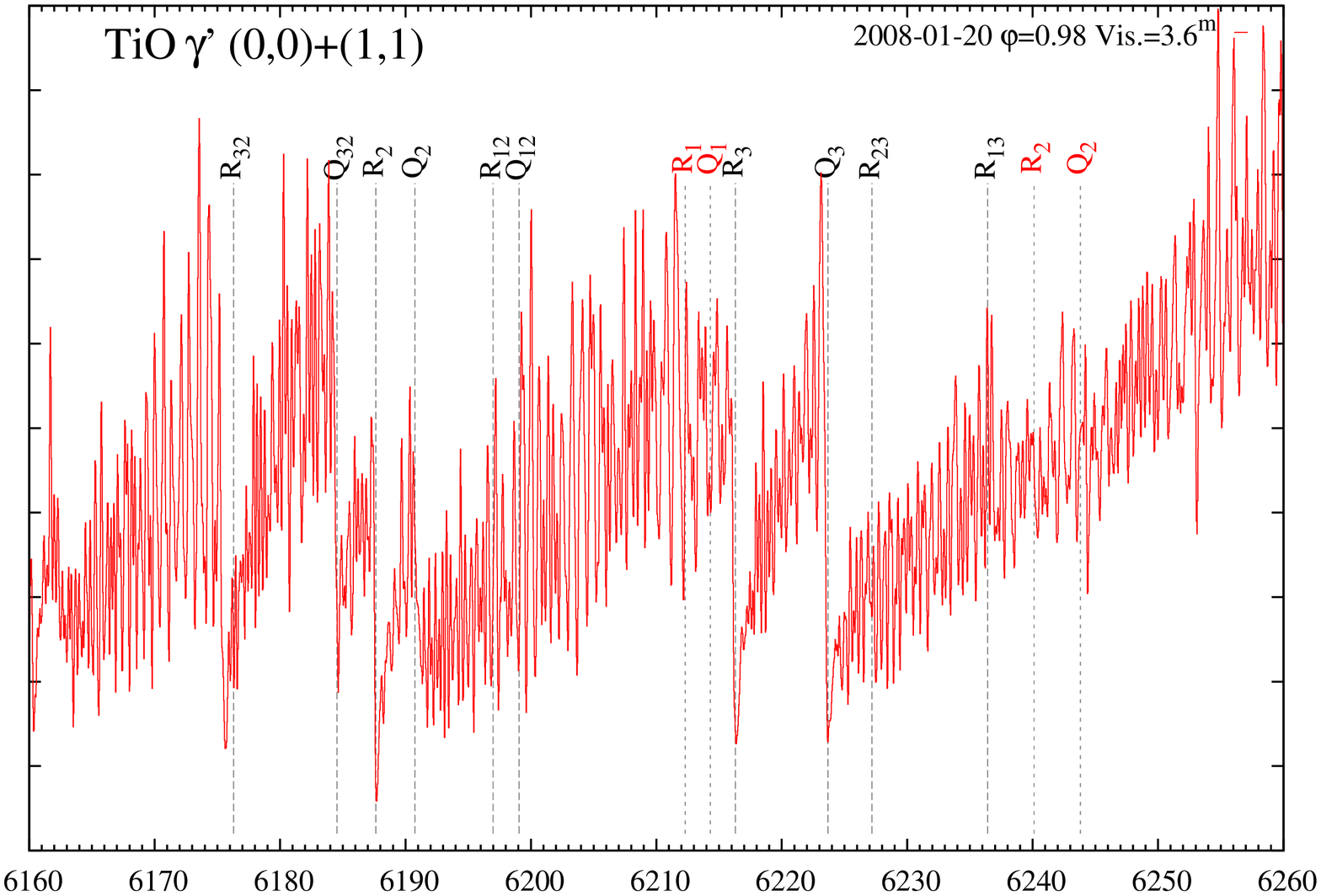}
\caption{Continued.}
\end{figure*}

  \setcounter{figure}{0}%

\begin{figure*} [tbh]
\centering
\includegraphics[angle=270,width=0.85\textwidth]{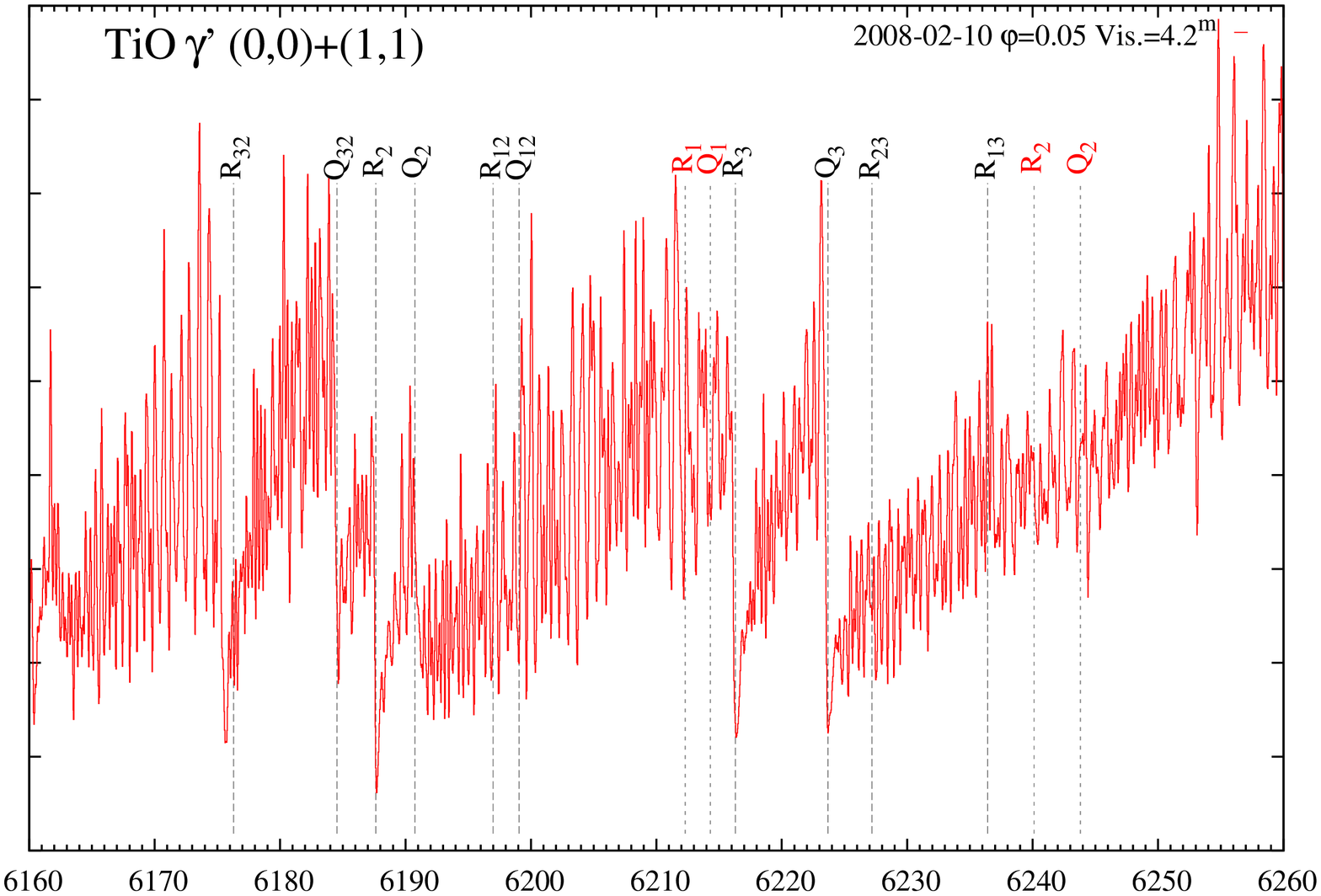}
\includegraphics[angle=270,width=0.85\textwidth]{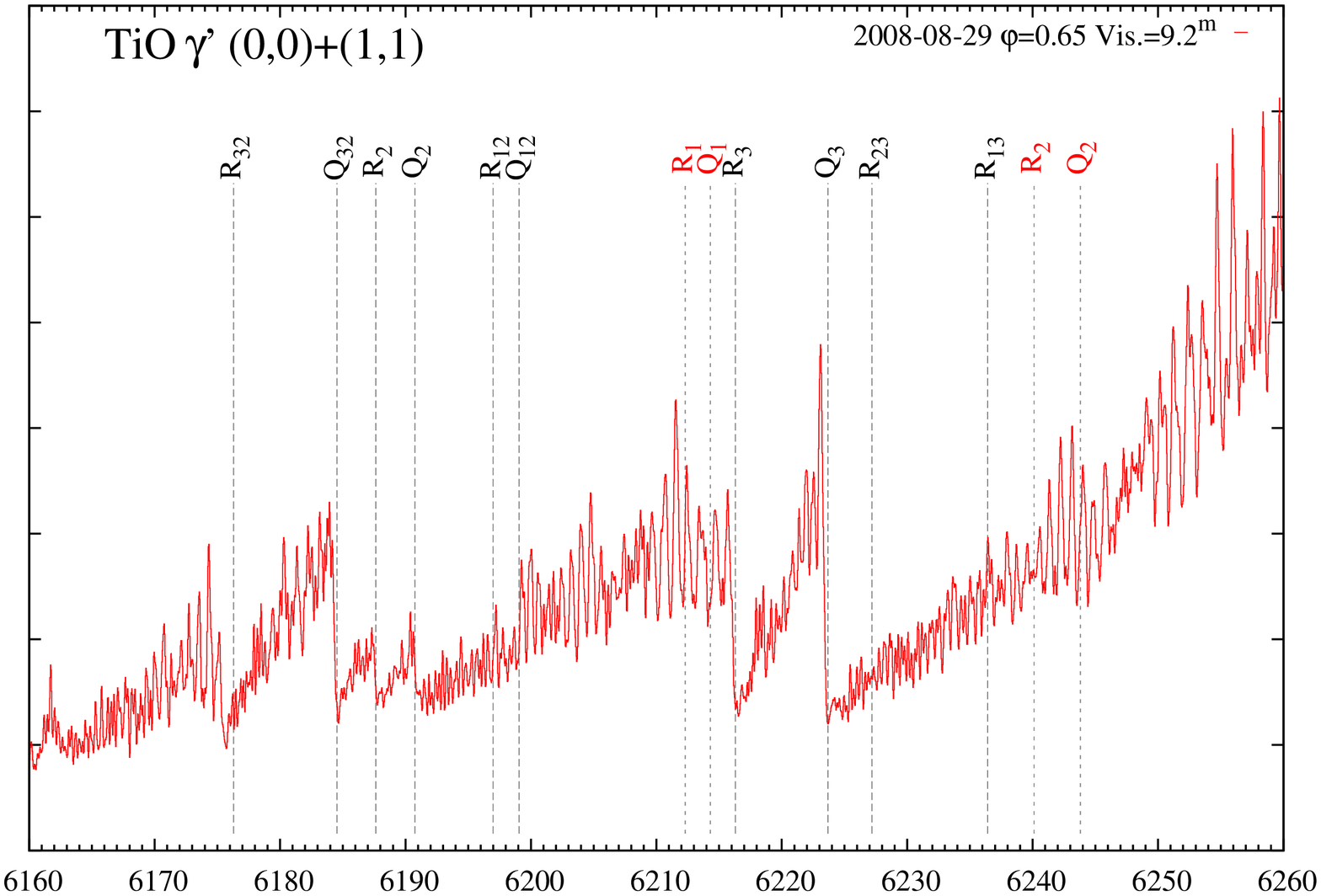}
\caption{Continued.}
\end{figure*}

  \setcounter{figure}{0}%

\begin{figure*} [tbh]
\centering
\includegraphics[angle=270,width=0.85\textwidth]{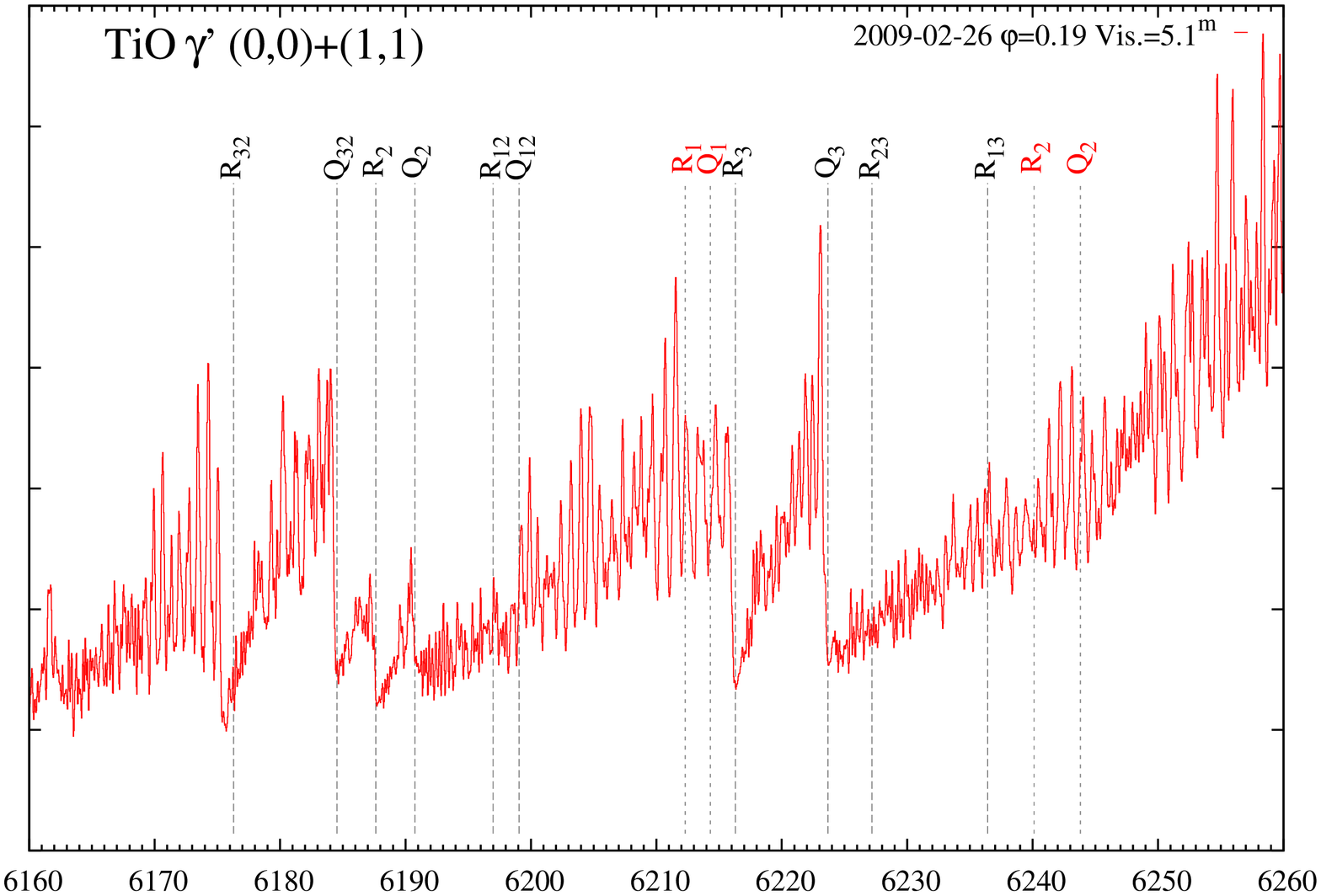}
\includegraphics[angle=270,width=0.85\textwidth]{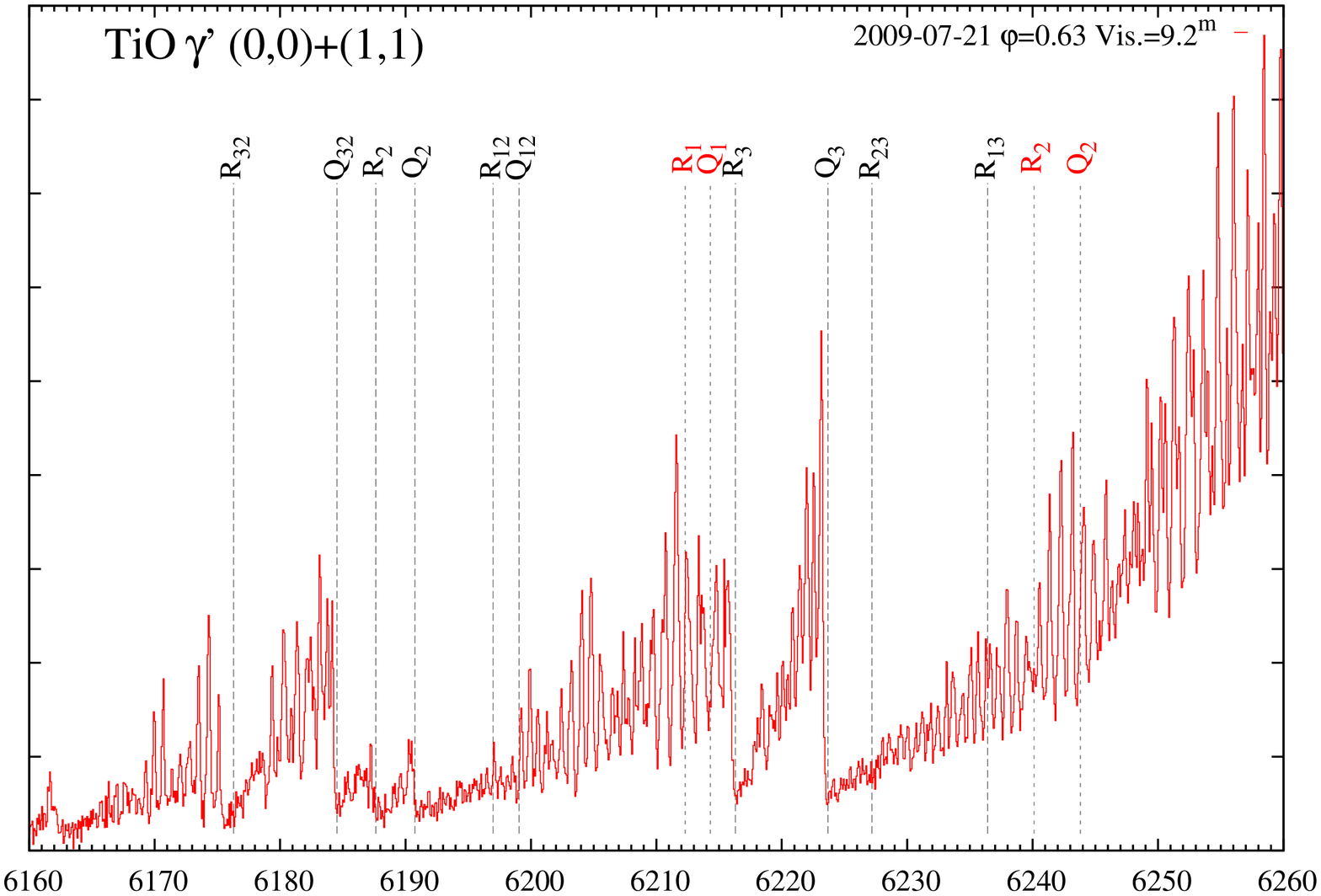}
\caption{Continued.}
\end{figure*}

  \setcounter{figure}{0}%

\begin{figure*} [tbh]
\centering
\includegraphics[angle=270,width=0.85\textwidth]{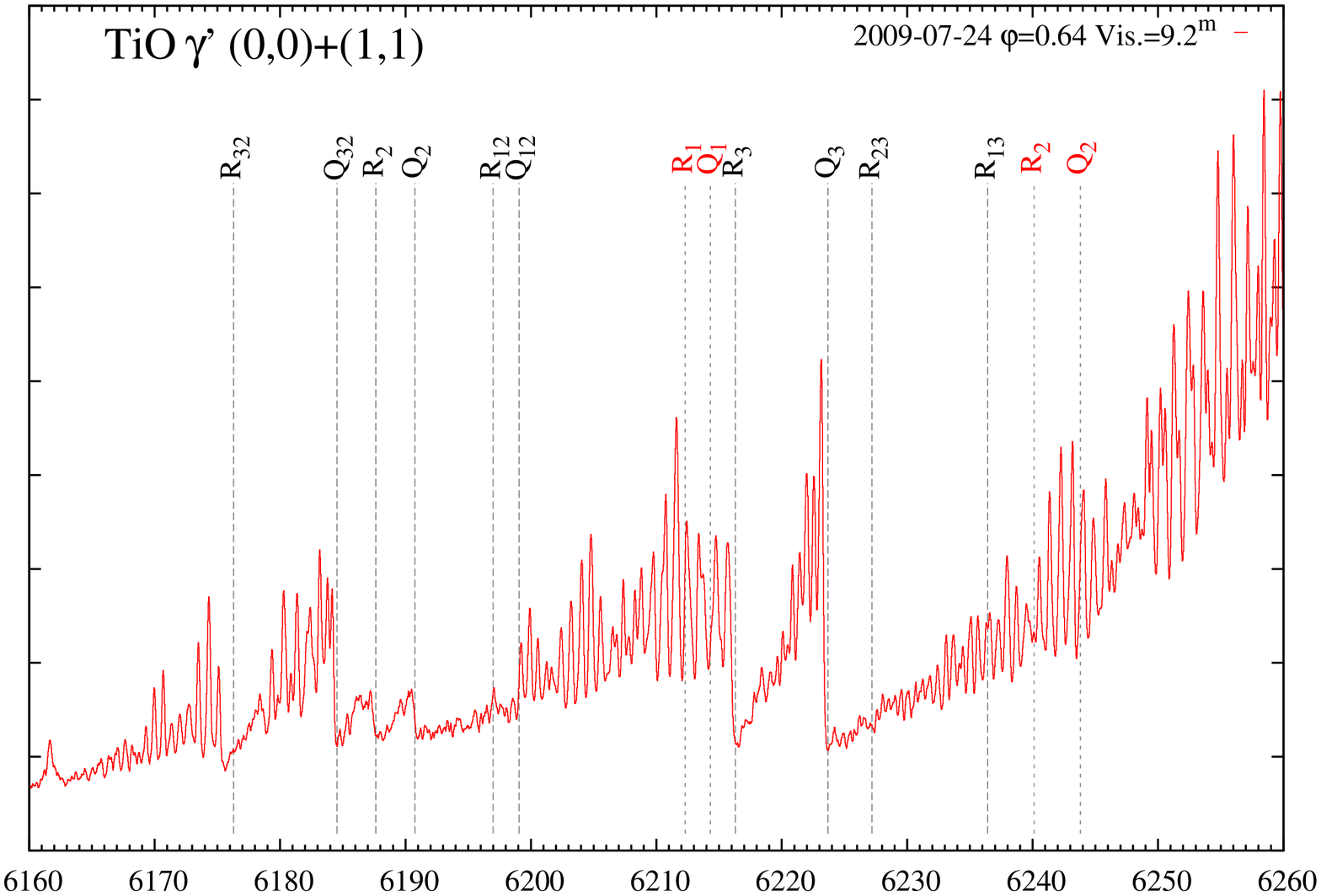}
\includegraphics[angle=270,width=0.85\textwidth]{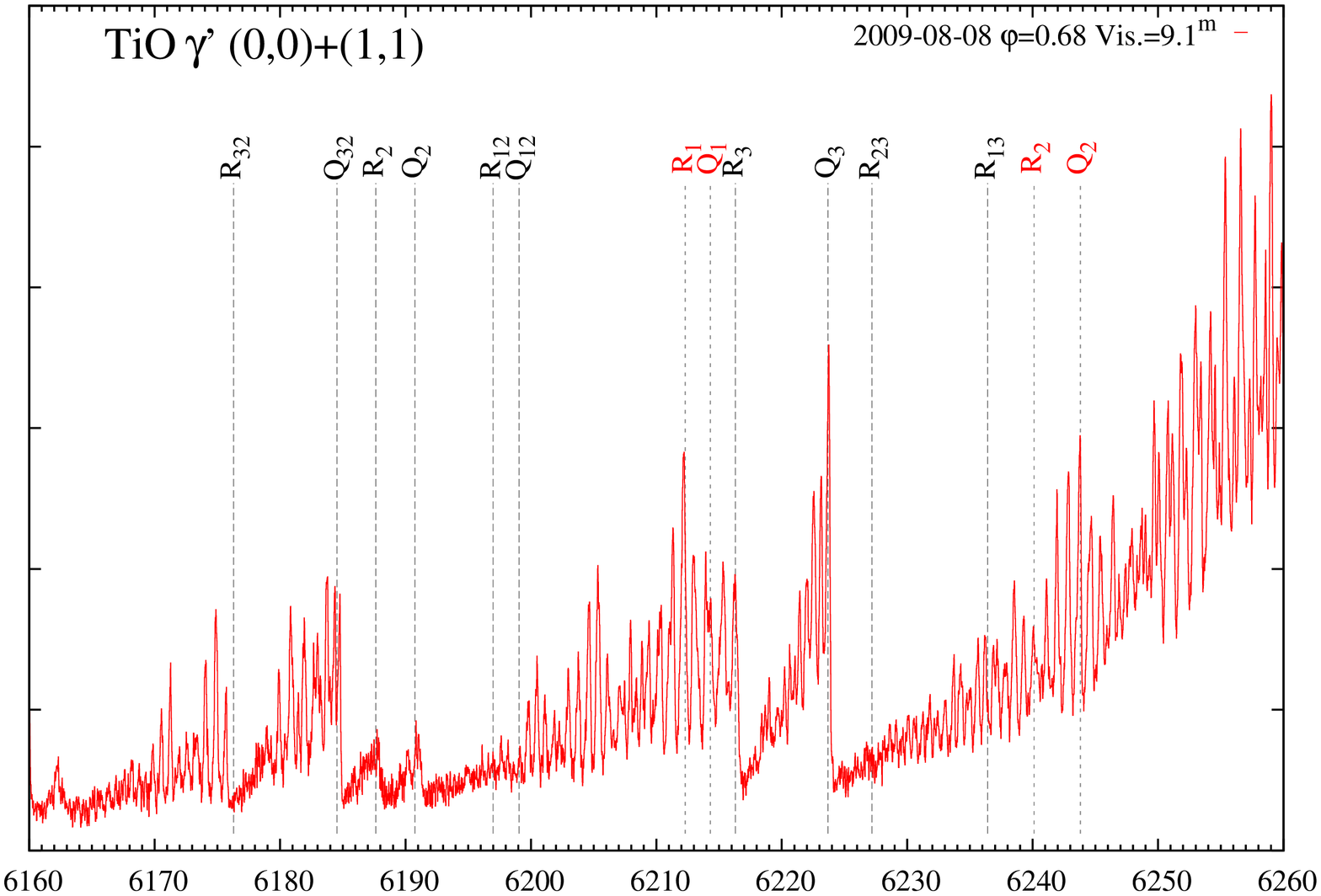}
\caption{Continued.}
\end{figure*}

  \setcounter{figure}{0}%

\begin{figure*} [tbh]
\centering
\includegraphics[angle=270,width=0.85\textwidth]{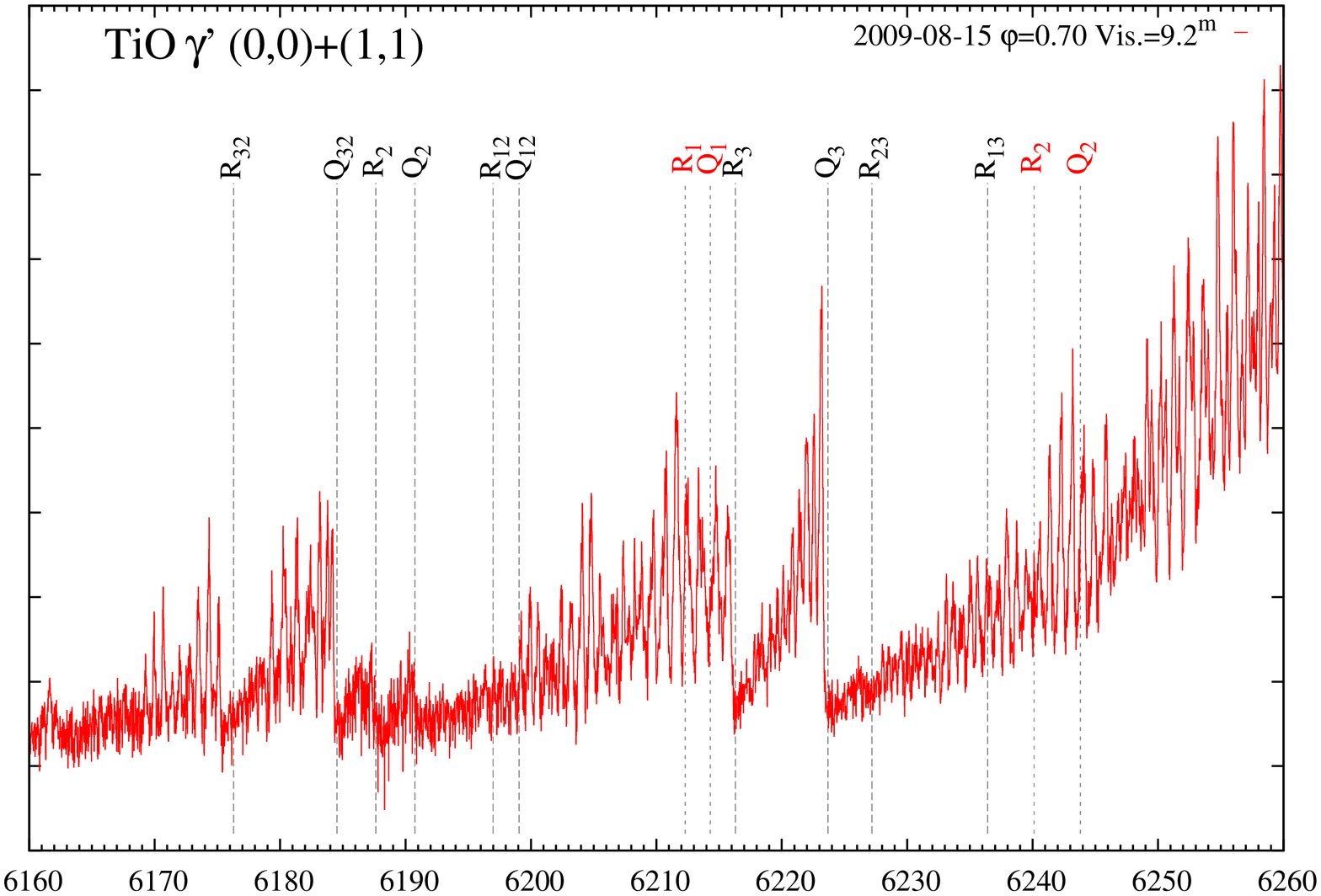}
\includegraphics[angle=270,width=0.85\textwidth]{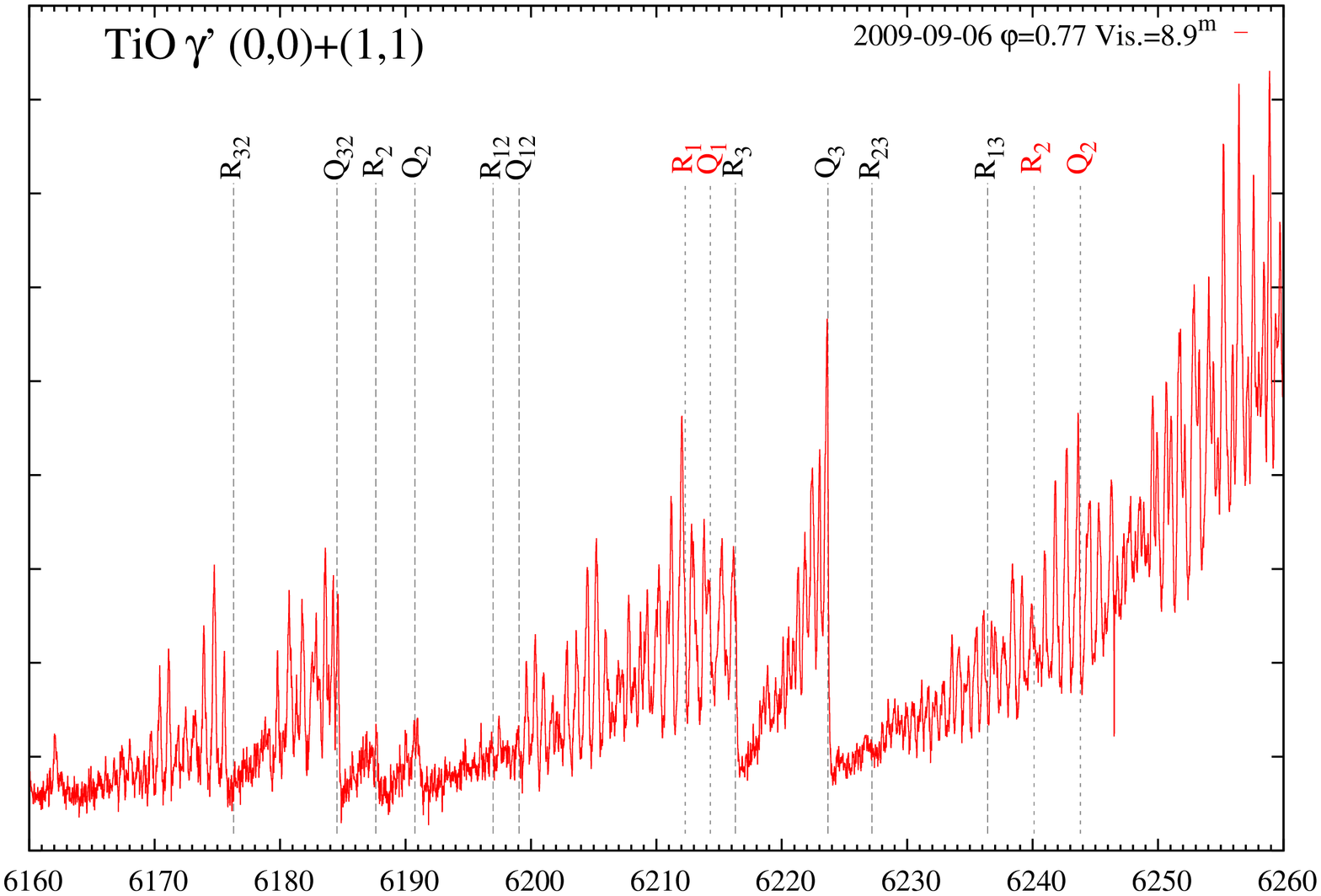}
\caption{Continued.}
\end{figure*}

  \setcounter{figure}{0}%

\begin{figure*} [tbh]
\centering
\includegraphics[angle=270,width=0.85\textwidth]{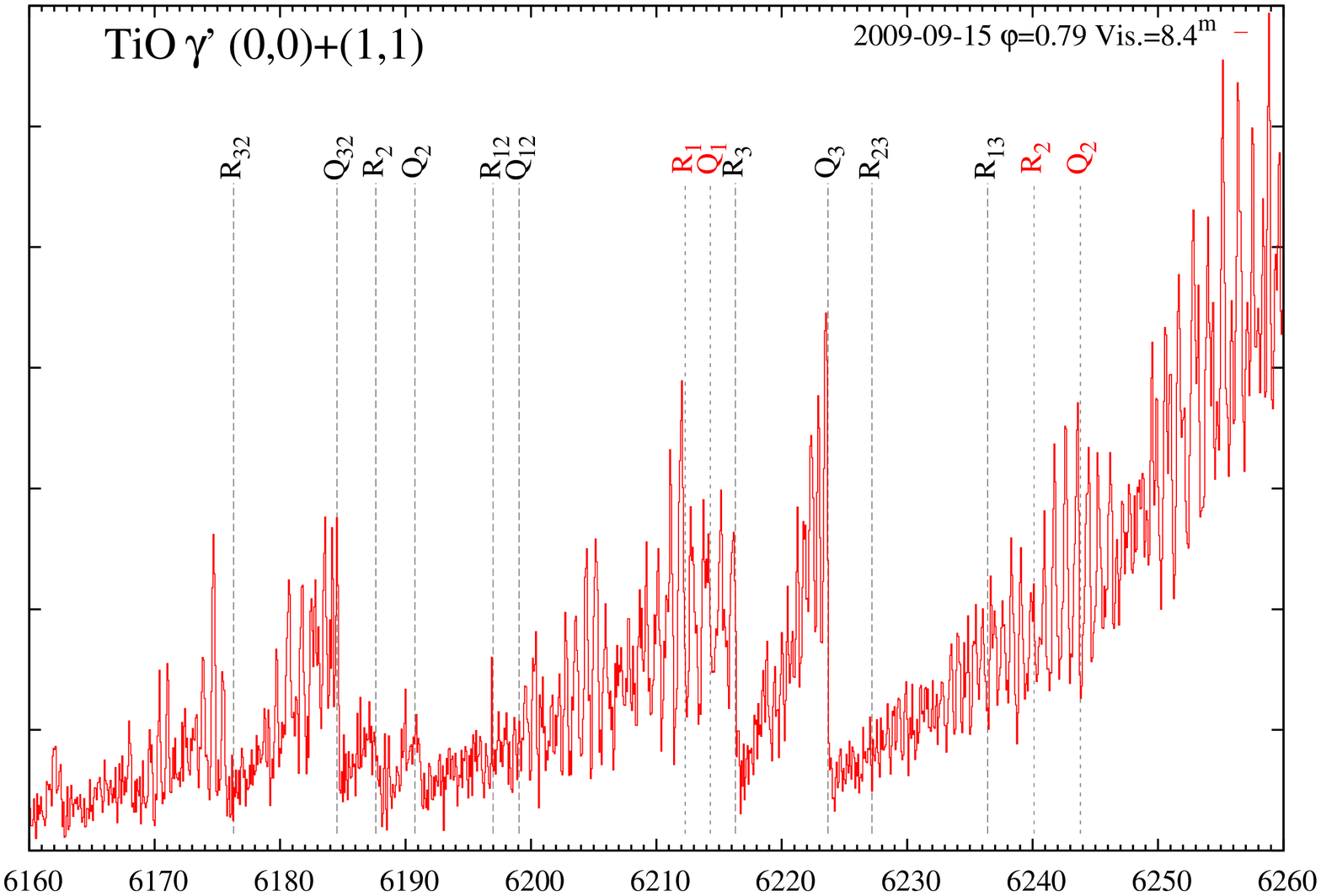}
\includegraphics[angle=270,width=0.85\textwidth]{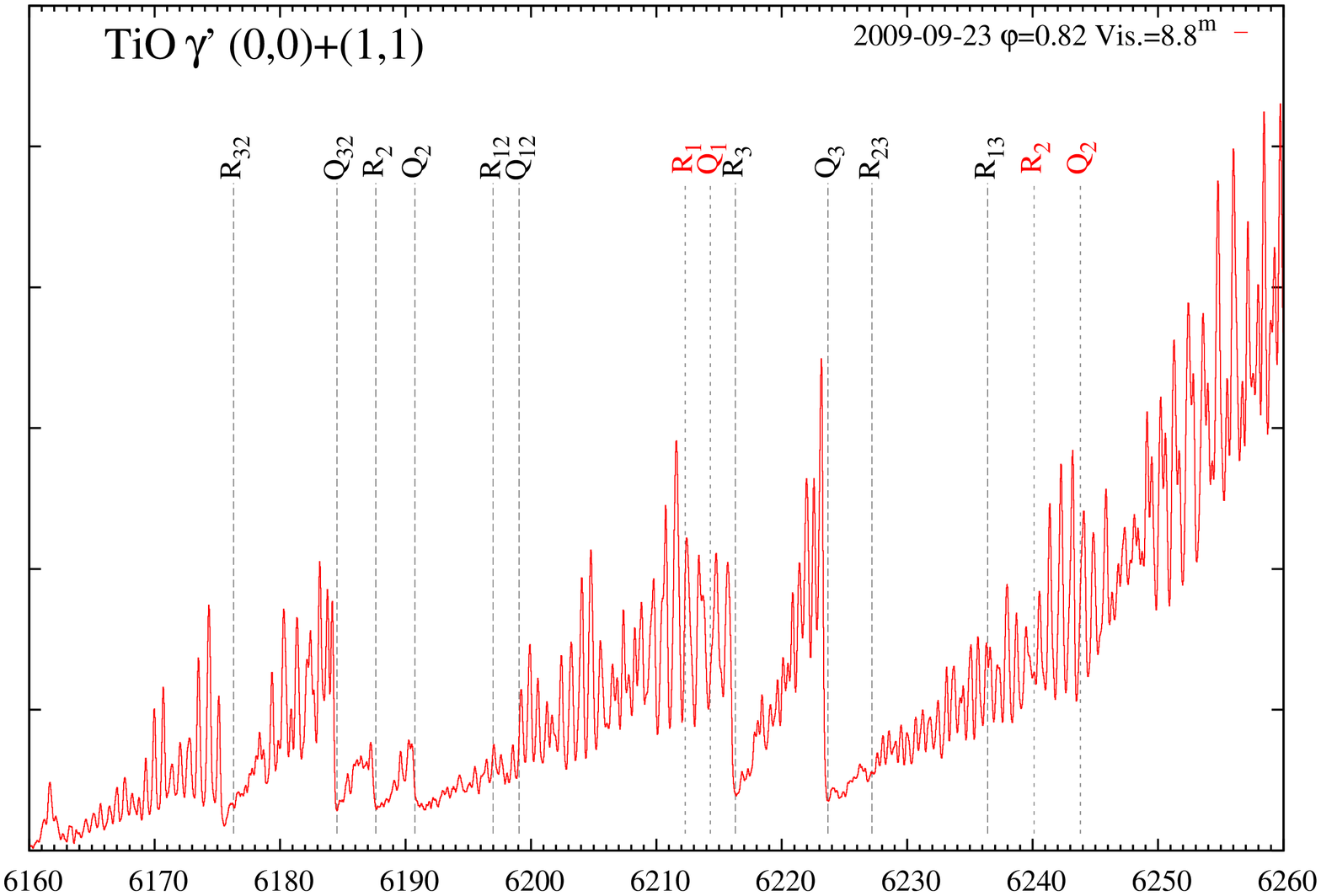}
\caption{Continued.}
\end{figure*}

  \setcounter{figure}{0}%

\begin{figure*} [tbh]
\centering
\includegraphics[angle=270,width=0.85\textwidth]{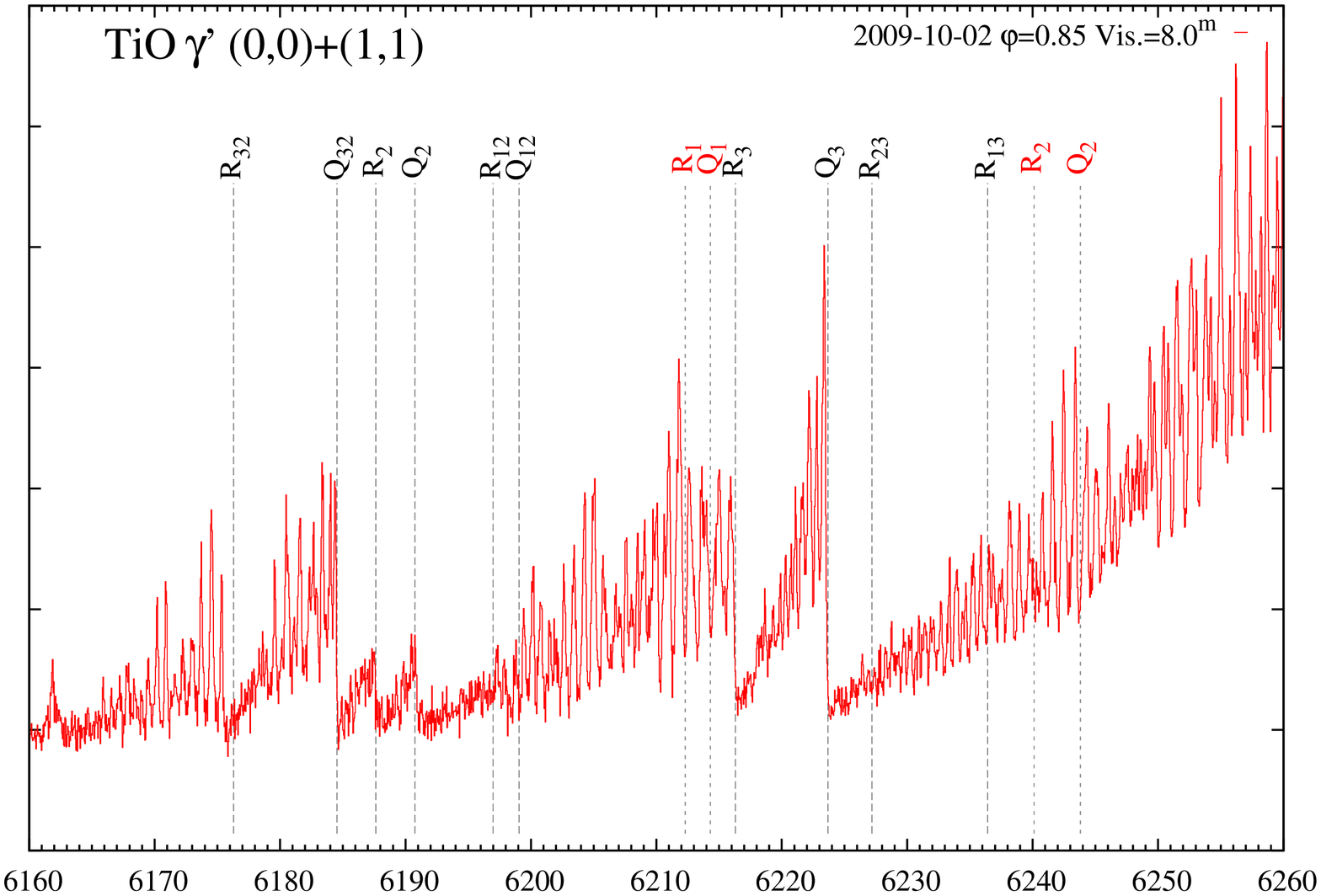}
\includegraphics[angle=270,width=0.85\textwidth]{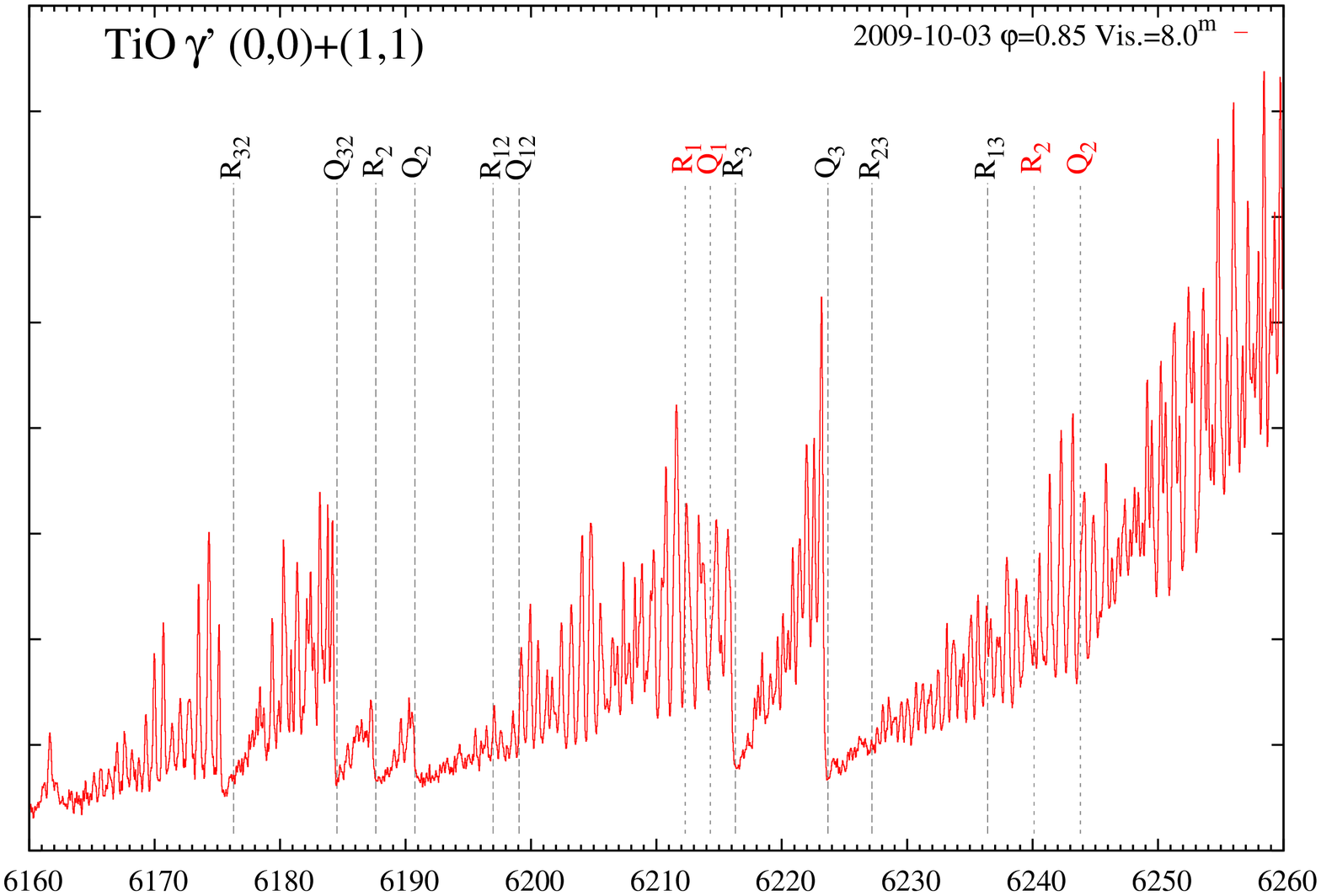}
\caption{Continued.}
\end{figure*}

  \setcounter{figure}{0}%

\begin{figure*} [tbh]
\centering
\includegraphics[angle=270,width=0.85\textwidth]{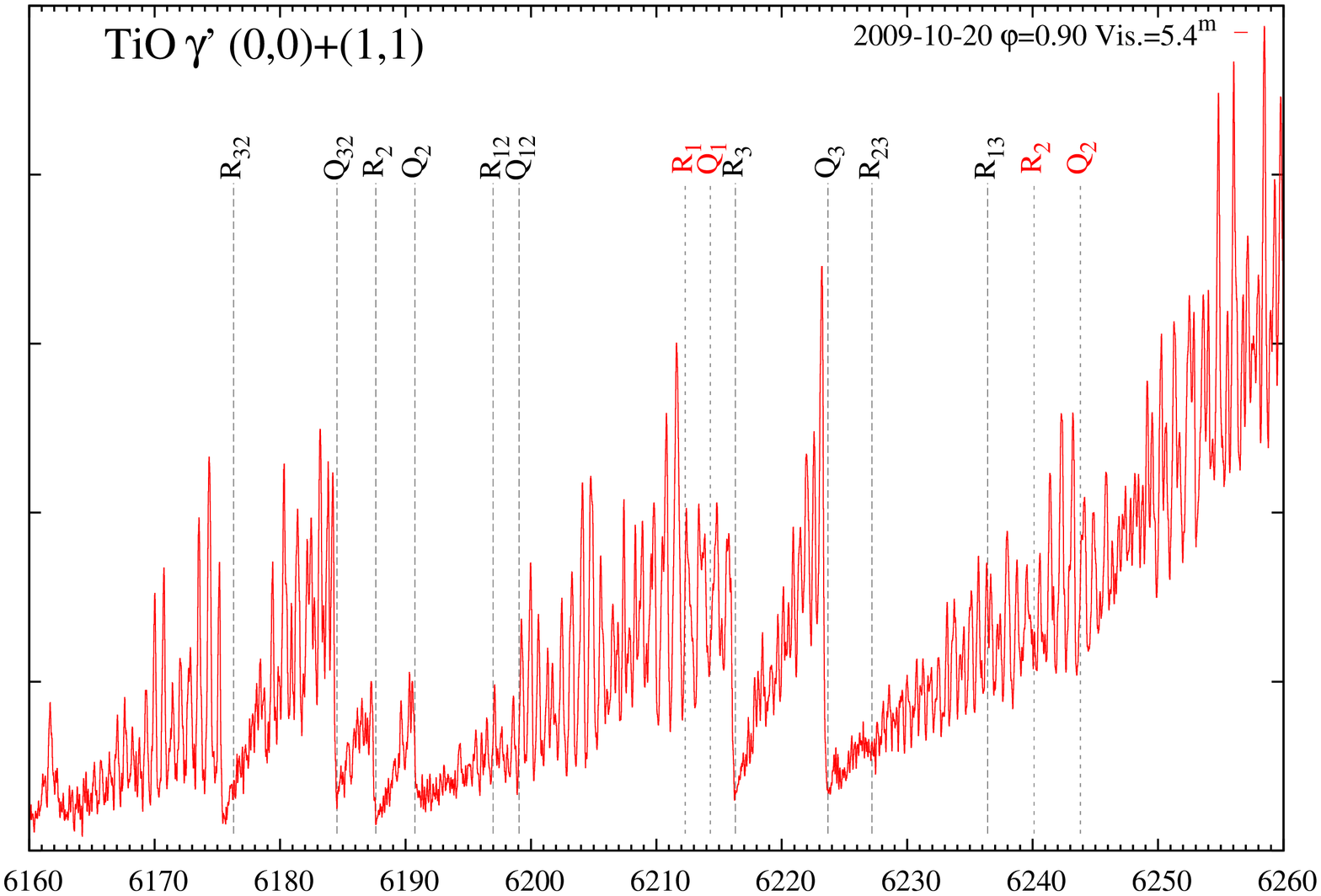}
\includegraphics[angle=270,width=0.85\textwidth]{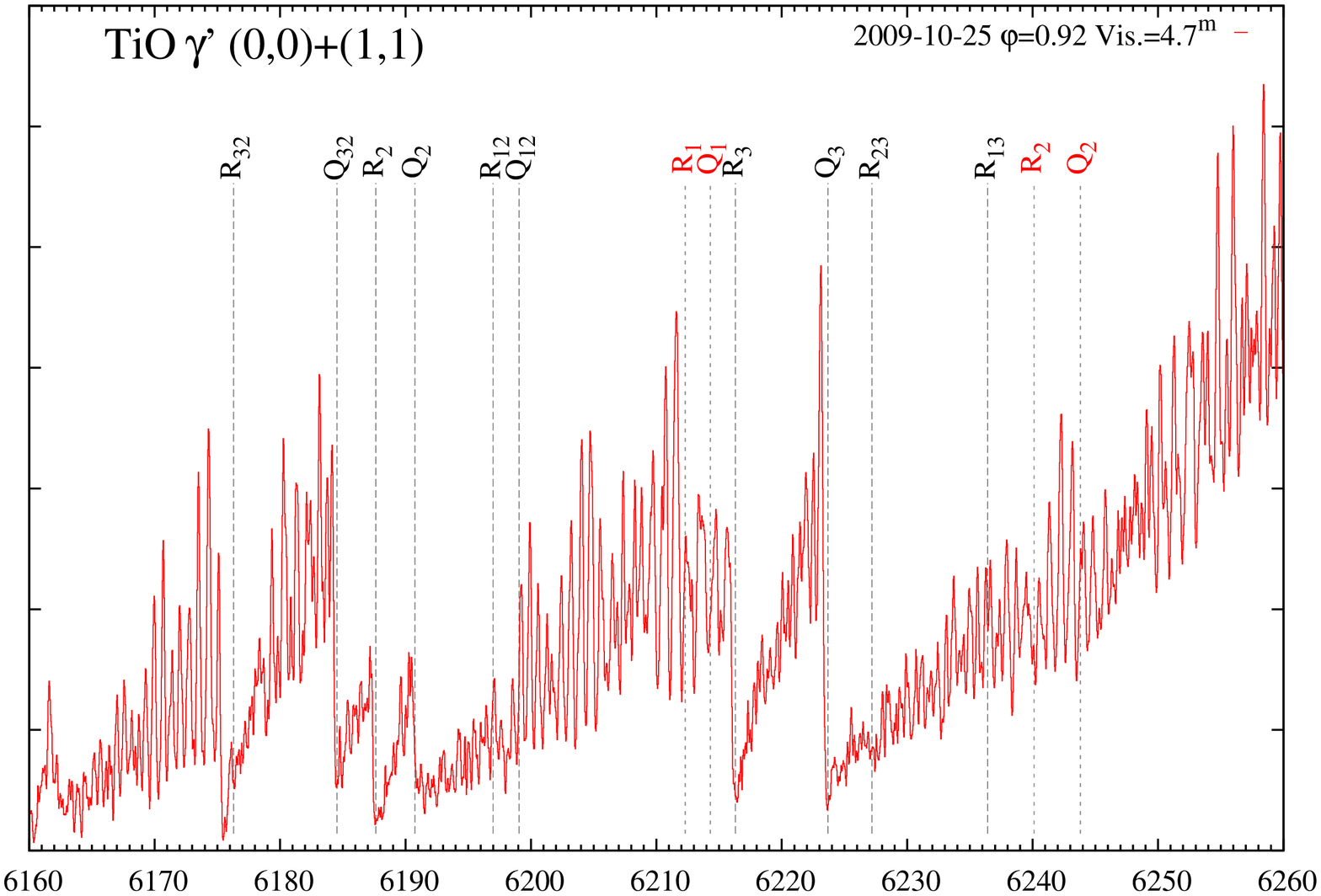}
\caption{Continued.}
\end{figure*}

  \setcounter{figure}{0}%

\begin{figure*} [tbh]
\centering
\includegraphics[angle=270,width=0.85\textwidth]{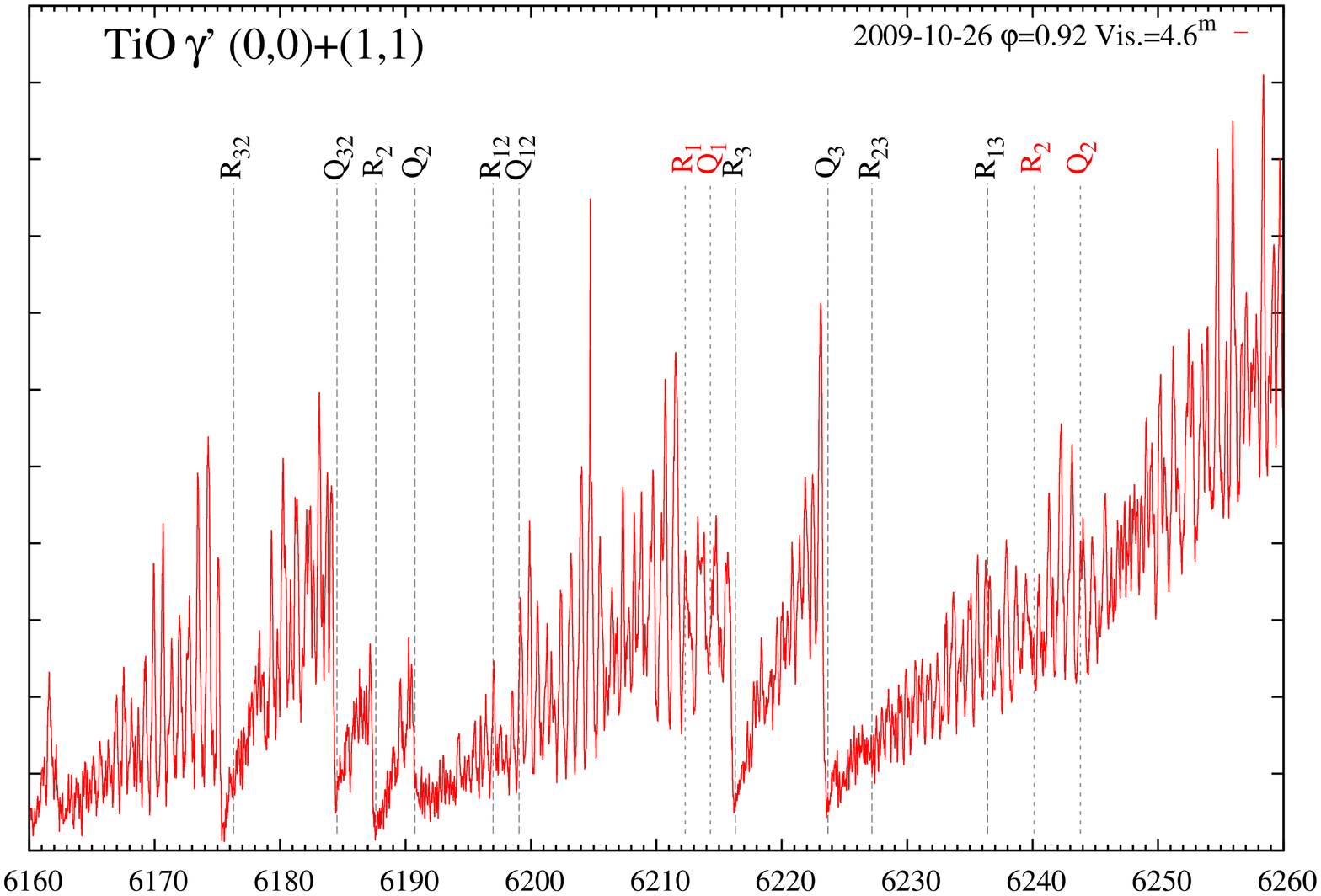}
\includegraphics[angle=270,width=0.85\textwidth]{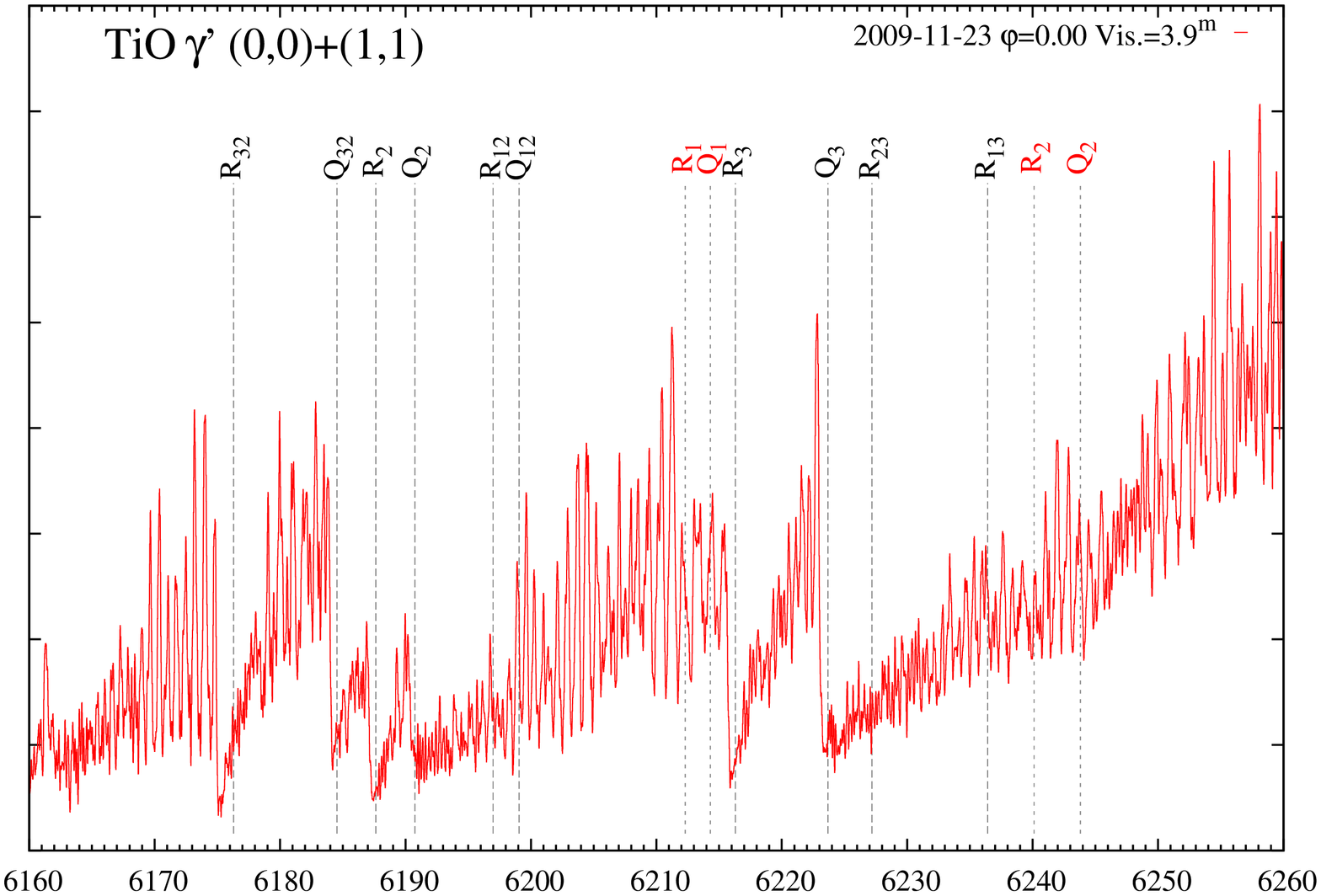}
\caption{Continued.}
\end{figure*}

  \setcounter{figure}{0}%

\begin{figure*} [tbh]
\centering
\includegraphics[angle=270,width=0.85\textwidth]{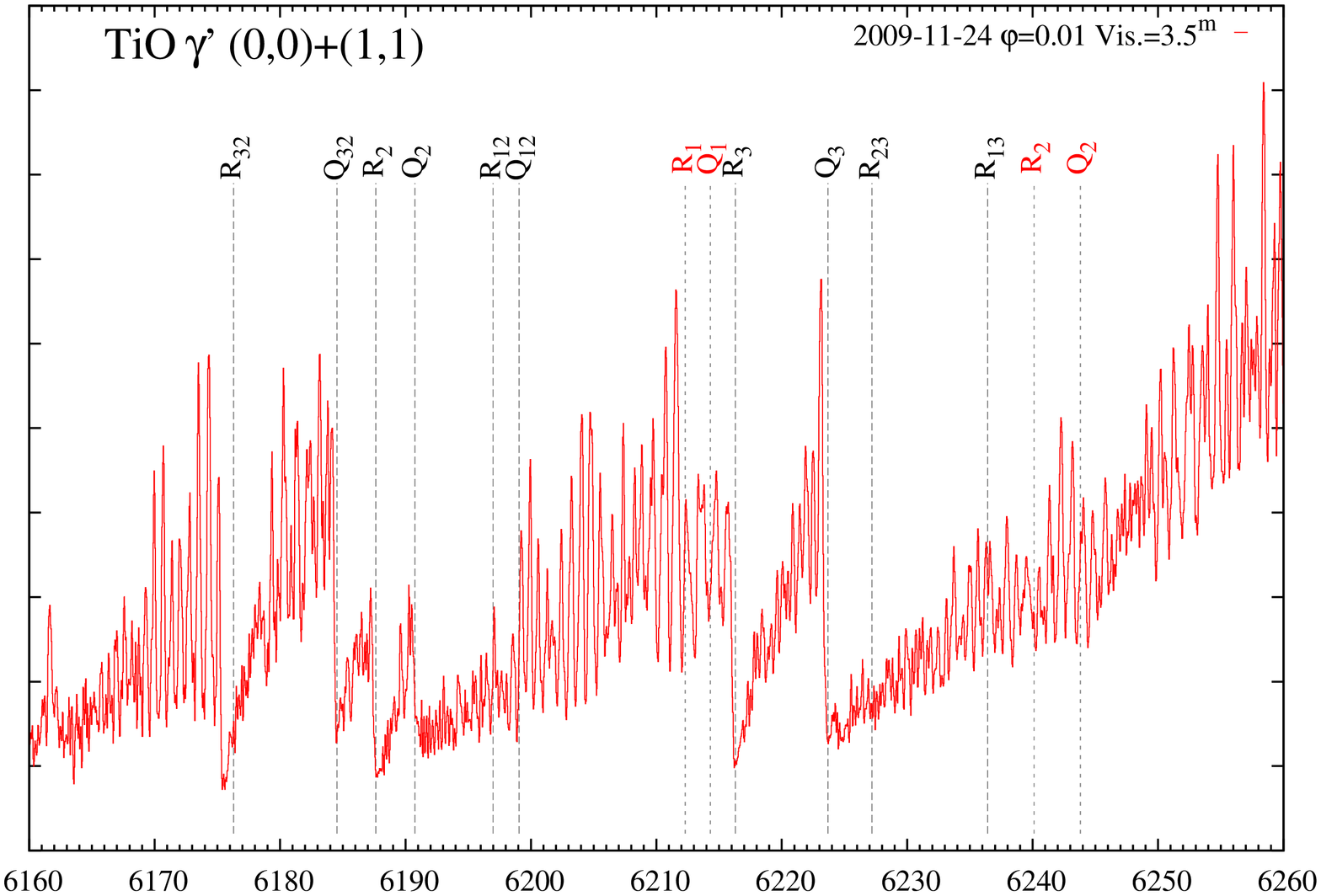}
\includegraphics[angle=270,width=0.85\textwidth]{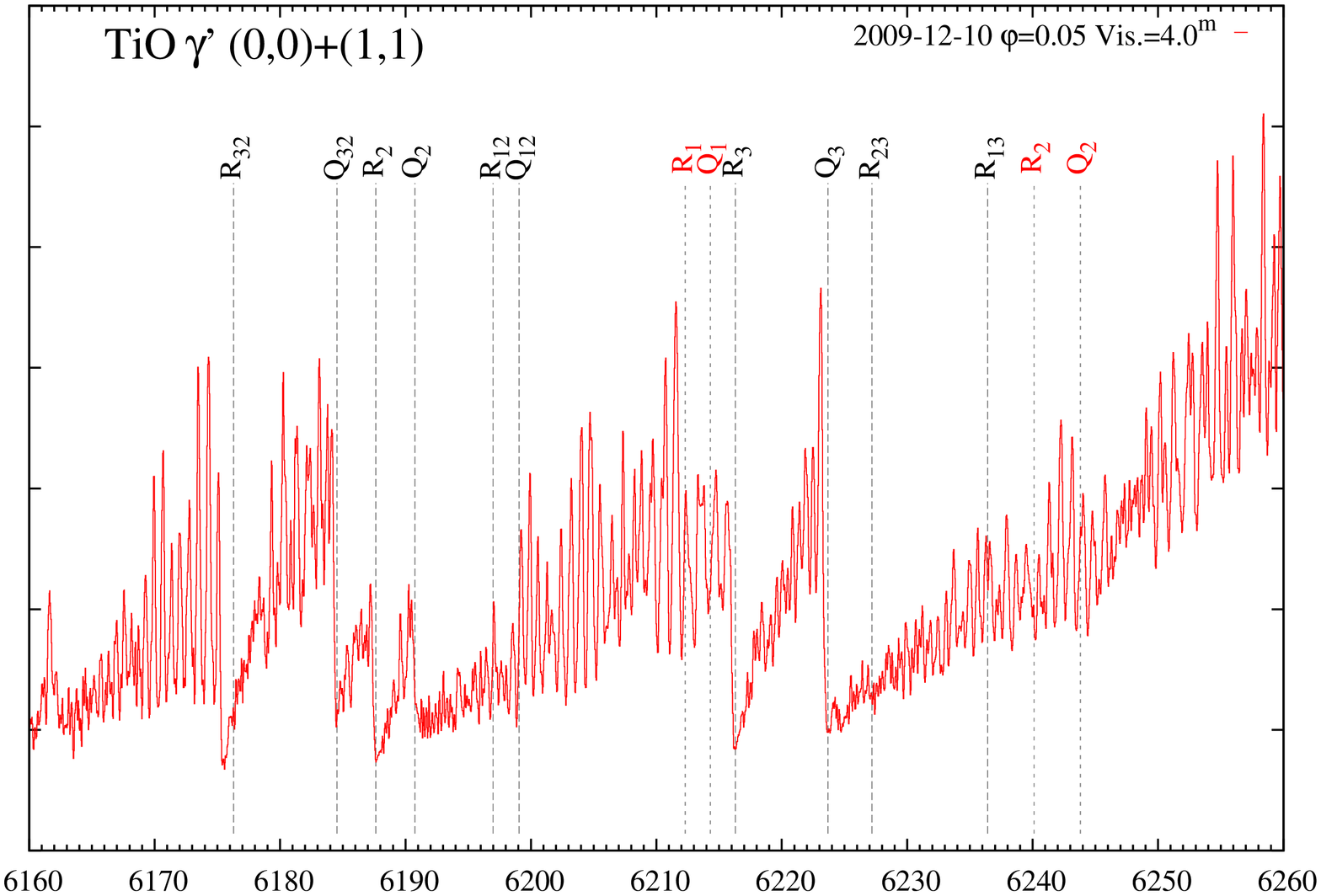}
\caption{Continued.}
\end{figure*}

  \setcounter{figure}{0}%

\begin{figure*} [tbh]
\centering
\includegraphics[angle=270,width=0.85\textwidth]{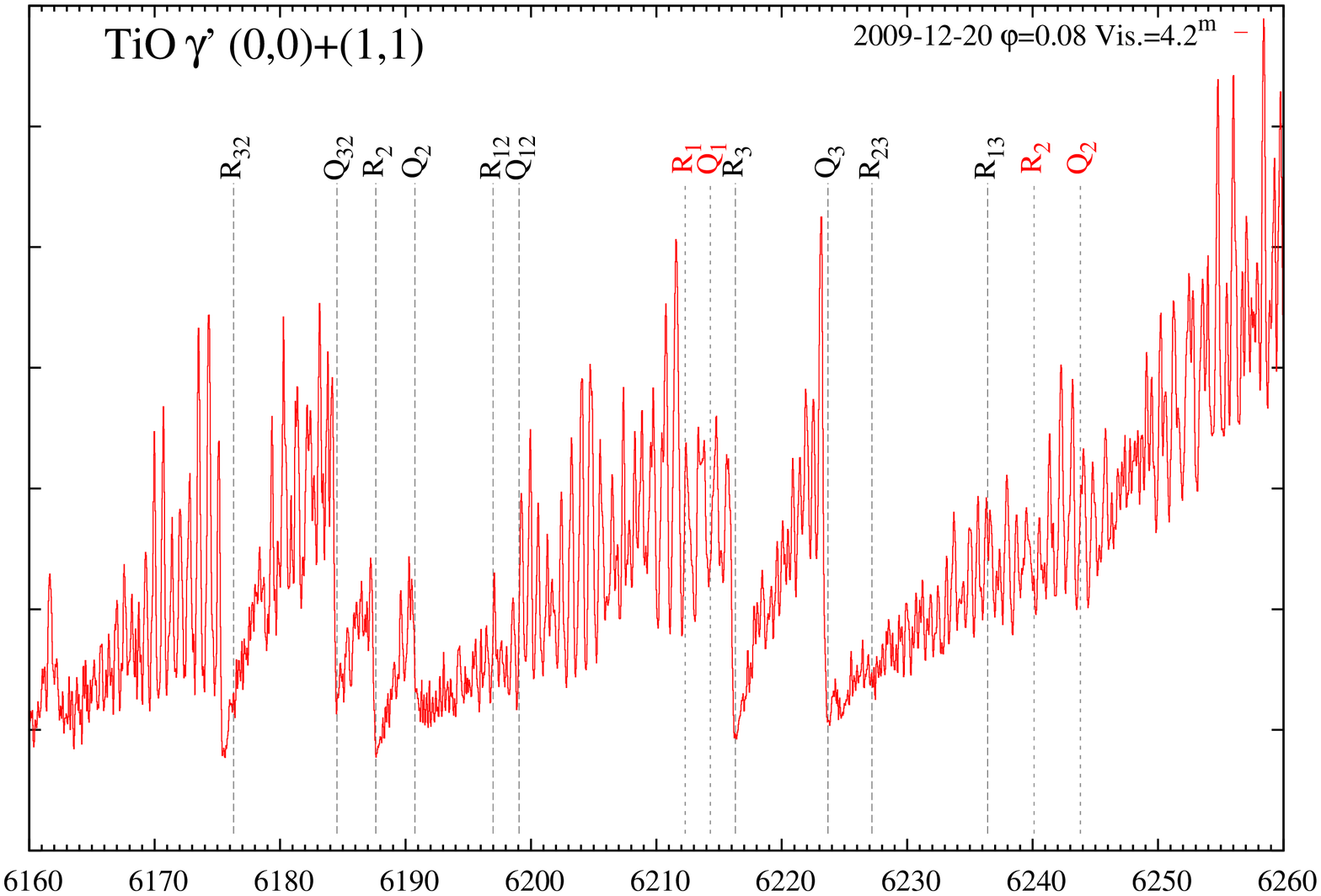}
\includegraphics[angle=270,width=0.85\textwidth]{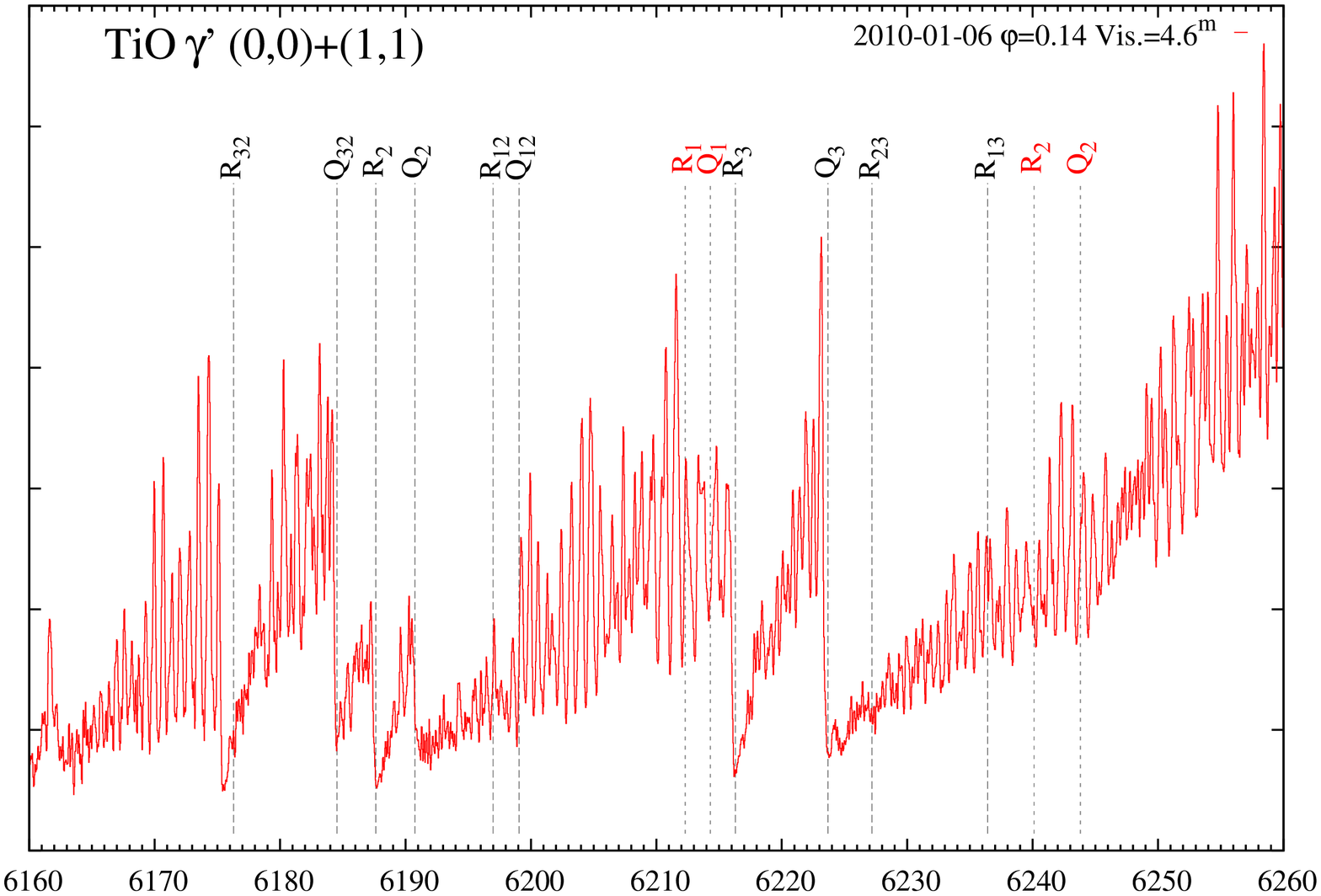}
\caption{Continued.}
\end{figure*}

  \setcounter{figure}{0}%

\begin{figure*} [tbh]
\centering
\includegraphics[angle=270,width=0.85\textwidth]{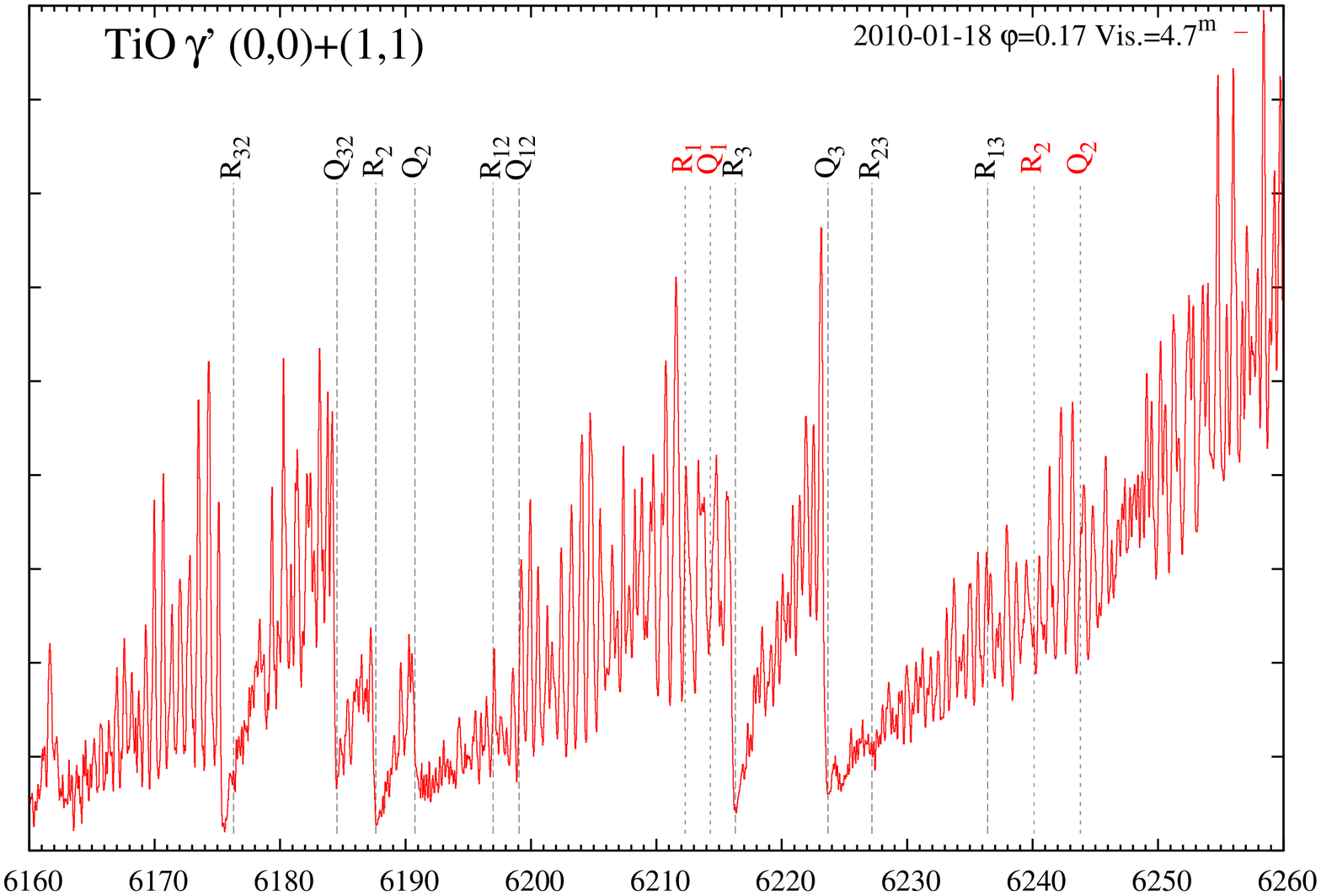}
\includegraphics[angle=270,width=0.85\textwidth]{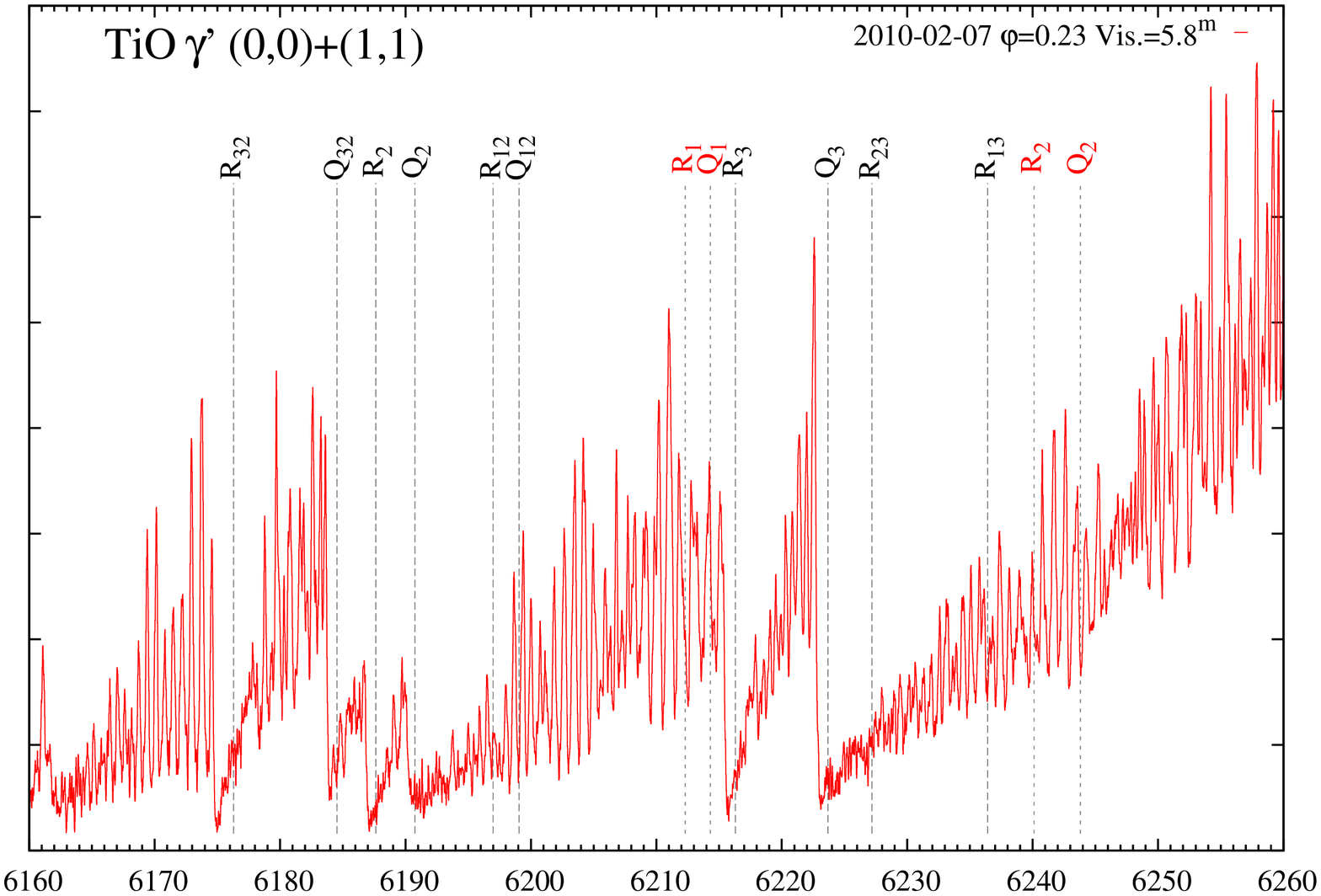}
\caption{Continued.}
\end{figure*}

  \setcounter{figure}{0}%

\begin{figure*} [tbh]
\centering
\includegraphics[angle=270,width=0.85\textwidth]{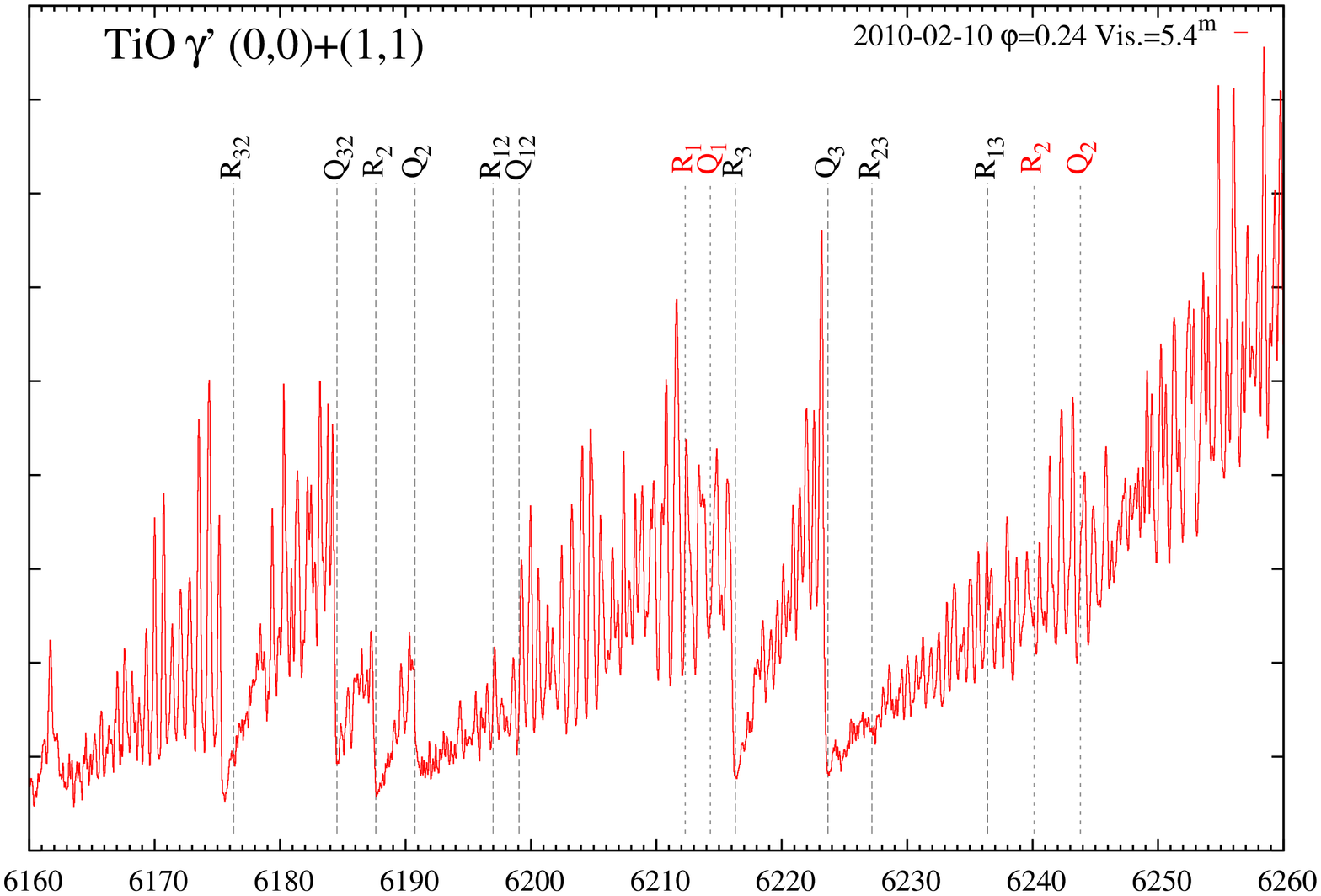}
\includegraphics[angle=270,width=0.85\textwidth]{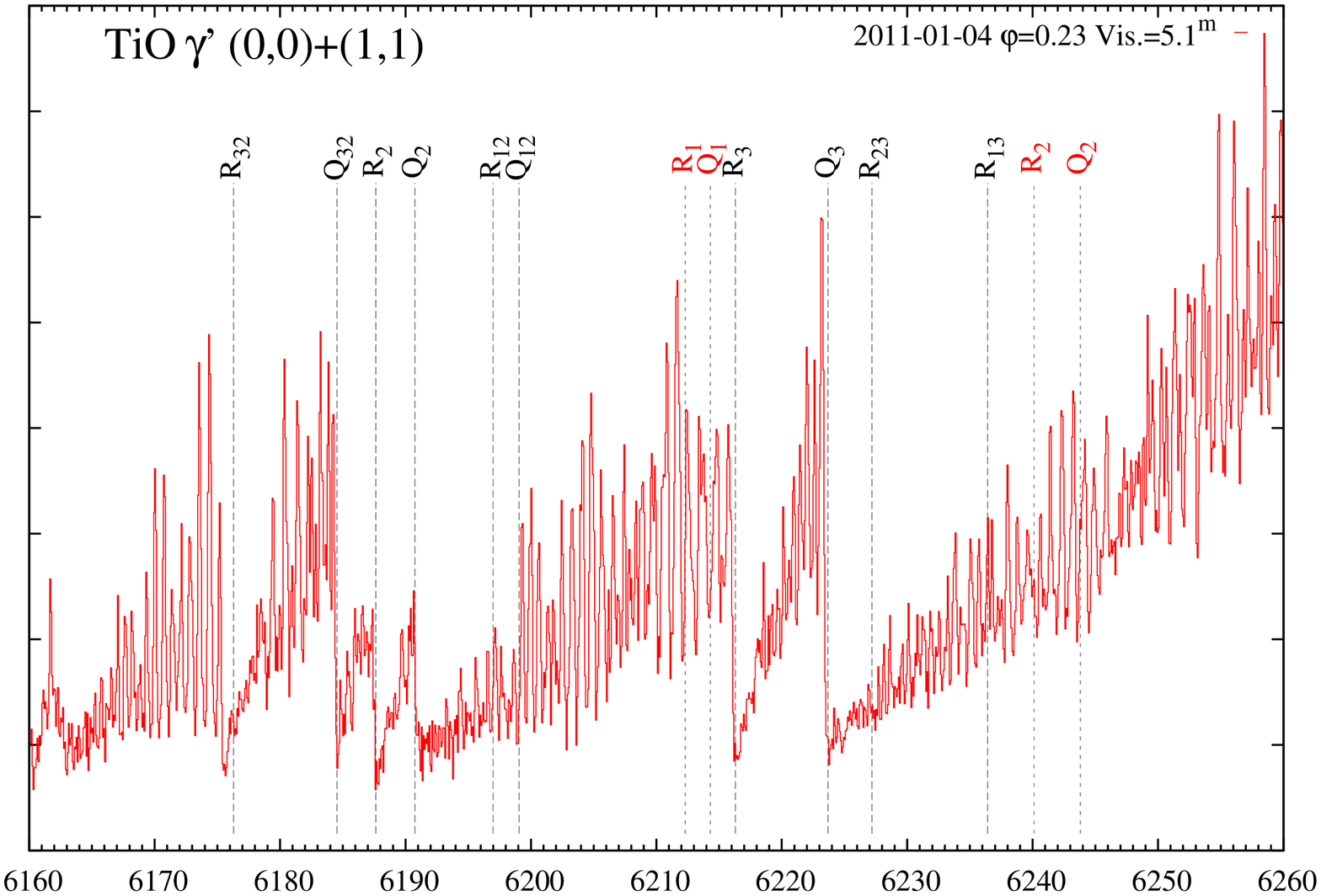}
\caption{Continued.}
\end{figure*}

  \setcounter{figure}{0}%

\begin{figure*} [tbh]
\centering
\includegraphics[angle=270,width=0.85\textwidth]{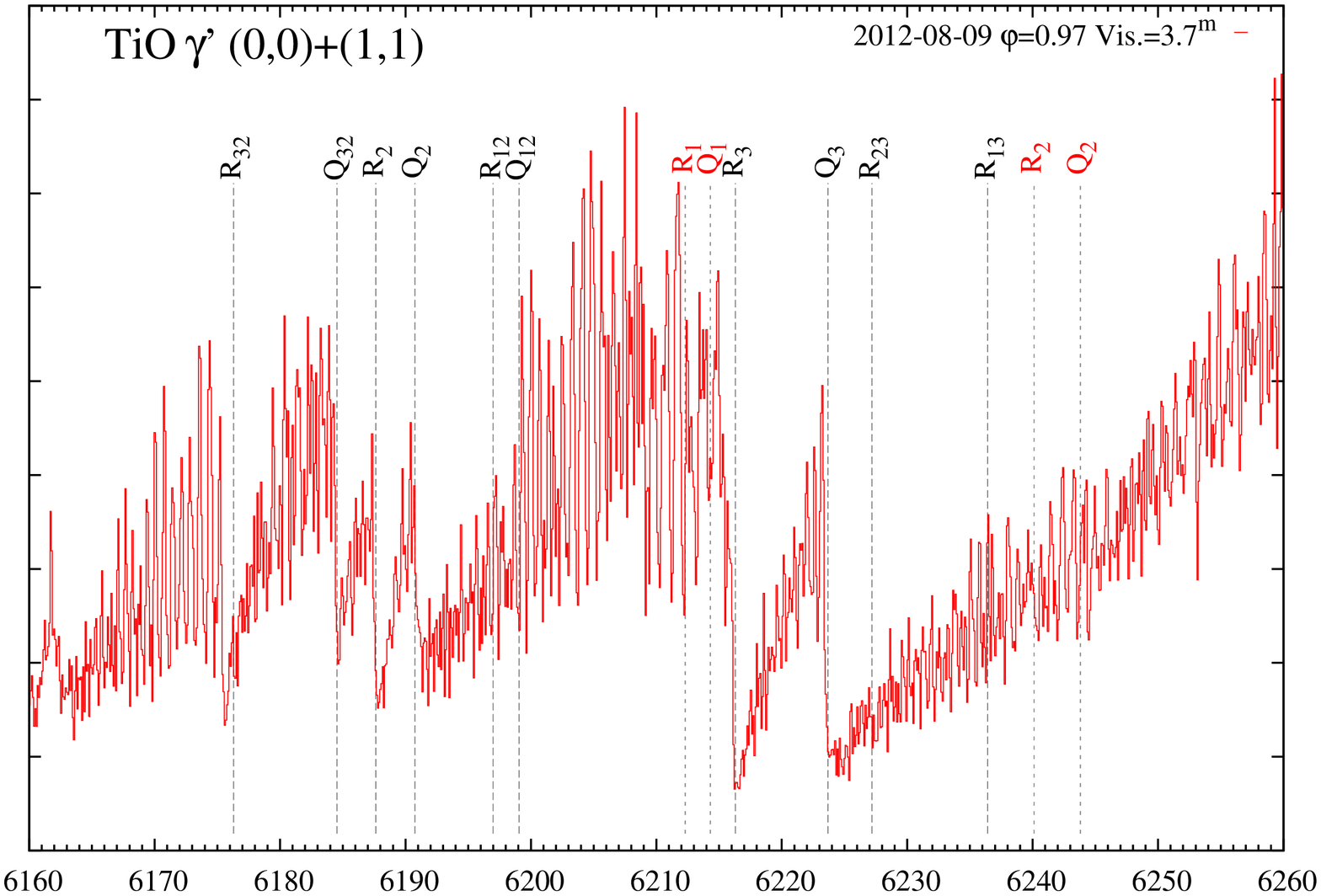}
\includegraphics[angle=270,width=0.85\textwidth]{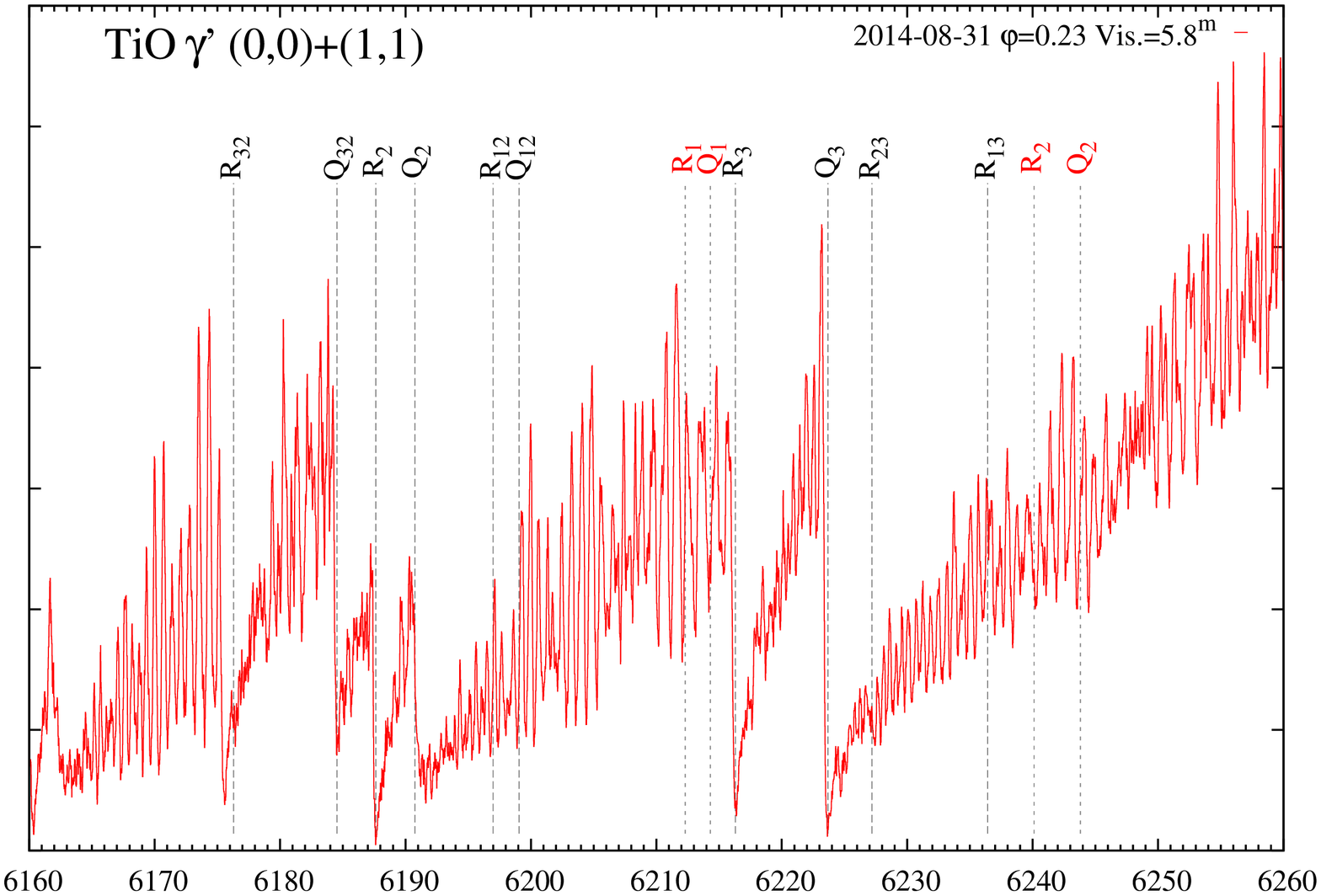}
\caption{Continued.}
\end{figure*}
\end{appendix}

\end{document}